\documentclass[aps,floatfix,showpacs,superscriptaddress,showkeys,preprint]
{revtex4}

\usepackage{titlesec}
\usepackage{graphicx,epsfig,epstopdf}
\usepackage{amssymb,amsmath,amsfonts,latexsym}

\usepackage{bm}
\usepackage{epstopdf}
\titleformat{\section}{\large\bfseries}{\thesection}{1em}{}

\newcommand{\bea}{\begin{eqnarray}}
\newcommand{\ena}{\end{eqnarray}}
\newcommand{\nn}{\nonumber\\}
\newcommand{\be}{\begin{equation}}
\newcommand{\en}{\end{equation}}

\newcommand{\ed}{\end{document}}

\newcommand{\bl}{\bigl}
\newcommand{\br}{\bigr}

\newcommand{\slp}{p\kern-5pt/}
\newcommand{\Tr}{\mbox{\rm{tr}}}

\begin{document}

\title{Exclusive decays $B \to \ell^-\bar\nu$ and $B \to D^{(\ast)}\ell^-\bar\nu$
in the covariant quark model}

\author {M. A. Ivanov}
\email{ivanovm@theor.jinr.ru}
\affiliation{
Bogoliubov Laboratory of Theoretical Physics,
Joint Institute for Nuclear Research,
141980 Dubna, Russia}

\author{J\"{u}rgen G. K\"{o}rner}
\email{koerner@thep.physik.uni-mainz.de}
\affiliation{PRISMA Cluster of Excellence, Institut f\"{u}r Physik, 
Johannes Gutenberg-Universit\"{a}t, 
D-55099 Mainz, Germany}

\author{C. T. Tran}
\email{ctt@theor.jinr.ru,tranchienthang1347@gmail.com}
\affiliation{
Bogoliubov Laboratory of Theoretical Physics,
Joint Institute for Nuclear Research,
141980 Dubna, Russia}
\affiliation{Advanced Center for Physics, Institute of Physics, Vietnam 
Academy of Science and Technology, 100000 Hanoi, Vietnam}
\affiliation{Department of General and Applied Physics, 
Moscow Institute of Physics and Technology, 141700 Dolgoprudny, Russia}

\date{\today}

\begin{abstract}
We study the exclusive leptonic and semileptonic $B$ decays 
$B \to \ell^- \bar\nu_{\ell}$ 
and  $B \to D^{(\ast)} \ell^-\bar\nu_{\ell}$ in the framework of the covariant 
quark model  with built-in infrared confinement. We compute the relevant 
form factors in the full kinematical momentum transfer region. 
The calculated form factors are used to evaluate branching fractions and 
polarization observables of the above transitions. 
We compare our results with experimental data and results from 
other theoretical studies.   
\end{abstract}


\maketitle

\section{Introduction}
\label{sec:intro} 

The decays $B \to \ell^-\bar\nu$ and $B \to D^{(\ast)} \ell^- \bar\nu$ 
($\ell=e,\mu,\tau$) play a prominent role in testing the Standard Model (SM) 
and looking for hints of New Physics (NP) in charged-current interactions. 
In the SM scenario a measurement of these decays provides a direct route 
to determine values of the $B$ meson decay constant $f_B$ and the semileptonic 
form factors. They also help to determine  the Cabibbo-Kobayashi-Maskawa (CKM) 
matrix elements $|V_{ub}|$ and $|V_{cb}|$ to a better precision. A puzzling 
feature of these decays is that there have been some recent hints that lepton
universality is broken in the tauonic modes of these decays.

The leptonic and semileptonic modes are difficult to measure experimentally 
due to 
the presence of a neutrino in the final state. Ideal in this regard
are B-factories where a $B$ meson pair is generated from 
the process $e^+e^- \to \Upsilon(4S) \to B\bar{B}$. One of the $B$ mesons 
($B_{\rm tag}$) is then reconstructed in hadronic or semileptonic modes, 
while signal decays of the other $B$ meson ($B_{\rm sig}$) are identified. 
A new player has entered the game in that the LHCb collaboration has been 
able to identify the semileptonic decays 
$\bar{B}^0 \to D^{\ast} \tau^- \bar{\nu}_{\tau}$ and
$\bar{B}^0 \to D^{\ast} \mu^- \bar{\nu}_{\mu}$ 
in hadronic collisions~\cite{Aaij:2015yra}. 

Since the first evidence reported by Belle collaboration in 
2006~\cite{Ikado:2006un}, many measurements of the branching fraction 
$ \mathcal{B}(B^- \to \tau^- \bar{\nu}_{\tau})$ have been reported by both Belle and BABAR collaborations. There had been a consistent excess compared 
to the SM prediction until Belle published their result of 
$ \mathcal{B}(B^- \to \tau^- \bar{\nu}_{\tau})
=[(7.2^{+2.7}_{-2.5} ({\rm stat}) \pm 1.1 ({\rm syst}))] \times 10^{-5}$  
with a significance of  $3.0 \sigma$ \cite{Hara:2012mm}. This result reduced 
the tension between theory and experiment and decreased the world average of 
the measured branching fraction to the recent value of 
$ \mathcal{B}(B^- \to \tau^- \bar{\nu}_{\tau})
=(11.4 \pm 2.2) \times 10^{-5}$~\cite{Bona:2009cj}, 
which is slightly larger than the SM expectation  $(8.1 \pm 0.7) \times 10^{-5}$ 
obtained from a global fit to CKM matrix elements~\cite{Bona:2009cj}.  
Note that the most recent result of 
$ \mathcal{B}(B^- \to \tau^- \bar{\nu}_{\tau})
=[12.5 \pm 2.8 ({\rm stat}) \pm 2.7 ({\rm syst})] 
\times 10^{-5}$~\cite{Abdesselam:2014hkd} 
reported by Belle in September 2014 is in good agreement with its previous 
result.

The SM calculation of the leptonic decays suffers from uncertainties in the 
input values of $f_B$ and $V_{ub}$. One can eliminate the $V_{ub}$ dependence by 
calculating the ratio of branching fractions
\be
R^{\tau}_{\pi} = \frac{\tau_{\bar{B}^0}}{\tau_{B^-}} 
\frac{\mathcal{B}(B^- \to \tau^- \bar{\nu}_{\tau})}
     { \mathcal{B}(\bar{B}^0 \to \pi^+ \ell^- \bar{\nu}_{\ell})},
\en
where $\ell = \mu, e$. The ratio is measured to be 
$(0.73 \pm 0.15)$~\cite{Fajfer:2012jt}, which exceeds the SM prediction of 
$R^{\tau}_{\pi} =0.31 \pm 0.06$~\cite{Fajfer:2012jt} 
by more than a factor of 2, while the measured value of 
$\mathcal{B}(\bar B^0 \to \pi^+ \ell^- \bar{\nu}_{\ell}) 
=(14.6\pm 0.7)\times 10^{-5}$
~\cite{delAmoSanchez:2010af,Ha:2010rf,Asner:2010qj} 
is consistent with the SM expectation.   

The semileptonic decays $B \to D^{(\ast)} \ell \nu$ have a much richer
structure than the leptonic decays. There is a large number of observables
in these decays, e.g., the forward-backward asymmetry of the charged
lepton. Recently there has been much interest in the ratios 
of branching fractions
\be
R(D^{(\ast)}) \equiv 
\frac{\mathcal{B}(\bar{B}^0 \to D^{(\ast)} \tau^- \bar{\nu}_{\tau})}
      {\mathcal{B}(\bar{B}^0 \to D^{(*)} \ell^- \bar{\nu}_{\ell})}.
\en
In taking these ratios some of the uncertainties in the form factors
are reduced. Furthermore, the dependence on the poorly known CKM matrix 
element $|V_{cb}|$ drops out in the ratio. 
Recently, three groups have reported measurements of these ratios
\[
\begin{array}{ll}
R(D)|_{\rm BABAR} = 0.440 \pm 0.072 \qquad & \qquad
R(D^\ast)|_{\rm BABAR} = 0.332 \pm 0.030 \quad \text{\cite{Lees:2012xj}} 
\\
R(D)|_{\rm BELLE}\,\, = 0.375 \pm 0.069 \qquad  & \qquad
R(D^\ast)|_{\rm BELLE} \,\,= 0.293 \pm 0.041 \quad 
\text{\cite{Huschle:2015rga} }
\\
 & \qquad R(D^\ast)|_{\rm LHCb} \;\;\; = 0.336 \pm 0.040 \quad 
\text{\cite{Aaij:2015yra}}
\\
\end{array}
\]
where the statistical and systematic uncertainties have been combined 
in quadrature. These measurements were combined in \cite{Rotondo:2015FPCP}
\be
R(D)|_{\rm expt} = 0.388 \pm 0.047 ,
\qquad
R(D^\ast)|_{\rm expt} = 0.321 \pm 0.021 ,
\label{eq:RD-expt}
\en
and compared with the SM  expectations given in 
\cite{Lees:2012xj,Tanaka:2010se,Fajfer:2012vx,Kamenik:2008tj}
\be
R(D)|_{\rm SM} = 0.297 \pm 0.017 ,
\qquad
R(D^\ast)|_{\rm SM} = 0.252 \pm 0.003 .
\label{eq:RD-SM}
\en
It is seen that there is a discrepancy of 1.8~$\sigma$
for $R(D)$ and 3.3~$\sigma$ for $R(D^\ast)$.

The deviation of leptonic and semileptonic 
tauonic $B$ meson decays from SM expectations has been the motivation of 
many theoretical studies in search for NP effects, including 
the two-Higgs-doublet models (2HDMs)~\cite{Hou:1992sy,Baek:1999ch,Crivellin:2015hha,Crivellin:2012ye}, 
the minimal supersymmetric standard model (MSSM)~\cite{Martin:1997ns}, and 
leptoquark models~\cite{Buchmuller:1986zs,Calibbi:2015kma}. In many studies, a general 
effective Lagrangian for the $b \to u \ell \nu$ and the $b \to c \ell \nu$ 
transitions in the presence of NP is imposed to investigate various NP 
operators and their coupling, together with their correlations
~\cite{Datta:2012qk,Tanaka:2012nw,Biancofiore:2013ki,Fajfer:2012vx}.

In this paper we focus on these decays within the SM framework using 
results from our covariant constituent quark model for the dynamics of the
transitions. 
Most of the theoretical studies on the semileptonic decays have been relying 
on elements of the Heavy Quark Effective Theory 
(HQET)~\cite{Neubert:1993mb,grozin2004heavy}, based on 
a systematic $1/m_Q$-expansion of the QCD Lagrangian. The leading order 
of the HQET-expansion corresponds to the Heavy Quark Symmetry when the heavy 
quark mass tends to infinity, simplifying the structure of the weak current 
transitions. 
The form factors of these transitions are then expressed through only 
a few universal functions. Unfortunately, HQET can give predictions 
only for the normalization of the form factors at zero recoil. As one moves 
away from the zero-recoil point one has to take recourse to full 
nonperturbative calculations. In this paper, we present a description of 
these decays that does not rely on HQET. We employ the covariant constituent 
quark model (CQM) with built-in infrared confinement which has 
been developed in several previous papers by our group 
(see~\cite{Branz:2009cd,Ivanov:2011aa} and references therein). 
In the CQM approach, the entire physical range of momentum transfer 
is accessible. This is one of those features that make the CQM different from other model approaches 
for the calculation of hadronic quantities. We mention that a similar study was done by  authors of~\cite{Ebert:2006hj,Ebert:2006nz,Faustov:2012nk} in the framework of a relativistic quark model based on the quasipotential approach, in which the full range of momentum transfer is also achievable.  Our aim is to give an independent 
calculation of these decays including the $q^{2}$ behavior of the 
transition form factors, the leptonic 
decay constants of the $B$ and $D$ mesons, the forward-backward asymmetry of 
the lepton and other polarization observables as well as ratios of branching 
fractions.  

\section{Model}
\label{sec:model}

The CQM is based on an effective Lagrangian describing the coupling of a 
hadron $H$ to its constituent quarks, the coupling strength of which is 
determined by the compositeness condition 
$Z_H = 0$~\cite{Salam:1962ap,Weinberg:1962hj}, where $Z_H$ 
is the wave function renormalization constant of the hadron $H$. 
Here $Z^{1/2}_H$ is the matrix element between a physical particle state and 
the corresponding bare state.  For $Z_H=0$ it then follows that 
the physical state does not contain  the bare one and is therefore described 
as a bound state. This does not mean that we can solve the QCD bound state 
equations but we are able to show that the compositeness condition 
provides an effective and self-consistent way to describe the coupling 
of a particle to its constituents. 

One starts with an effective Lagrangian written down in terms of quark and 
hadron variables \cite{Efimov:1988yd,Efimov:zg}. Then, by using 
Feynman rules, the $S$-matrix elements describing hadronic interactions 
are derived from a set of quark diagrams. In particular, the compositeness 
condition enables one to avoid a double counting of hadronic degrees of 
freedom. This approach is self-consistent and all calculations of physical 
observables are straightforward. There is a small set of model parameters: 
the constituent quark masses, the scale parameters 
that define the size of the constituent quarks distribution inside 
a given hadron, and the infrared cutoff parameter $\lambda$.

The coupling of a meson $M$ to its constituent quarks $q_1$ and $\bar q_2$ 
is given by the Lagrangian
\be
\label{eq:lag}
{\cal L}_{\rm int}(x)=g_M\,M(x)\cdot J_M (x) + {\rm H.c.},
\en
where $g_M$ denotes the coupling strength of the meson with its constituent 
quarks. The interpolating quark current in~(\ref{eq:lag}) is taken to be

\be\label{eq:current}
J_M (x)=\int\!\! dx_1\!\! \int\!\! dx_2\, F_M (x;x_1,x_2)\, 
\bar{q}_2 (x_2)\,\Gamma_M\, q_1(x_1),
\en
where the Dirac matrix $\Gamma_M$ projects onto the relevant meson state, 
i.e., $\Gamma_M=I$ for a scalar meson, $\Gamma_M=\gamma^5$ for a 
pseudo-scalar meson, and 
$\Gamma_M=\gamma^{\mu}$ for a vector meson. The vertex function $F_M$ is 
related to the scalar part of the Bethe-Salpeter amplitude and characterizes 
the finite size of the meson. We adopt the following form for the vertex
function
\be
F_M (x;x_1,x_2)=\delta(x - w_1 x_1 - w_2 x_2) \Phi_M((x_1-x_2)^2),
\label{eq:vertex}
\en 
where $w_i = m_{q_i}/(m_{q_1}+m_{q_2})$ so that $w_1+w_2=1$. 
This form of $F_M$ is invariant under 
the translation $F_M(x+a;x_1+a,x_2+a)=F_M(x;x_1,x_2)$, which is a necessary
condition to provide  the Lorentz invariance of the Lagrangian~(\ref{eq:lag}).

In order to simplify the calculations we adopt a Gaussian form for the vertex 
function as follows:
\be
\widetilde\Phi_M(-p^2) = \int\! dx\, e^{ipx} \Phi_M(x^2)
= e^{p^2/\Lambda^2_M},
\label{eq:Gauss}
\en
where the parameter $\Lambda_M$ characterizes the meson size.
Calculations of Feynman diagrams proceed
in the Euclidean region where $p^2=-p^2_E$, in which
the vertex function has the appropriate falloff
behavior to provide for the ultraviolet convergence of the loop integral.

In the evaluation of the quark-loop diagrams we use
the free local fermion propagator of the constituent quark
\be
\label{eq:prop}
S_q(k) = \frac{1}{ m_q-\not\! k -i\epsilon } = 
\frac{m_q + \not\! k}{m^2_q - k^2  -i\epsilon }
\en 
with an effective constituent quark mass $m_q$.

\begin{figure}[htbp]
\includegraphics[scale=0.6]{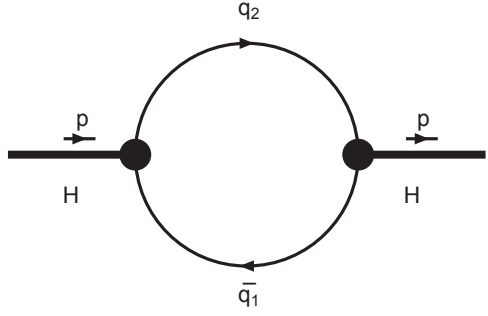}
\caption{One-loop self-energy diagram for a meson.}
\label{fig:mass}
\end{figure}

For the evaluation of the compositeness condition, we consider the meson mass 
function defined by the diagram in Fig.~\ref{fig:mass}.
One has
\bea
\widetilde\Pi_P(p^2) &=& N_c g_{P}^{2}
\int\frac{d^4k}{(2\pi)^4i} \widetilde\Phi^2_P(-k^2)
\Tr\Big(\gamma^5 S_1(k+w_1 p)\gamma^5 S_2(k-w_2 p)\Big),
\label{eq:massP}\\[2ex]
\widetilde\Pi^{\mu\nu}_V(p^2) &=&N_c g_{V}^{2} 
\int\frac{d^4k}{(2\pi)^4i} \widetilde\Phi^2_V(-k^2)
\Tr\Big(\gamma^\mu S_1(k+w_1 p)\gamma^\nu S_2(k-w_2 p)\Big),
\label{eq:massV}
\ena
where $N_c=3$ is the number of colors. Since the vector meson is
on its mass-shell one has $\epsilon_V\cdot p=0$ and
one needs only the part of the vector meson function proportional
to $g_{\mu\nu}$. It is given by 
\be
\widetilde\Pi_V(p^2) = \frac13 \bl(g_{\mu\nu}-\frac{p_\mu p_\nu}{p^2} \br)
                       \widetilde\Pi^{\mu\nu}_V(p).  
\en
The coupling constant $g_M$ in Eq.~(\ref{eq:lag}) is determined 
by the compositeness condition which is written in the form
\be
Z_M = 1 - \widetilde\Pi'_M(m^2_M) = 0,
\label{eq:Z=0}
\en
where $ \widetilde\Pi'_M(p^2)$ is the derivative of the mass operator
taken on the mass-shell $p^2=m^2_M$.
It is convenient to calculate the derivatives of the meson mass functions
by using the following identities
\bea
\frac{d}{dp^2} \widetilde\Pi_M(p^2) &=& 
\frac{1}{2p^2} p^\mu\frac{d}{dp^\mu} \widetilde\Pi_M(p^2),
\nn
p^\mu\frac{d}{dp^\mu} S(k + w p)&=& w\, S(k + w p)\not\! p\, S(k + w p).
\label{eq:identity}
\ena
Accordingly the  derivatives of the meson mass functions 
can be written as
\bea
\widetilde\Pi'_P(p^2)&=&
\frac{1}{2p^2}\,\frac{3g^2_P}{4\pi^2}\int\!\! \frac{dk}{4\pi^2i}
\widetilde\Phi^2_P \left(-k^2\right)
\nn
&\times& 
\Big\{ w_1\,\Tr\left[ S_1(k+w_1p)\not\!p \,S_1(k+w_1p)\gamma^5 
S_2(k-w_2p)\gamma^5\right]
\nn
&& - w_2\, \Tr\left[ S_1(k+w_1p)\gamma^5 S_2(k-w_2p)\not\!p \,
S_2(k-w_2p)\gamma^5\right]
\Big\},
\label{eq:primeP}\\[1.5ex]
\widetilde\Pi'_V(p^2)&=&
\frac{1}{2p^2}\,\frac13\left(g^{\mu\nu} - \frac{p^\mu p^\nu}{p^2}\right)
\frac{3g^2_V}{4\pi^2}
\int\!\! \frac{dk}{4\pi^2i}\widetilde\Phi^2_V \left(-k^2\right) 
\nn
&\times& 
\Big\{ w_1\,\Tr\left[ S_1(k+w_1p)\not\!p \,S_1(k+w_1p)\gamma_\mu 
S_2(k-w_2p)\gamma_\nu\right]
\nn
&& -w_2\,\Tr\left[ S_1(k+w_1p)\gamma_\mu S_2(k-w_2p)\not\!p \,
S_2(k-w_2p)\gamma_\nu\right]\Big\}.
\label{eq:primeV}
\ena

The loop integrations in Eqs.~(\ref{eq:primeP}) and ~(\ref{eq:primeV})
are done with the help of the Fock-Schwinger representation of quark
propagators
\bea
S_q (k+w p) &=& \frac{1}{ m_q-\not\! k- w \not\! p } 
=  \frac{m_q + \not\! k + w \not\! p}{m^2_q - (k+w p)^2}
\nn
&=& (m_q + \not\! k + w \not\! p)\int\limits_0^\infty \!\!d\alpha\, 
e^{-\alpha [m_q^2-(k+w p)^2]}.
\label{eq:Fock}
\ena
As will be described later, the use of the Fock-Schwinger representation allows 
one to do tensor loop integrals in a very efficient way since one can
convert loop momenta into derivatives of the exponent function 
(see, e.g.,~\cite{Faessler:2001mr,Anastasiou:1999bn,Fiorentin:2015vha}). 

As mentioned above, all loop integrations are carried out in Euclidean space. 
The transition from Minkowski space to Euclidean space is performed
by using the Wick rotation
\be
k_0=e^{i\frac{\pi}{2}}k_4=ik_4
\label{eq:Wick}
\en
so that $k^2=k_0^2-\vec{k}^2=-k_4^2-\vec{k}^2=-k_E^2 \leq 0.$
Simultaneously one has to rotate all external momenta, i.e.
 $p_0 \to ip_4$ so that $p^2=-p_E^2 \leq 0$.
Then the quadratic form in Eq.~(\ref{eq:Fock}) becomes positive-definite,
\[
m^2_q-(k+w p)^2=m^2_q + (k_E+w p_E)^2>0,
\]
and the integral over $\alpha$ is absolutely convergent.
We will keep the Minkowski notation to avoid
excessive relabeling. We simply imply that
 $k^2 \leq 0$ and $p^2 \leq 0$.

Collecting the representations of the vertex functions
and quark propagators given by Eqs.~(\ref{eq:Gauss})
and (\ref{eq:Fock}), respectively, one can perform the Gaussian
integration in the derivatives of the mass functions 
in  Eqs.~(\ref{eq:massP}) and (\ref{eq:massV}).
The exponent has the form $ak^2+2kr+z_0$, where $r=b\,p$. 
Using the following properties ($k$ is the loop momentum)
\be
\left.
\begin{aligned}
    k^\mu\, \exp(ak^2+2kr+z_0) &=\frac{1}{2}\frac{\partial }
{\partial r_\mu }\exp(ak^2+2kr+z_0)\\
    k^\mu k^\nu\, \exp(ak^2+2kr+z_0) &=
      \frac{1}{2}\frac{\partial }{\partial r_\mu } 
      \frac{1}{2} \frac{\partial }{\partial r_\nu }         \exp(ak^2+2kr+z_0)
\\
\text{etc.}&
\end{aligned}
\right\},
\label{eq:change-to-r}
\en
one can replace
$\not\! k $ by $ {\not\! \partial}_r 
= \gamma^\mu\frac{\partial}{\partial r_\mu}$
which allows one to exchange the tensor integrations
for a differentiation of the Gaussian exponent. For example,
Eq.~(\ref{eq:massP}) now has the form
\be
\widetilde\Pi_P(p^2) = \frac{3g^2_P}{16\pi^2}
\int\limits_0^\infty\!\! \int\limits_0^\infty\!
\frac{d\alpha_1 d\alpha_2}{a^2} 
\,
\Tr\left[\gamma^5 (m_1+{\not\! \partial}_r+w_1\not\!p)\gamma^5 
(m_2+{\not\! \partial}_r-w_2\not\!p)\right]
e^{-\frac{r^2}{a}+z_0}.
\label{eq:Pmass2}\,
\en
The $r$-dependent Gaussian exponent $e^{-r^2/a}$ can be moved to the left 
through the differential operator $\not\! \partial_r$ by using the following 
properties
\bea
\frac{\partial}{\partial r_\mu}\,e^{-r^2/a} &=& e^{-r^2/a}
\left[-\frac{2r^\mu}{a}+\frac{\partial}{\partial r_\mu}\right],
\nn[1.2ex]
\frac{\partial}{\partial r_\mu}\,
\frac{\partial}{\partial r_\nu}\,e^{-r^2/a} &=& e^{-r^2/a}
\left[-\frac{2r^\mu}{a}+\frac{\partial}{\partial r_\mu}\right]\cdot
\left[-\frac{2r^\nu}{a}+\frac{\partial}{\partial r_\nu}\right],
\nn[1.2ex]
\text{etc.}&&
\label{eq:dif}
\ena
Finally, one has to move the derivatives to the right by using
the commutation relation
\be
\left[\frac{\partial}{\partial r_\mu},r^\nu \right]
=g^{\mu\nu}.
\label{eq:comrel}
\en
The last step has been done by using a \textsc{form} code which
works for any numbers of loops and propagators.
In the remaining integrals over the Fock-Schwinger parameters 
$0\le \alpha_i<\infty$
we introduce an additional integration which converts the set of 
Fock-Schwinger parameters into a simplex. Using the transformation
\be
\prod\limits_{i=1}^n\int\limits_0^{\infty} 
\!\! d\alpha_i f(\alpha_1,\ldots,\alpha_n)
=\int\limits_0^{\infty} \!\! dtt^{n-1}
\prod\limits_{i=1}^n \int\!\!d\alpha_i 
\delta\left(1-\sum\limits_{i=1}^n\alpha_i\right)
  f(t\alpha_1,\ldots,t\alpha_n)
\label{eq:simplex}  
\en
one finds
\bea
\widetilde\Pi'_M(p^2) &=& \frac{3g^2_M}{4\pi^2}\int\limits_0^{\infty} \!\! 
\frac{dt\,t}{a_M^2} \int\limits_0^1\!\!d\alpha\,
e^{-t\,z_0 + z_M}\,f_M(t,\alpha),
\label{eq:prime_fin}\\[2ex]
z_0 &=&  \alpha m^2_{q_1} +(1-\alpha)m^2_{q_2} - \alpha(1-\alpha) p^2,
\nn 
z_M &=& \frac{2s_Mt}{2s_M+t} (\alpha-w_2)^2 p^2,
\nn[2ex] 
a_M &=& 2s_M+t , \qquad b = (\alpha-w_2)t. 
\nonumber
\ena
The function $f_M(t,\alpha)$ arises from the trace evaluation.
Further, we have introduced the parameter  $s_M=1/\Lambda^2_M$.

It is readily seen that the integral over $t$ in Eq.~(\ref{eq:prime_fin})
is well defined and convergent if $z_0>0$, i.e.
below the threshold  $p^2< (m_{q_1} + m_{q_2})^2$.  
The convergence of the integral in the case of negative
values of $z_0\le 0$, i.e. above  threshold  
$p^2\ge (m_{q_1} + m_{q_2})^2$,
is guaranteed by the addition of a small imaginary to the quark
mass, i.e. $m_q\to m_q - i\epsilon, \quad \epsilon>0$
in the quark propagator Eq.~(\ref{eq:prop}). It allows one to rotate
the integration variable $t$ to the imaginary axis $t\to i t$. 
As a result the integral Eq.~(\ref{eq:prime_fin}) 
becomes convergent but obtains an imaginary part corresponding to
quark pair production.

However, by cutting the scale integration at the upper limit 
corresponding to the introduction of an infrared cutoff
\be
\int\limits_0^\infty dt (\ldots) \to \int\limits_0^{1/\lambda^2} dt (\ldots),
\label{eq:conf}
\en
one can remove all possible thresholds present in the initial quark
diagram \cite{Branz:2009cd}. Thus the infrared cutoff parameter 
$\lambda$ effectively guarantees the confinement of quarks within hadrons. 
This method is quite general and can be used for diagrams with an arbitrary 
number of loops and propagators. 
In the CQM the infrared cutoff parameter $\lambda$ is taken to be universal 
for all physical processes.

\section{Leptonic B-meson decays}
\label{sec:lepdecays}

The model parameters are determined by fitting calculated quantities 
of basic processes to available experimental data or 
lattice simulations (for details, see Ref.~\cite{Branz:2009cd}, 
where a different set of weak and electromagnetic decays has been used).
In this paper we will use the updated least-squares fit
performed in Refs.~\cite{Gutsche:2015mxa,Ganbold:2014pua,Issadykov:2015iba}.
In this fit we have also updated some of the theoretical/experimental 
input values. The infrared cutoff parameter $\lambda$ of the model has 
been kept fixed. The numerical values of the constituent quark masses
and the parameter $\lambda$ are given by (all in GeV)
\be
\def\arraystretch{1.2}
\begin{array}{cccccc}
     m_u        &      m_s        &      m_c       &     m_b & \lambda     
\\\hline
 \ \ 0.241\ \   &  \ \ 0.428\ \   &  \ \ 1.67\ \   &  \ \ 5.04\ \   & 
\ \ 0.181\ \    
\end{array}
.
\label{eq: fitmas}
\en
Our prime goal is to study the pure leptonic B meson decays as well as
the semileptonic $B\to D^{(\ast)} \ell\bar\nu_\ell$ decays.
The most recent results of the fit for those parameters 
involved in this paper are taken from
our papers \cite{Gutsche:2015mxa,Ganbold:2014pua,Issadykov:2015iba} (all in GeV): 
\be
\def\arraystretch{1.2}
\begin{tabular}{c c c c c c c c c  }
 \ \ $\Lambda_{D^*}$ \ \ & \ \  $\Lambda_{D^*_s}$ \ \  &
 \ \   $\Lambda_D$  \ \  & \ \ $\Lambda_{D_s}$ \ \  & \ \  $\Lambda_{B^*_s}$ \ \  
 & \ \  $\Lambda_{B^*}$  \ \  &  \ \ $\Lambda_B$  \ \ &  
\ \ $\Lambda_{B_s}$ \ \  & \ \  $\Lambda_{B_c}$ \ \   
\\\hline
1.53 & 1.56 & 1.60 & 1.75 & 1.79 & 1.81 & 1.96 & 2.05 & 2.73 \\
\end{tabular}
.
\label{eq:sizeparam}
\en

\begin{figure}[htbp]
\includegraphics[scale=0.6]{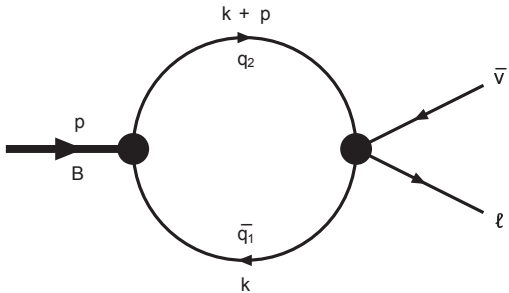}
\caption{Quark model diagram for the B meson leptonic decay.}
\label{fig:leptonic}
\end{figure}

The matrix elements of the leptonic decays are described by
the Feynman diagram shown in Fig.~\ref{fig:leptonic}.
The leptonic decay constants of the pseudoscalar and vector mesons are 
defined~by
\bea
N_c\, g_P\! \int\!\! \frac{d^4k}{ (2\pi)^4 i}\, \widetilde\Phi_P(-k^2)\,
{\rm tr} \biggl[O^{\,\mu} S_1(k+w_1 p) \gamma^5 S_2(k-w_2 p) \biggr] 
&=&f_P p^\mu ,
\nn
N_c\, g_V\! \int\!\! \frac{d^4k}{ (2\pi)^4 i}\, \widetilde\Phi_V(-k^2)\,
{\rm tr} \biggl[O^{\,\mu} S_1(k+w_1 p)\not\!\epsilon_V  S_2(k-w_2 p) \biggr] 
&=& m_V f_V \epsilon_V^\mu,
\label{eq:lept}
\ena
where $N_c=3$ is the number of colors,  and $O^\mu=\gamma^\mu(1-\gamma_5)$ is 
the weak Dirac matrix with left chirality. The mesons are taken on 
their mass-shells. The calculation of the matrix elements (\ref{eq:lept})
proceeds in a way similar to the case of the mass functions.

Our results for the leptonic decay constants of $B^{(*)}_{(s)}$ and 
$D^{(*)}_{(s)}$ 
mesons are given in Table~\ref{tab:decayconst}. For comparison, we also list 
the values of these constants obtained from experiments, Lattice and QCD sum 
rules. Our results show good agreement (within $10\%$) with results of 
the other studies. We mention that early attempts to account for flavor symmetry 
breaking in pseudoscalar meson decay constants were done 
in~\cite{Gershtein:1976aq,Khlopov:1978id}. 
\begin{table}[htbp]
\begin{tabular}{c c c c}
\hline\hline
{} & This work & Other & Ref.\\
\hline
$f_B$ & 193.1 & 190.6$\pm$4.7 & PDG~\cite{Agashe:2014kda}\\
$f_{B_s}$ & 238.7 & 242.0(9.5) & LAT~\cite{Bazavov:2011aa}\\
{} & {} & 259(32) & HPQCD LAT~\cite{Gray:2005ad}\\
{} & {} & 193(7) & LAT~\cite{DellaMorte:2007ij}\\
$f_{B_c}$ & 489.0 & $489\pm4\pm3$ & LAT~\cite{Chiu:2007km}\\
$f_{B^*}$ & 196.0 & $196(24)^{+39}_{-2}$ & LAT~\cite{Becirevic:1998ua}\\
{} & {} & $186.4\pm3.2$ & QCDSR~\cite{Lucha:2014nba}\\
$f_{B^*_s}$ & 229.0 & $229(20)^{+41}_{-16}$ & LAT~\cite{Becirevic:1998ua} \\
{} & {} & $215.2\pm3.0$ & QCD SR~\cite{Lucha:2014nba}\\
$f_{B_s}/f_B$ & 1.236 & 1.20(3)(1) & HPQCD LAT~\cite{Gray:2005ad} \\
{} & {} & 1.229(26) & LAT~\cite{Bazavov:2011aa}\\
$f_D$ & 206.1 & 204.6$\pm$5.0 & PDG~\cite{Agashe:2014kda} \\
$f_{D^*}$ & 244.3 & $278\pm13\pm10$ & LAT~\cite{Becirevic:2012ti}\\
{} & {} & $245(20)^{+3}_{-2}$ & LAT~\cite{Becirevic:1998ua} \\
{} & {} & $252.2\pm22.3\pm4$ & QCD SR~\cite{Lucha:2014xla}\\
$f_{D_s}$ & 257.5 & 257.5$\pm$4.6 & PDG~\cite{Agashe:2014kda}\\
$f_{D^*_s}$ & 272.0 & 311$\pm$9 & LAT~\cite{Becirevic:2012ti}\\
{} & {} & $272(16)^{+3}_{-20}$ & LAT~\cite{Becirevic:1998ua}\\
{} & {} & $305.5\pm26.8\pm5$ & QCD SR~\cite{Lucha:2014xla}\\
$f_{D_s}/f_D$ & 1.249 & 1.258$\pm$0.038 & PDG~\cite{Agashe:2014kda}\\
\hline\hline
\end{tabular}
\caption{Results for the leptonic decay constants $f_H$ (in MeV).}
\label{tab:decayconst}
\end{table}

In the SM, the purely leptonic decays $B^- \to \ell^- \bar{\nu}_{\ell}$ 
proceed via the annihilation of the quark-pair into an off-shell $W$ boson.
The branching fraction for the leptonic decays is given by
\be
 \mathcal{B}(B^- \to \ell^- \bar{\nu}_{\ell})=
\frac{G_F^2}{8\pi}m_Bm_{\ell}^2\left(1-\frac{m_{\ell}^2}{m_B^2}\right)^2
f_B^2|V_{ub}|^2\tau_B,
\en
where $G_F$ is the Fermi coupling constant, $m_B$ and $m_{\ell}$ are the $B$ 
meson and lepton masses, respectively, and $\tau_B$ is the $B$ meson life time. 
The expected branching fractions are $O(10^{-4})$, $O(10^{-7})$, and 
$O(10^{-11})$ for $\ell = \tau, \mu,$ and $e$, respectively. 
The different lepton masses affect the values of the
branching fractions through the helicity flip factor 
$\left(1-m_\ell^2/m_B^2\right)^2$.
\begin{table}
\begin{center}
\def\arraystretch{1.1}
\begin{tabular}{lccr}
\hline\hline
 & This work & Data & Ref.\\
\hline
\quad $B^- \to e^- \bar{\nu}_e $ \quad 
 & \quad $ 1.16\cdot 10^{-11}  $ \quad  
 & \quad $< 9.8 \cdot 10^{-7}  $ \quad 
 & \quad PGD~\cite{Agashe:2014kda} \quad 
\\
\quad\quad   &  \quad \quad   
&\quad  $ (0.88 \pm 0.12)\cdot 10^{-11} $ \quad 
& \quad  UTfit~\cite{Bona:2009cj}\quad 
\\
\quad \quad  & \quad\quad     
& \quad $ (0.85\pm 0.27 )\cdot 10^{-11} $\quad  
&\quad  CKMfitter~\cite{Charles:2004jd} \quad 
\\[1.2ex]
\quad $B^- \to {\mu}^- \bar{\nu}_{\mu}$ \quad 
 & \quad   $ 0.49\cdot 10^{-6} $ \quad 
 & \quad  $< 1.0$ $\cdot 10^{-6}$ \quad 
 & \quad  PGD~\cite{Agashe:2014kda} \quad 
\\ 
\quad\quad   &\quad \quad     
 &\quad  $ (0.38 \pm 0.05)\cdot 10^{-6} $\quad  
 & \quad  UTfit~\cite{Bona:2009cj}\quad 
\\
\quad\quad   &  \quad \quad   
 &\quad  $ (0.37 \pm 0.02)\cdot 10^{-6} $ \quad 
 & \quad  CKMfitter~\cite{Charles:2004jd}\quad 
\\[1.2ex]
\quad $B^- \to {\tau}^- \bar{\nu}_{\tau}$ \quad 
 & \quad  $ 1.10 \cdot 10^{-4} $ \quad 
 &  \quad $ (1.14 \pm 0.27)\cdot 10^{-4}$ \quad 
 &  \quad PGD~\cite{Agashe:2014kda} \quad \\
\hline\hline
\end{tabular}
\end{center}
\caption{Leptonic B-decay branching fractions.}
\label{tab:leptBr}
\end{table}

\section{Form factors of semileptonic B-meson decays}

\begin{figure}[htbp]
\includegraphics[scale=0.5]{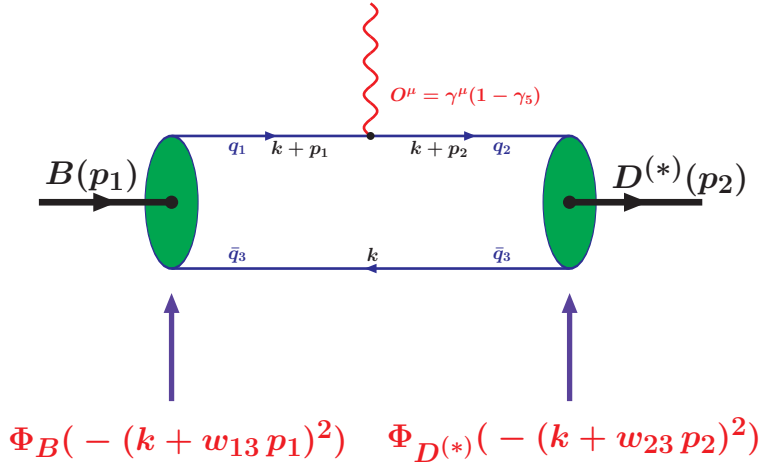}
\caption{Quark model diagram for B meson semileptonic decay.}
\label{fig:semilept}
\end{figure}

The invariant matrix element of the semileptonic decays 
$B\to D^{(\ast)} \ell^- \bar\nu_\ell$  can be written as
\be
M(B\to D^{(\ast)} \ell^- \bar\nu_\ell )  =  
\frac{G_F}{\sqrt{2}} V_{cb}
 <D^{(\ast)}\,|\,\bar{c}\,O^\mu\, b\,|\,B>\,
\bar\ell O_\mu \nu_\ell  ,
\en 
where the matrix elements of the semileptonic $B \to D^{(\ast)}$ transitions 
in the covariant quark model are defined by the diagram 
in Fig.~\ref{fig:semilept} 
and are written as
\bea
&&
T^\mu \equiv
\langle D(p_2)\,
|\,\bar c\, O^{\,\mu}\, b\,
| B(p_1) \rangle
\,=\,
\nn[1.2ex]
&=&
N_c\, g_B\,g_D\!\!  \int\!\! \frac{d^4k}{ (2\pi)^4 i}\, 
\widetilde\Phi_B\Big(-(k+w_{13} p_1)^2\Big)\,
\widetilde\Phi_{D}\Big(-(k+w_{23} p_2)^2\Big)
\nn
&\times&
\Tr \biggl[
O^{\,\mu}\, S_1(k+p_1)\, \gamma^5\, S_3(k)\, \gamma^5\, S_2(k+p_2) 
\biggr]
\nn[1.2ex]
 & = & F_+(q^2)\, P^{\,\mu} + F_-(q^2)\, q^{\,\mu},\quad \text{and}
\label{eq:PP'}\\[1.5ex]
&&
\epsilon^\dagger_{2\,\alpha} T^{\mu\alpha} \equiv 
\langle D^\ast(p_2,\epsilon_2)\,
|\,\bar c\, O^{\,\mu}\,b\, |\,B(p_1)
\rangle 
\,=\,
\nn[1.2ex]
&=&
N_c\, g_B\,g_{D^\ast} \!\! \int\!\! \frac{d^4k}{ (2\pi)^4 i}\, 
\widetilde\Phi_B\Big(-(k+w_{13} p_1)^2\Big)\,
\widetilde\Phi_{D^\ast}\Big(-(k+w_{23}p_2)^2\Big)
\nn
&\times&
\Tr \biggl[ 
O^{\,\mu} \,S_1(k+p_1)\,\gamma^5\, S_3(k) \not\!\epsilon_2^{\,\,\dagger} \,
S_2(k+p_2)\, \biggr]
\nn[1.2ex]
 & = &
\frac{\epsilon^{\,\dagger}_{2\,\alpha}}{m_1+m_2}\,
\left( - g^{\mu\alpha}\,Pq\,A_0(q^2) + P^{\,\mu}\,P^{\,\alpha}\,A_+(q^2)
       + q^{\,\mu}\,P^{\,\alpha}\,A_-(q^2) 
+ i\,\varepsilon^{\mu\alpha P q}\,V(q^2)\right).
\label{eq:PV}
\ena
Here, $P=p_1+p_2$, \,$q=p_1-p_2$, and $\epsilon_2$ is the polarization vector
of the $D^\ast$ meson so that $\epsilon_2^\dagger\cdot p_2=0$.
The particles are on their mass-shells: $p_1^2=m_1^2=m_B^2$ and
 $p_2^2=m_2^2=m_{D^{(\ast)}}^2$.
Altogether there are three flavors of quarks involved in these
processes. We therefore introduce a notation with two subscripts
$w_{ij}=m_{q_j}/(m_{q_i}+m_{q_j})$ $(i,j=1,2,3)$ such that $w_{ij}+w_{ji}=1$. 
In our case one has $q_1=b$,  $q_2=c$, and $q_3=d$.

Our numerical results for the form factors are well represented
by a double--pole parametrization
\be
F(q^2)=\frac{F(0)}{1 - a s + b s^2}, \quad s=\frac{q^2}{m_1^2}. 
\label{eq:DPP}
\en 
The double--pole approximation is quite accurate. The error
relative to the exact results is less than $1\%$ over the entire $q^{2}$ 
range.
For the $B \to D^{(\ast)}$ transition the parameters of the dipole
approximation are given by
\be
\begin{array}{c|rr|rrrr}
 & \quad F_+ \quad & \quad  F_- \quad & \quad A_0 \quad  & \quad A_+ \quad & 
  \quad A_- \quad & \quad V \quad \\[1.1ex]
\hline
F(0) &  0.78   & -0.36 &  1.62 & 0.67  & -0.77 & 0.77
\\[1ex] 
a    &  0.74   &  0.76 &  0.34 & 0.87  &  0.89 & 0.90
\\[1ex] 
b    &  0.038  & 0.046 & -0.16 & 0.057 & 0.070 & 0.075.
\\[1.1ex] 
\hline
\end{array}
\label{eq:ff_param}
\en
\vspace{1.2ex}

\begin{figure}[htbp]
\begin{tabular}{lr}
\includegraphics[scale=0.6]{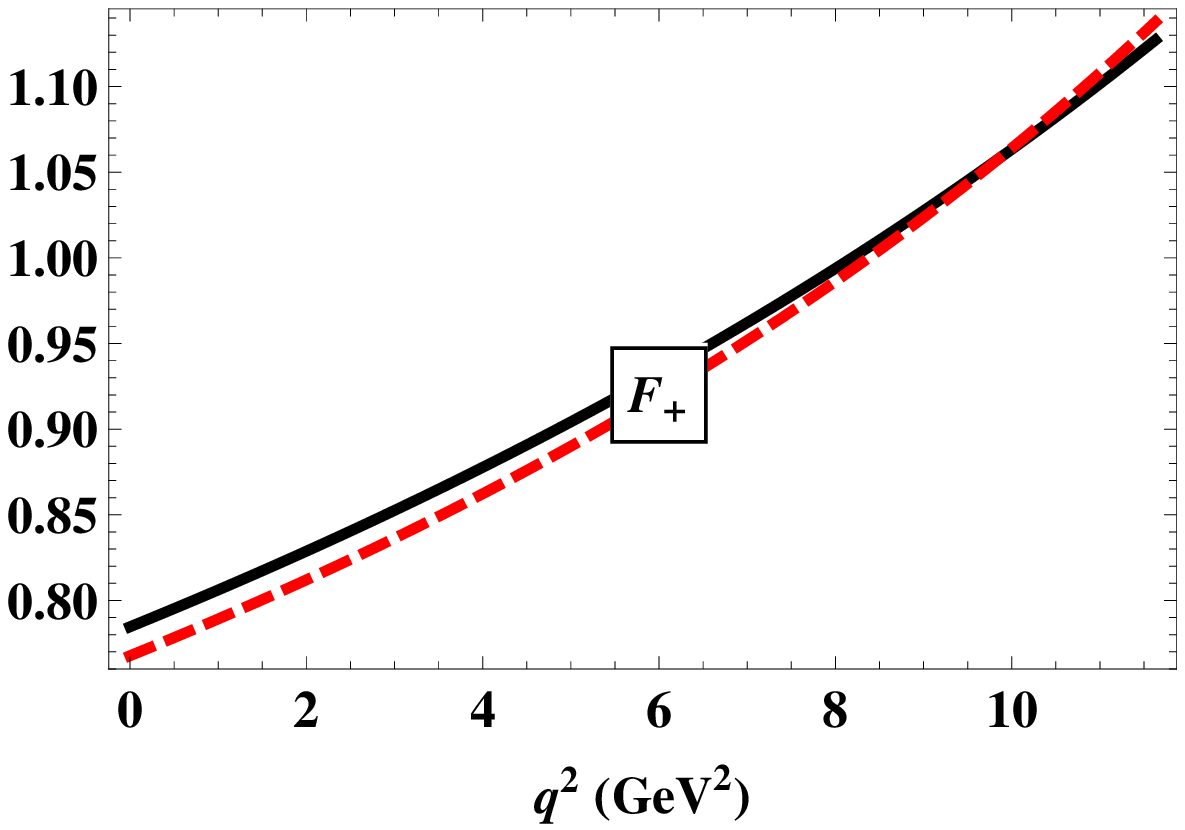} &
\includegraphics[scale=0.6]{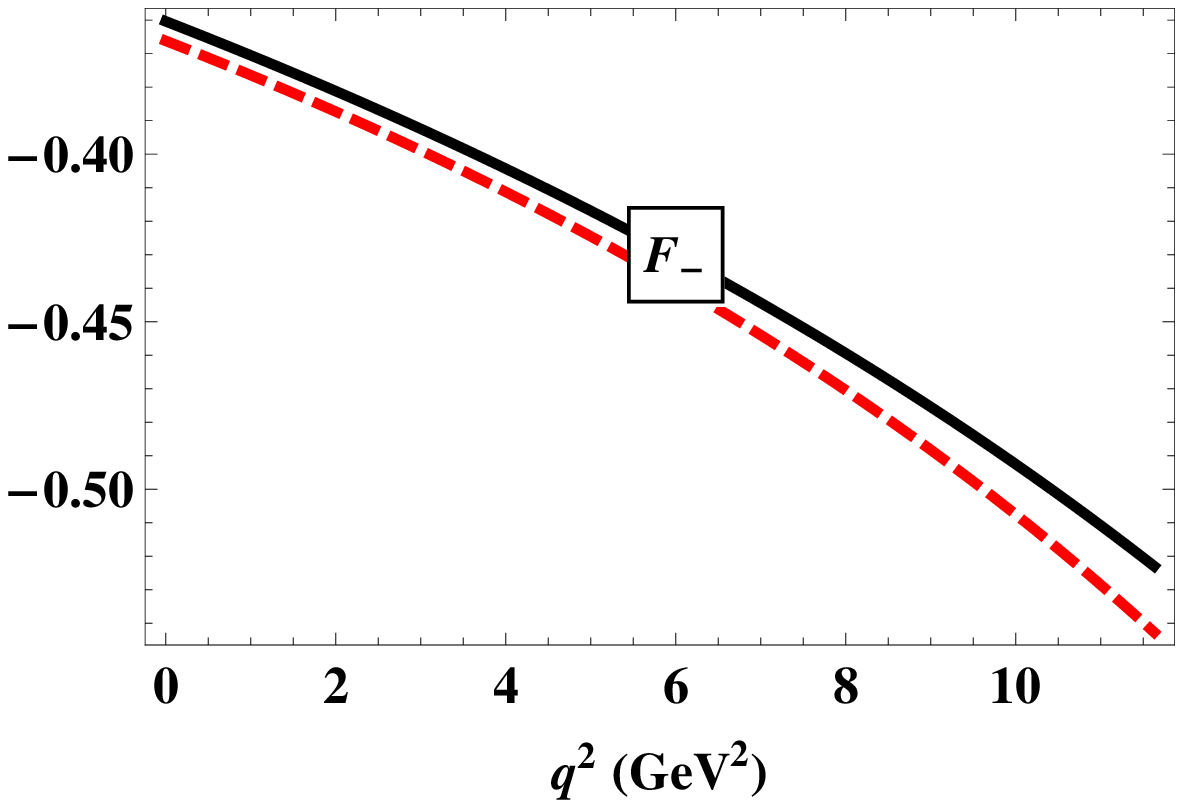}\\
\includegraphics[scale=0.6]{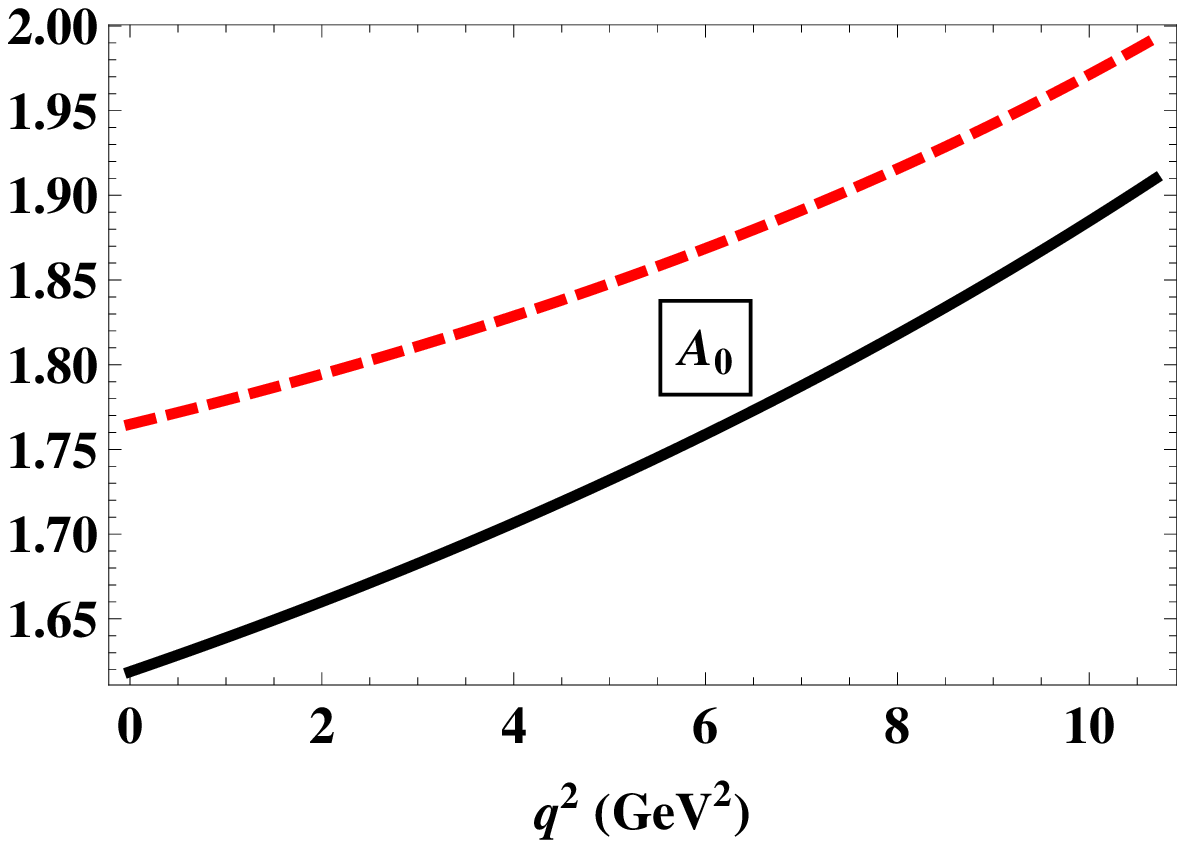}&
\includegraphics[scale=0.6]{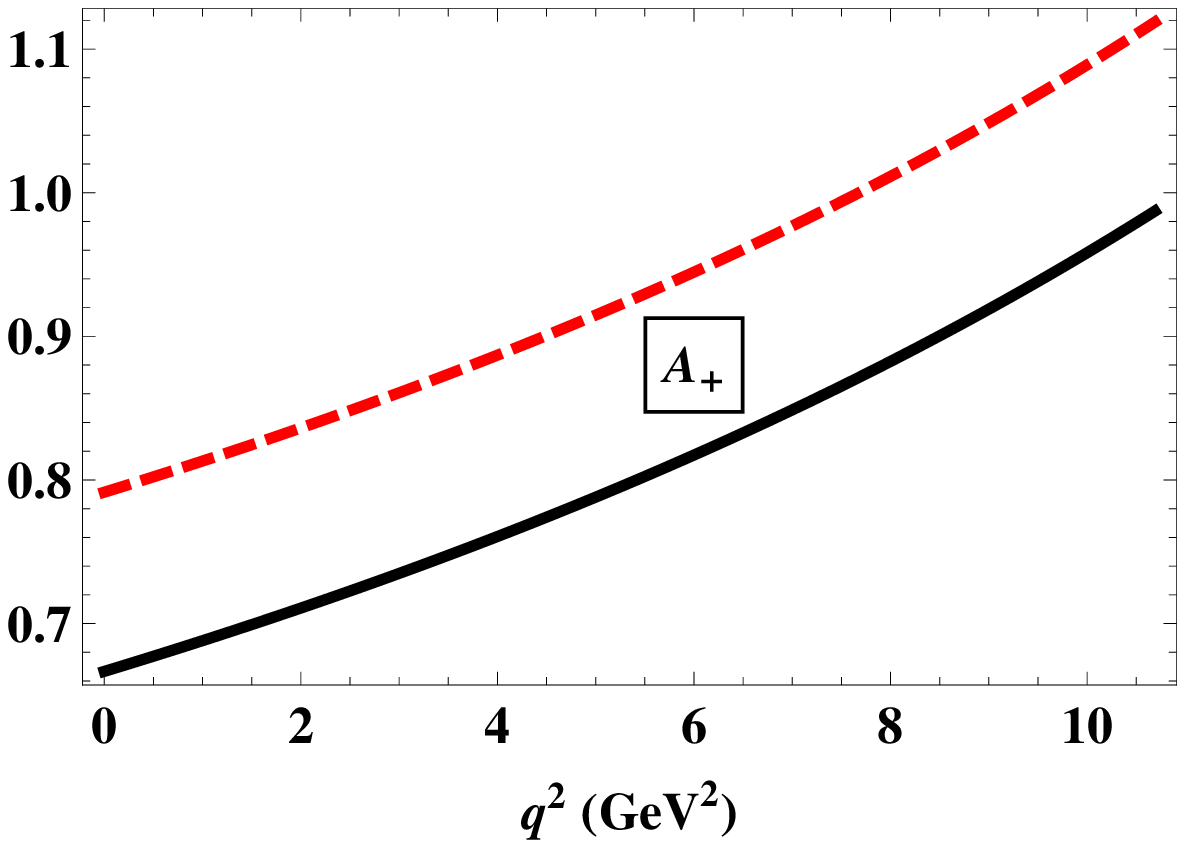}\\
\includegraphics[scale=0.6]{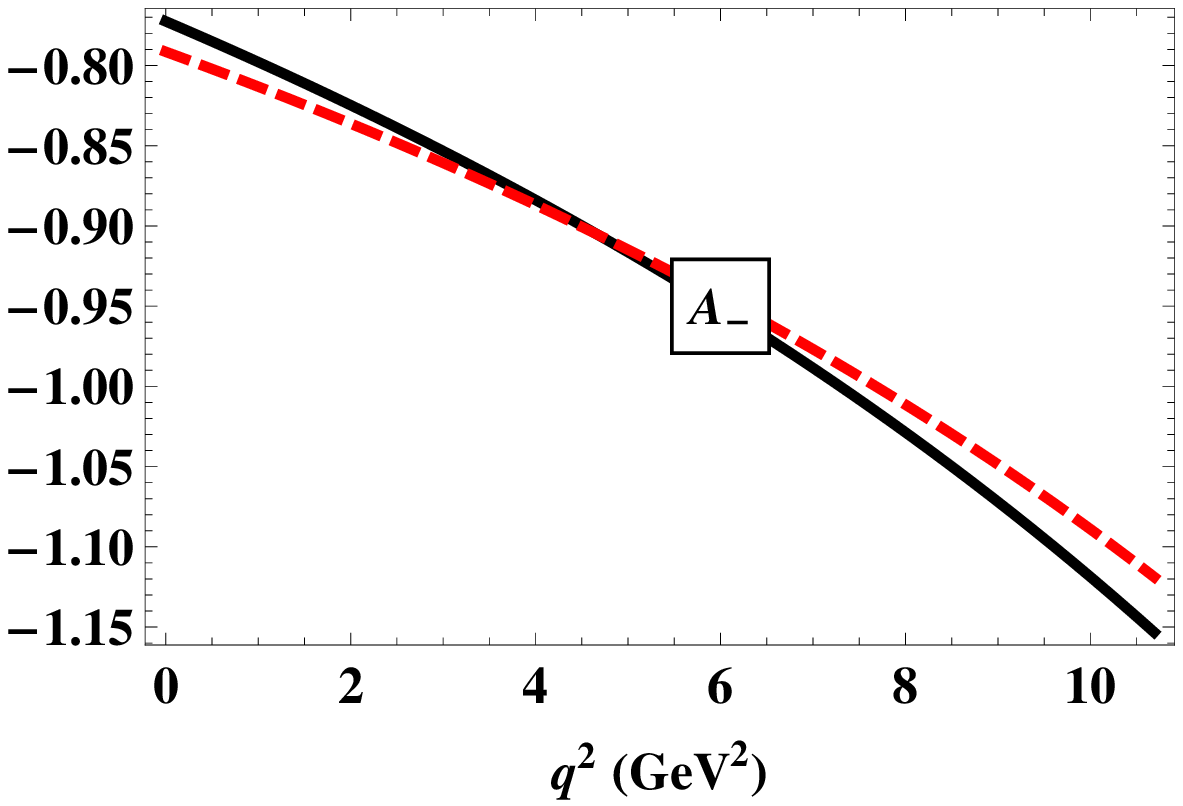}&
\includegraphics[scale=0.6]{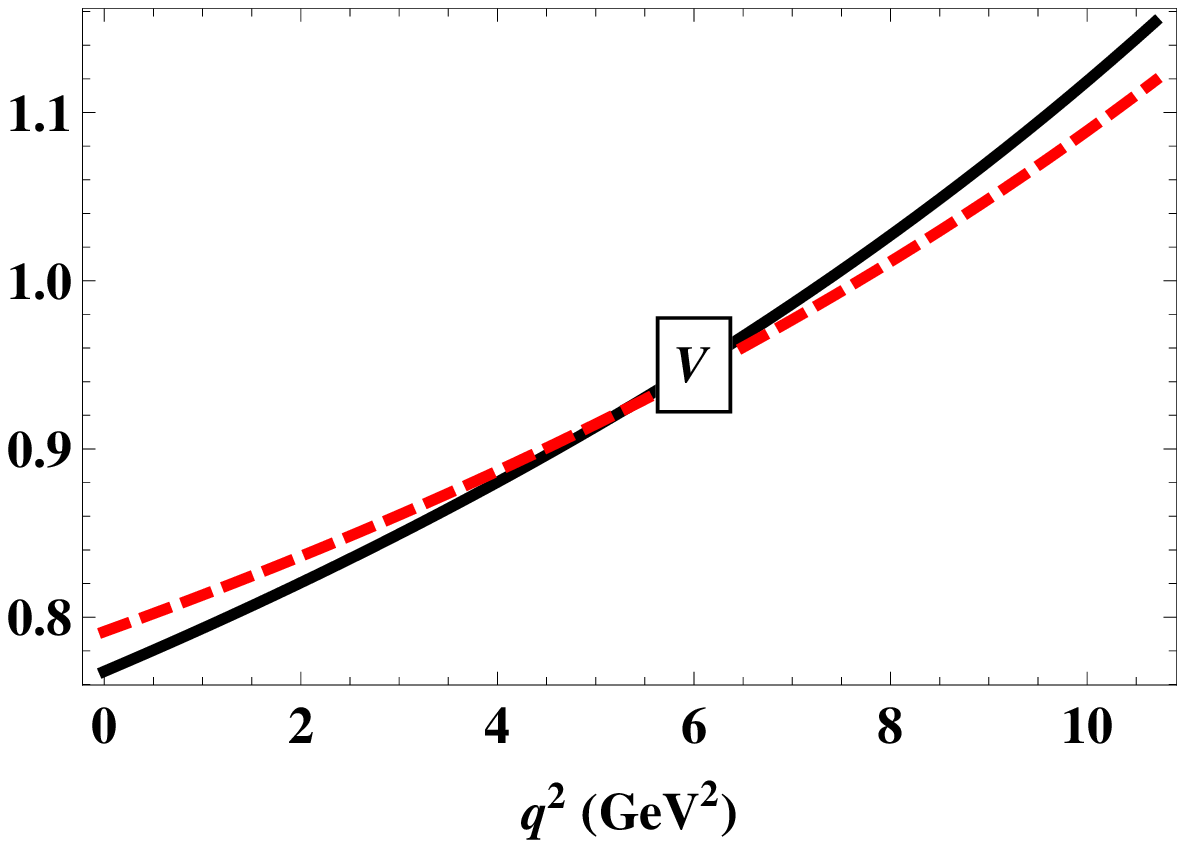}\\
\end{tabular}
\caption{Form factors of the decays $B \to D^{(*)} \ell \nu$.
The solid lines are the results of exact calculations in our approach,
the dashed lines are the form factors obtained in the heavy quark limit. }
\label{fig:formfactors}
\end{figure}

Since $b/a$ is quite small for the form factors $F_{+},\,F_{-},\,A_+,\,A_-$, 
and  $V$, these form factors show a monopole-like falloff behavior 
whereas $A_{0}$ has a substantial $(q^{2})^{-2}$ contribution.
In Fig.~\ref{fig:formfactors} we present our results for the semileptonic 
form factors within the full range of momentum transfer 
$0\le q^2 \le q^2_{\rm max}$, where $q^2_{\rm max}=(m_B-m_{D^{(*)}})^2$. 
The results of the exact calculations are shown by solid lines
whereas the results obtained in the heavy quark limit are
shown by dashed lines. We will discuss the heavy quark limit in
the next section.
It is interesting to note that the QCD counting rules prescribe a 
$(q^{2})^{-1}$ and a $(q^{2})^{-2}$ falloff behavior for the form factors 
$F_{+},\,F_{-},\,A_{0}$ and $A_+,\,A_-,V$, respectively.

\begin{figure}[htbp]
\includegraphics[scale=0.6]{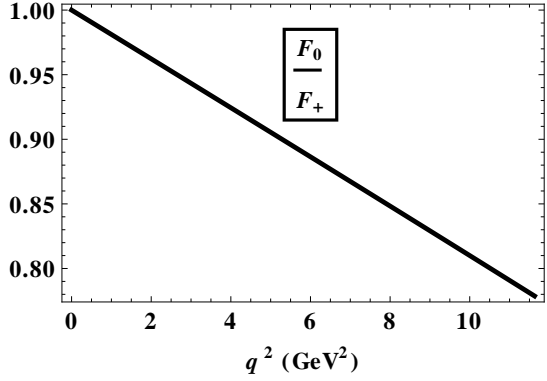}
\caption{Ratio $F_0(q^2)/F_+(q^2)$.}
\label{fig:formfactorRate}
\end{figure}

As recently noticed in~\cite{Becirevic:2012jf}, the ratio 
$F_0(q^2)/F_+(q^2)$ exhibits a linear $q^2$ behavior
\be
F_0(q^2) =  F_+(q^2) + \frac{q^2}{Pq}\,F_-(q^2),
\qquad
\frac{F_0(q^2)}{F_+(q^2)} = 1-\alpha q^2,
\label{FFlinear}
\en
where the slope $\alpha=0.020(1)~\text{GeV}^{-2}$ was determined precisely 
based on lattice values of the two form factors. 
 We also plot the $q^2$ dependence of the ratio 
$F_0(q^2)/F_+(q^2)$ in Fig.~\ref{fig:formfactorRate}, 
which shows a linear behavior as mentioned. Our value for the slope
is $\alpha=0.019~\text{GeV}^{-2}$ which very well agrees with 
the lattice result.
\clearpage

\section{Heavy quark limit} 

It is instructive to explore the heavy quark limit (HQL) in the heavy-to-heavy
transition $B\to D(D^\ast)$. In the HQL one takes
the limit $m_B=m_b + E,\,\,\,m_b\to\infty $ and
          $m_D=m_{D^\ast}=m_c + E,\,\,\,m_c\to\infty $ 
in the expressions for the coupling constants and form factors. 
In this limit the heavy quark propagators are reduced to the static form
\bea
S_b(k+p_1)&=&\frac{1}{m_b-\!\not\! k - \!\not\! p_1 }\to 
\frac{1+\not\! v_1}{-2kv_1-2 E} + O\left(\frac{1}{m_b}\right),
\nn
S_c(k+p_2)&=&\frac{1}{m_c-\!\not\! k - \!\not\! p_2 }\to 
\frac{1+\not\! v_2}{-2kv_2-2 E} + O\left(\frac{1}{m_c}\right),
\label{eq:prop_HQL}
\ena
where $p_i$ and $v_i=p_i/m_i$ ($i=1,2$) are the momenta and
the four-velocities of the initial and final states.
Moreover, we have to keep the size parameters of heavy hadrons equal to
each other in order to provide the correct normalization
of the Isgur-Wise function at zero recoil. 
By using technique developed in our previous papers, see, for instance,
\cite{Ivanov:1992wx,Ivanov:1998ms}, one can arrive at the following
expressions for the semileptonic heavy-to-heavy transitions
defined by Eqs.~(\ref{eq:PP'}) and ~(\ref{eq:PV})
\bea
T^\mu_{\rm HQL} &=& 
\xi(w)\cdot
\tfrac14\Tr\Big[O^\mu(1+\not\! v_1)\gamma^5\cdot\gamma^5(1+\not\! v_2)\Big]
=\xi(w)\cdot(v_1^\mu + v_2^\mu),
\label{eq:PP'-HQL}\\[1.5ex]
\epsilon^\dagger_{2\,\nu} T^{\mu\nu}_{\rm HQL} &=&  
\xi(w)\cdot
\tfrac14\Tr\Big[O^\mu(1+\not\! v_1)\gamma^5\cdot
\not\!\epsilon_2^{\,\dagger}(1+\not\! v_2)\Big]
\nn
&=&\xi(w)\cdot\epsilon^\dagger_{2\,\nu} 
( -g^{\mu\nu}(1+w) + v_1^\mu v_2^\nu + v_1^\nu v_2^\mu
- i\,\varepsilon^{\mu\nu v_1 v_2} ).
\label{eq:PV-HQL}
\ena
Here, $w=v_1v_2$, and the Isgur-Wise function is equal to
\be
\xi(w) = \frac{J_3(E,w)}{J_3(E,1)}, \qquad
J_3(E,w) = \int\limits_0^1 \frac{d\tau}{W}
\int\limits_0^\infty\!\! du\, \widetilde\Phi^2(z) 
\left( \sigma_S(z) + \sqrt{\frac{u}{W}} \sigma_V(z) \right),
\label{eq:IW}
\en
where $W=1+2\tau(1-\tau)(w-1)$, $z=u - 2 E \sqrt{u/W}$, and
\[
\widetilde\Phi(z) = \exp(-z/\Lambda^2), \qquad
 \sigma_S(z) = \frac{m_u}{m_u^2 +z},  \qquad
 \sigma_V(z) = \frac{1}{m_u^2 +z}.
\]
By using the definition of the form factors given by 
Eqs.~(\ref{eq:PP'}) and ~(\ref{eq:PV}) one can easily obtain
the expressions of the form factors in the HQL. One finds
\bea
F_\pm(q^2) &=& \pm\frac{m_1 \pm m_2}{2\sqrt{m_1m_2}}\,\xi(w),
\nn
A_0(q^2) &=& \frac{\sqrt{m_1 m_2}}{m_1 - m_2} (1+w)\xi(w),
\quad 
A_+(q^2) = - A_-(q^2) = V(q^2) = \frac{m_1 + m_2}{2\sqrt{m_1m_2}}\,\xi(w),
\ena
where $w=(m_1^2+m_2^2-q^2)/(2m_1m_2)$. We use the physical masses
of the heavy hadrons in the numerical calculations. For the size
parameter we adopt the average value 
$\Lambda=(\Lambda_B+\Lambda_D+\Lambda_{D^\ast})/3= 1.70$~GeV.
The parameter $E$ characterizes the difference in mass between the heavy hadron and
the corresponding heavy quark. We use its minimal value $E=m_D-m_c=0.20$~GeV
in order to avoid the complication with confinement.

In Fig.~\ref{fig:formfactors} we display the heavy-to-heavy transition 
form factors calculated in the HQL and compare them with the results
of exact calculations. One can see that the two results obtained with and without use of the HQL behave very similar to each other which demonstrates the fidelity of HQET.

One can also consider the near zero-recoil behavior of the form factors
in a similar way as we did in our paper on the semileptonic decay
$\Lambda_b\to \Lambda_c + \tau\bar\nu_\tau$~\cite{Gutsche:2015mxa}.
The standard parametrization of the $(w-1)$ expansion takes the form
\[
F(q^2(w)) = F(q^2_{\rm max})\,\Big[1 - \rho^2 (w-1)+c\,(w-1)^2 + \ldots
      \Big],
\]
where $\rho^2$ is called the slope parameter and $c$ the convexity parameter.
The numerical results are given below
\be
\begin{array}{c|rr|rrrr}
 & \quad F_+ \quad & \quad  F_- \quad & \quad A_0 \quad  & \quad A_+ \quad & 
  \quad A_- \quad & \quad V \quad \\[1.1ex]
\hline
F(q^{2}_{\rm max}) & 1.12  & -0.52  & 1.91  & 0.99  & -1.15   & 1.16
\\[1ex] 
\rho^2          &  0.72  & 0.74  & 0.42  & 0.93   &   0.95  & 0.96
\\[1ex] 
 c              &  0.49  & 0.51  & 0.28  & 0.82   &   0.85  & 0.86
\\[1.1ex] 
\hline
\end{array}
\label{eq:zerorecoil}
\en
\vspace{1.2ex}
which may be compared with the results obtained
for the monopole form factor of a $B_c$-resonance contribution:
$\rho^2$=0.71 and $c=0.51$. 

It is interesting to compare the zero-recoil values of our exact form factors 
with the predictions of leading order HQET at $w=1$ where $\xi(1)=1$. 
One has
\bea
F_{+}&=&\frac{m_{1}+m_{2}}{2\sqrt{m_{1}m_{2}}}=1.138, \qquad
F_{-}=-\frac{m_{1}-m_{2}}{2\sqrt{m_{1}m_{2}}}=-0.543,
\nn[1.2ex]
A_+ &=& -A_{-} = V =\frac{m_{1}+m_{2}}{2\sqrt{m_{1}m_{2}}}=1.119, \qquad
A_0 = \frac{2\sqrt{m_{1}m_{2}}}{m_{1}-m_{2}}=1.993. 
\ena
The zero-recoil values of our model form factors can be seen to be quite 
close to the corresponding HQET values except for the form factor $A_{+}$
where our form factor value exceeds the HQET result by $\sim 13\, \%$. 

\section{Helicity amplitudes and two-fold distributions}

\begin{figure}[htbp]
\begin{center}
\epsfig{figure=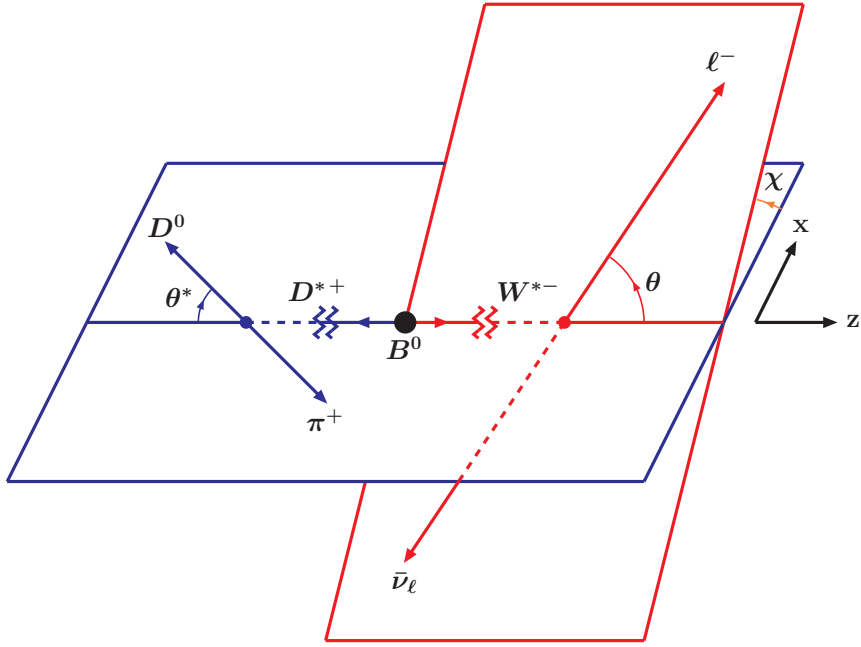,scale=.6}
\caption{Definition of angles $\theta$, $\theta^\ast$, and $\chi$ in
the cascade decay $\bar B^0\to D^{\ast\,+}(\to D^0\pi^+)\ell^-\bar\nu_\ell$.}
\label{fig:bkangl}
\end{center}
\end{figure}

Let us first consider the polar angle differential decay distribution 
in the momentum transfer squared $q^2$. The polar angle is defined
by the angle between $\vec q=\vec p_1-\vec p_2$ and the three-momentum of
the charged lepton $\vec k_1$ in the ($\ell^-\bar\nu_\ell$) rest frame as 
shown in Fig.~\ref{fig:bkangl}. One has 
\be
\frac{d^2\Gamma}{dq^2 d\cos\theta} = 
\frac{|{\bf p_2}| \, v}{(2\pi)^3\,32\,m_1^2}
\cdot\sum\limits_{\rm pol}|M|^2
=\frac{G^2_F}{(2\pi)^3}\,|V_{cb}|^2 
\frac{|{\bf p_2}|\,v }{64 m_1^2}
H^{\mu\nu} L_{\mu\nu}\,,
\label{eq:2-fold-dis}
\en
where  $|{\bf p_2}|=\lambda^{1/2}(m_1^2,m_2^2,q^2)/2m_1 $
is the momentum of the daughter meson and where we have introduced the 
velocity-type parameter $v=1-m_\ell^2/q^2$ as well as the contraction 
of hadron and lepton tensors $H^{\mu\nu}L_{\mu\nu}$.

As discussed in some detail in~\cite{Gutsche:2015mxa}
the covariant contraction $H^{\mu\nu} L_{\mu\nu}$ can be converted to a sum
of bilinear products of hadronic and leptonic helicity amplitudes using the 
completeness relation for the polarization four-vectors of the process.
A synopsis of the necessary steps in this transformation is provided in 
the Appendix. 

One needs to relate the mesonic helicity amplitudes to the 
invariant form factors defined in Eqs.~(\ref{eq:PP'}) and (\ref{eq:PV}).
To do so one requires explicit representations of the polarization
four-vectors $\epsilon^{\mu}(\lambda_{W})$. They read
\be
\epsilon^\mu(t) =
\frac{1}{\sqrt{q^2}}(q_0,0,0,|{\bf p_2}|\,),\quad 
\epsilon^\mu(\pm) = 
\frac{1}{\sqrt{2}}(0,\mp 1,-i,0), \quad
\epsilon^\mu(0) =
\frac{1}{\sqrt{q^2}}(\,|{\bf p_2}|,0,0,q_0). 
\en
The linear relations between the two sets of form factors can then
be calculated in the following way. \\[1.2ex]

\noindent
{\boldmath{$B\to D$} \,\,\,\bf transition:}\\
The helicity amplitudes are defined by 
$H_{\lambda_{W}} = \epsilon^{\dagger \mu}(\lambda_{W})T_\mu$.
One obtains
\be
H_t   = \frac{1}{\sqrt{q^2}}(Pq\, F_+ + q^2\, F_-),
\qquad
H_\pm = 0,
\qquad
H_0   = \frac{2\,m_1\,|{\bf p_2}|}{\sqrt{q^2}} \,F_+.
\label{eq:hel_pp}
\en
Note the zero-recoil relation $H_{0}=0$. At the other end of the spectrum
at maximal recoil $q^{2}=0$ one has $H_{t}=H_{0}$.
In the Appendix we describe how to obtain the differential 
$(q^2,\cos\theta)$ distribution. One has  
\bea
\lefteqn{\frac{d\Gamma(B\to D \ell^-\bar\nu_\ell)}{dq^2d\cos\theta}
  \ =\ \,\frac{G_F^2 |V_{cb}|^2 |{\bf p_2}| q^2 v^2}{32 (2\pi)^3 m_1^2}}\nn
&& \hspace{2cm} \times \Big\{
  2\,\sin^2\theta\,{\cal H}_L 
+ 2\,\delta_\ell \left( 
2\,\cos^2\theta\,{\cal H}_L
+ 2\,{\cal H}_S - 4\,\cos\theta\,{\cal H}_{SL} \right)
\Big\},
\label{eq:distr2}
\ena
where we have introduced the helicity flip penalty factor 
$\delta_\ell = m^2_\ell/2q^2$ and the helicity
structure functions ${\cal H}_L    = |H_{ 0}|^2 $,
${\cal H}_S    = |H_{ t}|^2$, 
and ${\cal H}_{SL} =  {\rm Re}(  H_{0} H_{t}^\dagger ) $. \\

\noindent
{\boldmath$B\to D^\ast$ \,\,\bf transition:}\\
The helicity amplitudes are defined by 
$H_{\lambda_{W}\,\lambda_{D^{\ast}}} = \epsilon^{\dagger \mu}(\lambda_{W})
\epsilon_2^{\dagger\alpha}(\lambda_{D^{\ast}})T_{\mu \alpha}$.
In addition to the $W_{\rm off-shell}$ polarization four-vectors
$\epsilon^{\mu}(\lambda_{W})$ 
one needs the polarization four-vectors 
$\epsilon^\alpha_2(\lambda_{D^{\ast}})$ of the $D^{\ast}$. They 
read ($E_{2}=m_{1}-q_{0}$)
\be
\epsilon^\alpha_2(\pm) = 
\frac{1}{\sqrt{2}}(0\,,\,\,\pm 1\,,\,\,-i\,,\,\,0\,)\,,
\qquad
\epsilon^\alpha_2(0) = 
\frac{1}{m_2}(|{\bf p_2}|\,,\,\,0\,,\,\,0\,,\,\,-E_2\,).
\label{eq:vect_pol}
\en
One obtains
\bea
H_{t0} &=& 
\epsilon^{\dagger \mu}(t)\epsilon_2^{\dagger \alpha}(0)T_{\mu\alpha}
\,=\,
\frac{1}{m_1+m_2}\frac{m_1\,|{\bf p_2}|}{m_2\sqrt{q^2}}
\left(Pq\,(-A_0+A_+)+q^2 A_-\right),
\nn[1.2ex]
H_{\pm1\pm1} &=& 
\epsilon^{\dagger \mu}(\pm)\epsilon_2^{\dagger \alpha}(\pm)T_{\mu\alpha}
\,=\,
\frac{1}{m_1+m_2}\left(-Pq\, A_0\pm 2\,m_1\,|{\bf p_2}|\, V \right),
\nn[1.2ex]
H_{00} &=&  
\epsilon^{\dagger \mu}(0)\epsilon_2^{\dagger \alpha}(0)T_{\mu\alpha}
\nn
&=&
\frac{1}{m_1+m_2}\frac{1}{2\,m_2\sqrt{q^2}} 
\left(-Pq\,(m_1^2 - m_2^2 - q^2)\, A_0 + 4\,m_1^2\,|{\bf p_2}|^2\, A_+\right).
\label{eq:hel_vv}
\ena
Note the zero-recoil relations $H_{t0}=0$ and $H_{\pm1\pm1}=H_{00}$.
At maximal recoil $q^{2}=0$ the dominating helicity amplitudes are
$H_{t0}$ and $H_{00}$ with $H_{t0}=H_{00}$. 

The differential $(q^2,\cos\theta)$ distribution finally reads (see the
Appendix)
\bea
\frac{d\Gamma(B\to D^{\ast} \ell^-\bar\nu_\ell)}{dq^2d(\cos\theta)} &=&\,
\frac{G_F^2 |V_{cb}|^2 |{\bf p_2}| q^2 v^2}{32 (2\pi)^3 m_1^2}
{\rm Br}(D^\ast\to D\pi)
\nn
&\times& \Big\{
(1+\cos^2\theta)\,{\cal H}_U + 2\,\sin^2\theta\,{\cal H}_L 
-2\,\cos\theta\,{\cal H}_P
\nn
&+& 2\,\delta_\ell \left( \sin^2\theta\,{\cal H}_U 
+ 2\,\cos^2\theta\,{\cal H}_L
+ 2\,{\cal H}_S - 4\,\cos\theta\,{\cal H}_{SL} \right)
\Big\}.
\label{eq:distrII}
\ena
We have used the zero width approximation for the $D^\ast$ intermediate state
which brings in the branching fraction ${\rm Br}(D^\ast\to D\pi)$.
The relevant bilinear combinations of the helicity amplitudes
are defined in Table~\ref{tab:bilinears}. We have dropped
a factor of ``3'' in the definition of ${\cal H}_S$ and ${\cal H}_{IS}$
compared to our paper \cite{Faessler:2002ut}.  Note that the
helicity structure functions satisfy the zero-recoil relations
$2{\cal H}_U={\cal H}_L={\cal H}_T={\cal H}_I$ and 
${\cal H}_P={\cal H}_A={\cal H}_S={\cal H}_{SA}={\cal H}_{ST}={\cal H}_{S}=0$.
Similar relations hold for the imaginary parts. At maximal recoil one has
${\cal H}_L={\cal H}_S={\cal H}_{SL}$ for the dominating helicity structure 
functions.
\begin{table}[htbp] 
\begin{center}
\caption{ Definition of helicity structure functions and their parity 
properties for the case \newline
$B\to D^{\ast} \ell^-\bar\nu_\ell$.}
\def\arraystretch{1}
\begin{tabular}{ll}
\hline
parity-conserving (p.c.) \qquad & \qquad  parity-violating (p.v.)  \\
\hline
${\cal H}_U   = |H_{+1 +1}|^2 + |H_{-1 -1}|^2$   \qquad &  \qquad
${\cal H}_P   = |H_{+1 +1}|^2 - |H_{-1 -1}|^2$   \\
${\cal H}_L    = |H_{0 0}|^2 $   \qquad &  \qquad
${\cal H}_{A} = \tfrac 12   {\rm Re}\left(  H_{+1 +1}  H_{0\, 0}^\dagger 
                         - H_{-1 -1} H_{0\,0}^\dagger \right)$  
\\
${\cal H}_{T} ={\rm Re}\left( H_{+1+1}H_{-1-1}^{\dagger}\right)$
\qquad &  \qquad
${\cal H}_{IA} = \tfrac 12   {\rm Im}\left(  H_{+1 +1}  H_{0\, 0}^\dagger 
                         - H_{-1 -1} H_{0\,0}^\dagger \right)$
\\ 
${\cal H}_{IT} ={\rm Im}\left( H_{+1+1}H_{-1-1}^{\dagger}\right)$
\qquad &  \qquad
${\cal H}_{SA} = \tfrac 12 \, {\rm Re}\left(  H_{+1 +1}  H_{0\, t}^\dagger 
                         - H_{-1 -1} H_{0\,t}^\dagger \right)$
\\ 
${\cal H}_{I}  =  \tfrac 12 \, {\rm Re}\left(  H_{+1 +1}  H_{0\, 0}^\dagger 
                         + H_{-1 -1} H_{0\,0}^\dagger \right)$ 
 \qquad &  \qquad 
${\cal H}_{ISA} = \tfrac 12 \, {\rm Im}\left(  H_{+1 +1}  H_{0\, t}^\dagger 
                         - H_{-1 -1} H_{0\,t}^\dagger \right)$
\\
${\cal H}_{II}  =  \tfrac 12 \, {\rm Im}\left(  H_{+1 +1}  H_{0\, 0}^\dagger 
                         + H_{-1 -1} H_{0\,0}^\dagger \right)$ 
 \qquad &  \qquad 
\\
${\cal H}_S    = |H_{0 t}|^2$   \qquad &  \qquad  
\\
 ${\cal H}_{ST} =  \tfrac 12 \,{\rm Re}\left(  H_{+1 +1}  H_{0\, t}^\dagger 
                         + H_{-1 -1} H_{0\,t}^\dagger \right)$
 \qquad &  \qquad  
\\ 
${\cal H}_{IST} =  \tfrac 12 \,{\rm Im}\left(  H_{+1 +1}  H_{0\, t}^\dagger 
                         + H_{-1 -1} H_{0\,t}^\dagger \right)$
 \qquad &  \qquad  
\\ 
${\cal H}_{SL} =  {\rm Re}\left(  H_{0\,0} H_{0\,t}^\dagger \right) $
 \qquad &  \qquad
 \\
${\cal H}_{ISL} =  {\rm Im}\left(  H_{0\,0} H_{0\,t}^\dagger \right) $
 \qquad &  \qquad
 \\
${\cal H}_{\rm tot} =  {\cal H}_U+{\cal H}_L+\delta_{\ell}\Big(
{\cal H}_U+{\cal H}_L+3{\cal H}_S \Big) $
 \qquad &  \qquad
 \\[2ex]
\hline
\end{tabular}
\label{tab:bilinears}
\end{center}
\end{table}

Let us begin discussing the $\cos\theta$ distribution for the 
$B \to D^{\ast}\ell^{-}\bar \nu_{\ell}$ case.
The distribution~(\ref{eq:distr2}) is described by a 
tilted parabola whose normalized form reads
\be
\widetilde W(\theta)=\frac{a+b\cos\theta+c\cos^{2}\theta}{2(a+c/3)}.
\en
The linear
coefficient $b/2(a+c/3)$ can be projected out by defining a 
forward-backward asymmetry given by \footnote{We take this opportunity to 
correct a typo in~\cite{Gutsche:2015mxa}. The factor $-3/2$ in~Eq.~(38) of
\cite{Gutsche:2015mxa} should read $-3/4$.  } 
\bea
\mathcal{A}_{FB}(q^2) = 
\frac{d\Gamma(F)-d\Gamma(B)}{d\Gamma(F)+d\Gamma(B)}
&=&
\frac{ \int_{0}^{1} d\!\cos\theta\, d\Gamma/d\!\cos\theta
      -\int_{-1}^{0} d\!\cos\theta\, d\Gamma/d\!\cos\theta }
     { \int_{0}^{1} d\!\cos\theta\, d\Gamma/d\!\cos\theta
      +\int_{-1}^{0} d\!\cos\theta\, d\Gamma/d\!\cos\theta} 
\nn[1.2ex]
&=& \frac{b}{2(a+c/3)}=-\frac34 \frac{{\cal H}_P\,
+4\,\delta_\ell\,{\cal H}_{SL}}{{\cal H}_{\rm tot}}.
\label{fbAsym}
\ena
In the $\tau$ mode there are two sources of the parity-odd forward-backward
asymmetry, namely, a purely parity-violating source from the VA interaction
leading to the ${\cal H}_P$ contribution, and a parity-conserving source from
the VV and AA interactions leading to the ${\cal H}_{SL}$ contribution.
The parity-conserving parity-odd contribution ${\cal H}_{SL}$ arises from the 
interference of the $(0^{+};1^{-})$ and $(0^{-};1^{+})$ components of the 
$VV$ and
$AA$ product of currents, respectively. In the case of the
$B \to D$ transition the forward-backward asymmetry arises solely from the
$(0^{+};1^{-})$ interference term of the $VV$ product of currents.

The coefficient $c/2(a+c/3)$ of the quadratic contribution is obtained
by taken the second derivative of $\widetilde W(\theta)$. Accordingly we
define a convexity parameter by writing 
\be
C_F^\ell(q^2) = \frac{d^{2}\widetilde W(\theta)}{d(\cos\theta)^{2}}
= \frac{c}{a+c/3} 
= \frac34 (1-2\delta_\ell)
\frac{ {\cal H}_U - 2 {\cal H}_L }{ {\cal H}_{\rm tot} }.
\label{eq:convex_lep}
\en 

When calculating the $q^{2}$ averages of the forward-backward asymmetry
and the convexity parameter one 
has to multiply the numerator and denominator of (\ref{fbAsym}) and
(\ref{eq:convex_lep}) by the 
$q^{2}$-dependent piece of the phase space factor in~(\ref{eq:distr2})
given by  
$
C(q^2) = |\mathbf{p_2}| q^2 v^2.
$
For example, the mean forward-backward asymmetry can then be calculated 
according to
\be
\langle \mathcal{A}_{FB}\rangle = -\frac34 \,\,
\frac{\int dq^{2} C(q^{2})\big({\cal H}_P\,+4\,\delta_\ell\,{\cal H}_{SL}\big)}
{\int dq^{2} C(q^{2}){\cal H}_{\rm tot}}.
\label{eq:FBint}
\en

Finally, integrating Eq.~(\ref{eq:distr2}) over $\cos\theta$ one obtains 
\be
\frac{d\Gamma(B\to D^{(\ast)} \ell^-\bar\nu_\ell)}{dq^2} =\,
\frac{G_F^2 |V_{cb}|^2 |{\bf p_2}| q^2 v^2}{12 (2\pi)^3 m_1^2}
\,\,{\rm Br}(D^\ast\to D\pi)\,\,\cdot{\cal H}_{\rm tot},
\label{eq:distr1}
\en
where ${\cal H}_{\rm tot} =  {\cal H}_U+{\cal H}_L+\delta_{\ell}\Big(
{\cal H}_U+{\cal H}_L+3{\cal H}_S \Big) $.

The discussion of the $\cos\theta$ distribution for the 
$B \to D\ell^{-}\bar \nu_{\ell}$ case proceeds in a similar way except that
one has to drop the contributions of the helicity structure functions
${\cal H}_U$ and ${\cal H}_P$.

\section{Four-fold angular decay distribution} 

The lepton-hadron correlation function $L_{\mu\nu}H^{\mu\nu}$ reveals 
even more structures when one uses the cascade decay 
$\bar{B}^0\to D^{\ast\,+}(\to D^0\pi^+) \ell^-\bar\nu_\ell$ to analyze the 
polarization of the $D^\ast$ meson. The derivation of the four-fold angular 
decay distribution is detailed in Appendix A. One has 
\be
\frac{d\Gamma(\bar{B}^0\to D^{\ast\,+}(\to D^0\pi^+) \ell^-\bar\nu_\ell)}
     {dq^2\,d\cos\theta\,d(\chi/2\pi)\,d\cos\theta^\ast} 
=
\frac{G_F^2}{(2\pi)^3}\frac{ |V_{cb}|^2 |{\bf p_2}| q^2 v^2}{12 m_1^2}\,
{\rm Br}(D^\ast\to D\pi)\,W(\theta^\ast,\theta,\chi),
\label{eq:four-width}
\en
where
\bea
W(\theta^\ast,\theta,\chi) &=&
\,\,  \frac{9}{32}\,(1+\cos^2\theta)\,\sin^2\theta^\ast\,{\cal H}_U
+ \frac{9}{8}\,\sin^2\theta\,\cos^2\theta^\ast\,{\cal H}_L
- \frac{9}{16}\,\cos\theta\,\sin^2\theta^\ast\,{\cal H}_P
\nn[1.2ex]
&&\,-\,\frac{9}{16}\,\sin^2\theta\,\sin^2\theta^\ast\,\cos 2\chi\, {\cal H}_T
    -\,\frac{9}{8}\, \sin\theta\,\sin 2\theta^\ast\,\cos\chi\, {\cal H}_A
\nn[1.2ex]
&&
    +\,\frac{9}{16}\,\sin 2\theta\,\sin 2\theta^\ast\,\cos\chi\, {\cal H}_I
    +\,\frac{9}{8}\,\sin\theta\,\sin 2\theta^\ast\,\sin\chi\, {\cal H}_{II}
\nn[1.2ex]
&&
    -\,\frac{9}{16}\,\sin 2\theta\,\sin 2\theta^\ast\,\sin\chi\, {\cal H}_{IA}
    +\,\frac{9}{16}\,\sin^2\theta\,\sin^2\theta^\ast\,\sin 2\chi\, 
{\cal H}_{IT}
\nn[1.2ex]
&+&\,\delta_\ell\,\Big[
  \,\,\, \frac{9}{4}\,\cos^2\theta^\ast\,{\cal H}_S
    -\,\frac{9}{2}\,\cos\theta\,\cos^2\theta^\ast\,{\cal H}_{SL} 
    +\,\frac{9}{4}\,\cos^2\theta\,\cos^2\theta^\ast\,{\cal H}_{L}
\nn[1.2ex]
&&
\qquad    +\,\frac{9}{16}\,\sin^2\theta\,\sin^2\theta^\ast\,{\cal H}_{U} 
    +\,\frac{9}{8}\,\sin^2\theta\,\sin^2\theta^\ast\,\cos 2\chi\, {\cal H}_{T}
\nn[1.2ex]
&&\qquad 
    +\,\frac{9}{4}\,\sin\theta\,\sin 2\theta^\ast\,\cos\chi\, {\cal H}_{ST}
    -\,\frac{9}{8}\,\sin 2\theta\,\sin 2\theta^\ast\,\cos\chi\, {\cal H}_{I}
\nn[1.2ex]
&&\qquad 
    -\,\frac{9}{4}\,\sin\theta\,\sin 2\theta^\ast\,\sin\chi\, {\cal H}_{ISA}
    +\,\frac{9}{8}\,\sin 2\theta\,\sin 2\theta^\ast\,\sin\chi\, {\cal H}_{IA}
\nn[1.2ex]
&&\qquad 
    -\,\frac{9}{8}\,\sin^2\theta\,\sin^2\theta^\ast\,\sin 2\chi\, {\cal H}_{IT}
\Big].
\label{eq:distr4}
\ena
In our quark model all helicity amplitudes are real, which implies the
vanishing of all terms proportional to $\sin\chi$ and $\sin2\chi$. 
The angular decay distribution for the remaining terms agrees with the results 
of~\cite{Korner:1987kd,Korner:1989qb,Korner:1989ve} when one takes into 
account the different definition of the polar angle $\theta$ used 
in~\cite{Korner:1987kd,Korner:1989ve,Korner:1989qb} such that 
$\theta \to 180^{\circ}-\theta$.

The four-fold distribution allows one to define a number of physical
observables which can be measured experimentally.
Integrating Eq.~(\ref{eq:distr4}) over $\cos\theta^\ast$ and 
$\chi$ one recovers the two-fold ($q^2,\cos\theta$) distribution of
Eq.~(\ref{eq:distr2}) that gives rise to the lepton side forward-backward 
asymmetry parameter $A_{FB}$ and the convexity parameter $C_F^\ell(q^2)$. 
Integrating 
Eq.~(\ref{eq:distr4}) over $\cos\theta$ and 
$\chi$ one obtains the hadron side $\cos\theta^\ast$ distribution described 
by a untilted parabola (without a linear term). The normalized form of the  
$\cos\theta^\ast$ distribution
reads $\widetilde {W} (\theta^\ast)=(a'+c'\cos^{2}\theta^\ast)/2(a'+c'/3)$,
which can again be characterized by its convexity 
parameter given by
\be
C_F^h(q^2) = \frac{d^{2}\widetilde W(\theta)}{d(\cos\theta^{\ast})^{2}}
=\frac{c'}{a'+c'/3}=-\,\frac32 
\frac{ {\cal H}_U - 2 {\cal H}_L 
      +\delta_\ell( {\cal H}_U - 2 {\cal H}_L -6{\cal H}_S )}{ 
{\cal H}_{\rm tot} }.
\label{eq:convex_had}
\en

We define a normalized angular decay distribution 
$\widetilde W(\theta^\ast,\theta,\chi)$ through
\be
\widetilde W(\theta^\ast,\theta,\chi)=\frac{W(\theta^\ast,\theta,\chi)}
{ {\cal H}_{\rm tot}}.
\label{eq:normdis}
\en
\noindent 
The normalized angular decay distribution 
$\widetilde W(\theta^\ast,\theta,\chi)$ obviously integrates to $1$ after
$\cos\theta^\ast,\,\cos\theta$, and $\chi/2\pi$ integration.

The remaining coefficient functions ${\cal H}_{T}(1 - 2\delta_\ell)$,
${\cal H}_{T}(1 - 2\delta_\ell)$, and 
$ ({\cal H}_{A} - 2 \delta_\ell{\cal H}_{ST})$ in Eq.(\ref{eq:distr4}) can be 
projected from the
three-fold angular decay distribution Eq.(\ref{eq:distr4}) by taking the
appropriate trigonometric moments of the normalized decay distribution
$\widetilde W(\theta^\ast,\theta,\chi)$. The trigonometric moments are
defined by
\be
W_{i} = \int d\cos\theta\,d\cos\theta^\ast\,d(\chi/2\pi)\,
M_{i}(\theta^\ast,\theta,\chi)\widetilde W(\theta^\ast,\theta,\chi) 
\equiv  <\, M_{i}(\theta^\ast,\theta,\chi) \,> ,
\en
where $M_{i}(\theta^\ast,\theta,\chi)$ defines the trigonometric moment that 
is being taken.
One finds 
\bea
W_T(q^2) &\equiv& <\,  \cos 2\chi \,>  
= -\,\tfrac 12\,(1 - 2\delta_\ell) \,\frac{{\cal H}_{T}}{{\cal H}_{\rm tot}},
\nn
W_I(q^2) &\equiv& <\,   
\cos\theta\cos\theta^{\ast}\cos \chi \,>  
= \frac{9\pi^2\,(1 - 2\delta_\ell)}{512}
\, \frac{{\cal H}_{I}}{{\cal H}_{\rm tot}},
\nn
W_A(q^2) &\equiv&<\,\sin\theta\cos\theta^{\ast}\cos \chi \,>   
= \,- \frac{3\pi}{16}\, 
\frac{ {\cal H}_{A} - 2 \delta_\ell{\cal H}_{ST} }{ {\cal H}_{\rm tot} }.
\label{eq:W}
\ena
The coefficient functions ${\cal H}_{T}(1 - 2\delta_\ell)$,
${\cal H}_{T}(1 - 2\delta_\ell)$, and 
$ ({\cal H}_{A} - 2 \delta_\ell{\cal H}_{ST})$ can also be projected out by
taking piecewise sums and differences of different sectors of the angular 
phase space \cite{Korner:1989qb}.

Finally, we consider the longitudinal and transverse polarizations of the
lepton where we consider only the angular average of the two polarization
states. For the longitudinal polarization one obtains 
\be
P^\ell_z(q^2) =\frac{\delta_{\ell}{\cal H}_{hf}-{\cal H}_{nf}}
{\delta_{\ell}{\cal H}_{hf}+{\cal H}_{nf}}= -\, 
\frac{ {\cal H}_U + {\cal H}_L 
      - \delta_\ell( {\cal H}_U + {\cal H}_L + 3 {\cal H}_S )}
{ {\cal H}_{\rm tot} }.
\label{eq:lpolarization}
\en
The transverse polarization can be calculated using the representation of the
polarized lepton tensor written down in the Appendix 
of~\cite{Gutsche:2015mxa}. One obtains
\be
P^\ell_x(q^2) = -\, \frac{3\pi\sqrt{\delta_\ell}}{4\sqrt{2}}
\frac{ {\cal H}_P - 2 {\cal H}_{SL}} { {\cal H}_{\rm tot} }.
\label{eq:tpolarization}
\en
For the decay $B \to D\ell^{-}\bar \nu_{\ell}$ one has to drop the 
transverse contributions ${\cal H}_U $ and ${\cal H}_P $ in
Eqs~(\ref{eq:lpolarization}) and (\ref{eq:tpolarization}).
It is interesting to note that for this decay
there exists a very simple relation connecting $P^\ell_x(q^2)$ and
$A_{FB}(q^{2})$ which reads 
\be
P^{\ell}_x(q^2) = -\frac{\pi\sqrt{q^2}}{2m_\tau} A_{FB}(q^2). 
\en
The polarization of the lepton depends on the frame in which it is defined.
The polarization components $P^{\ell}_{z}$ and $P^{\ell}_{x}$ in
(\ref{eq:lpolarization}) and (\ref{eq:tpolarization}) are calculated in the
$(\ell^{-}\bar \nu_{\tau})$ rest frame. The corresponding polarization
components in the $B$ rest frame have been calculated 
in~\cite{Hagiwara:1989cu}. 

\section{Results and discussion}
\label{sec:result}

\begin{figure}[htbp]
\begin{tabular}{lr}
\includegraphics[scale=0.6]{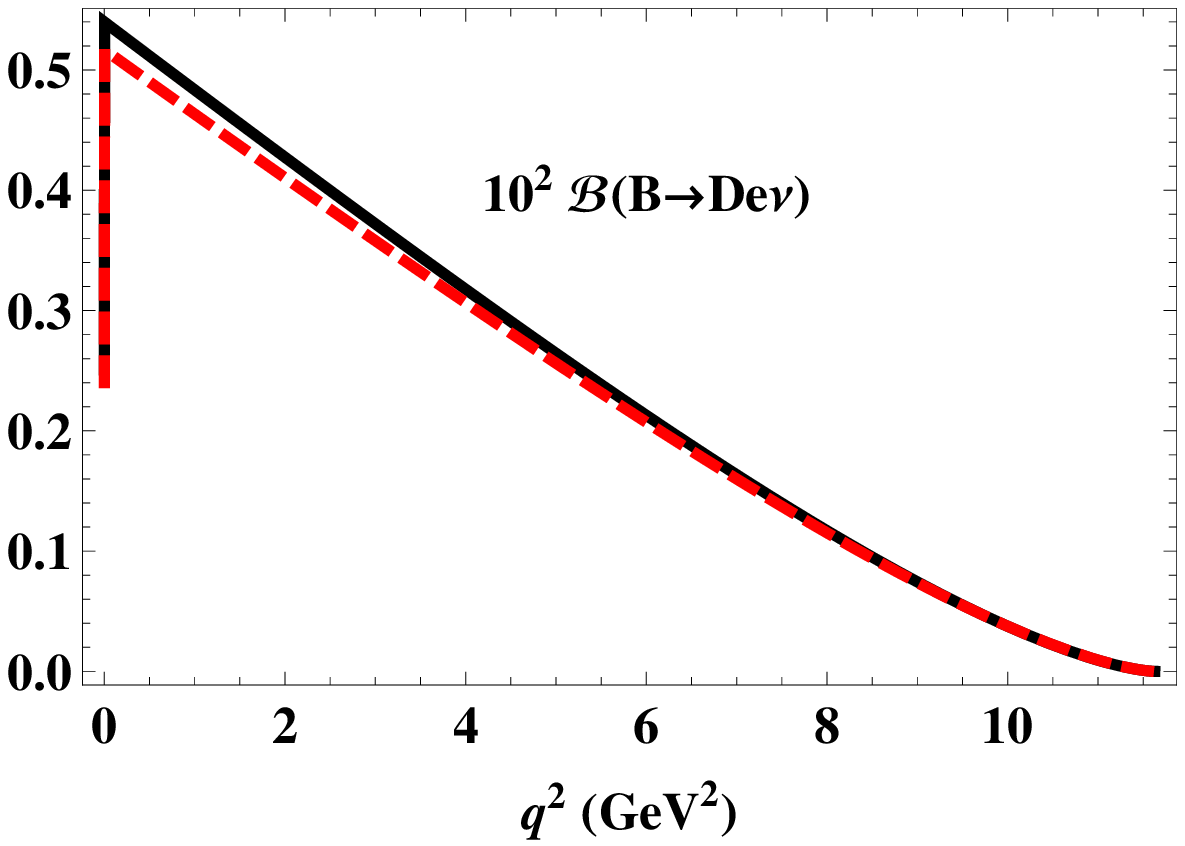} &
\includegraphics[scale=0.6]{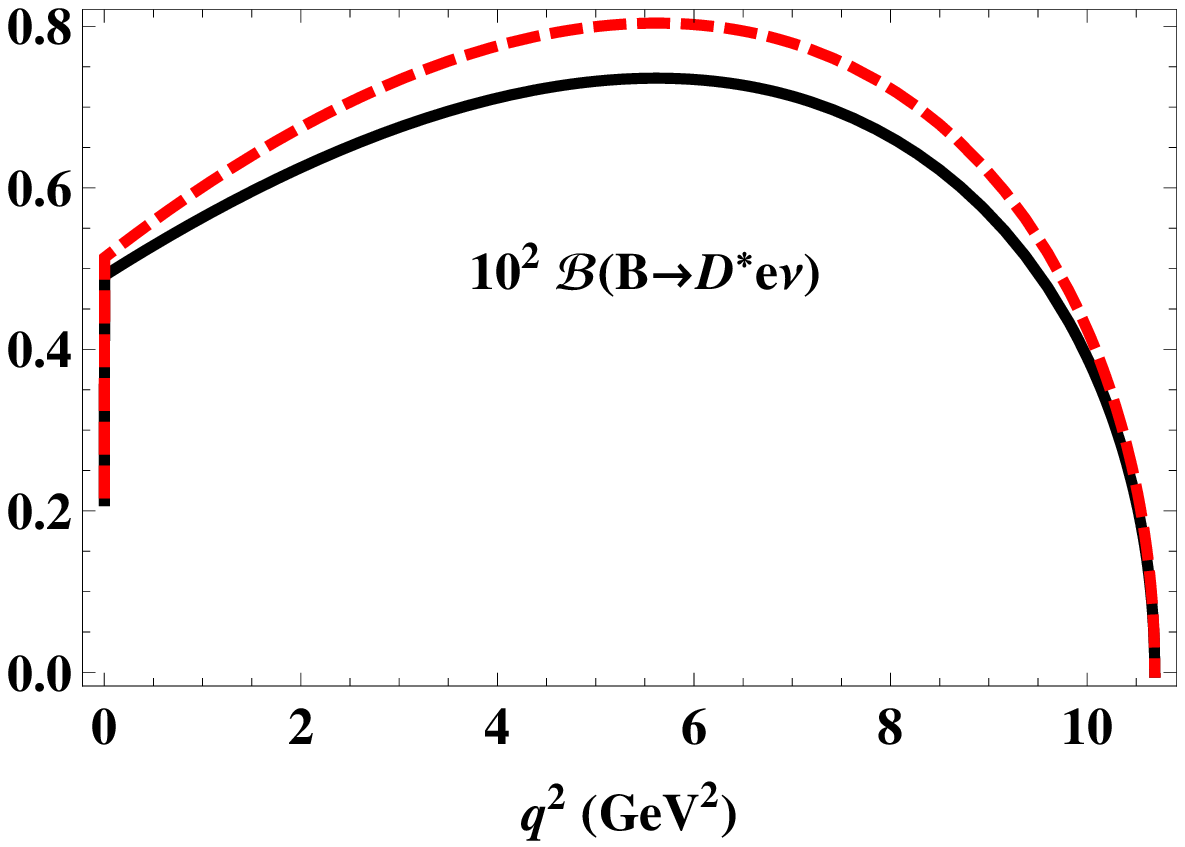}\\
\includegraphics[scale=0.6]{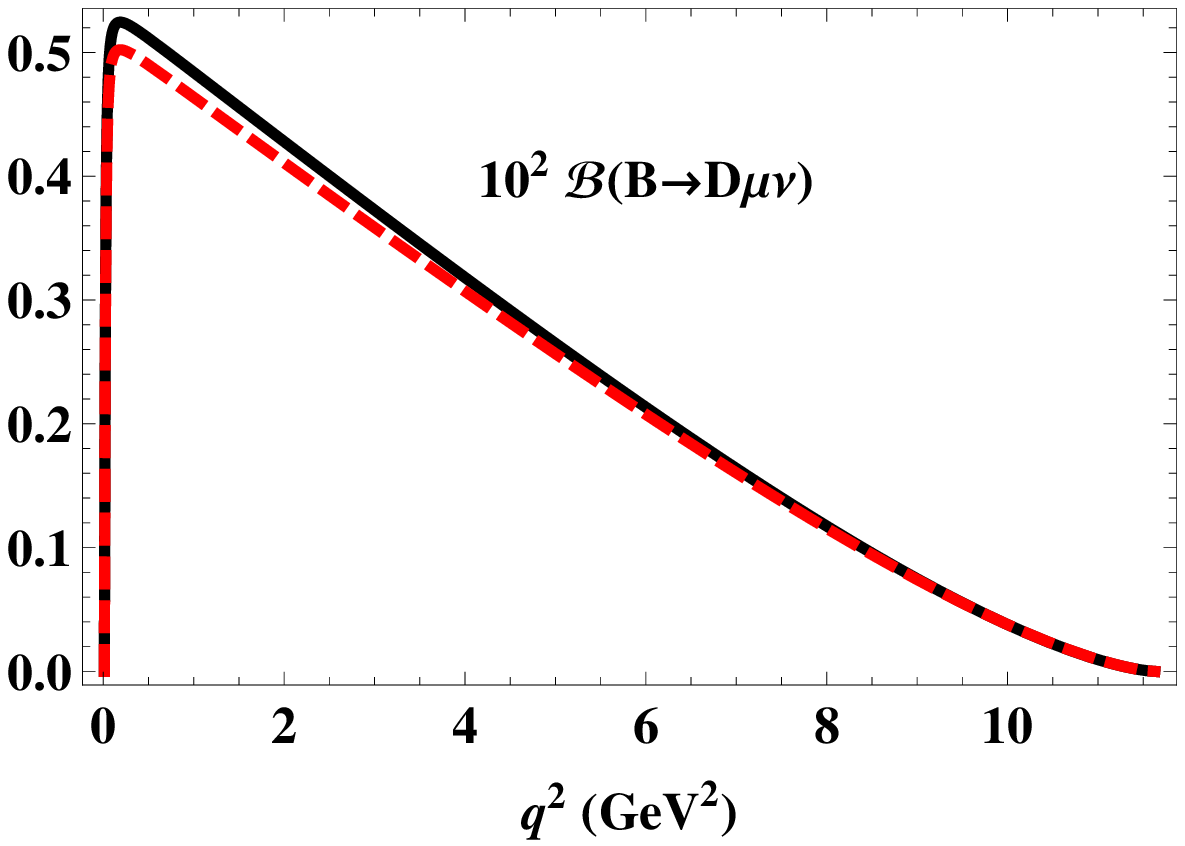}&
\includegraphics[scale=0.6]{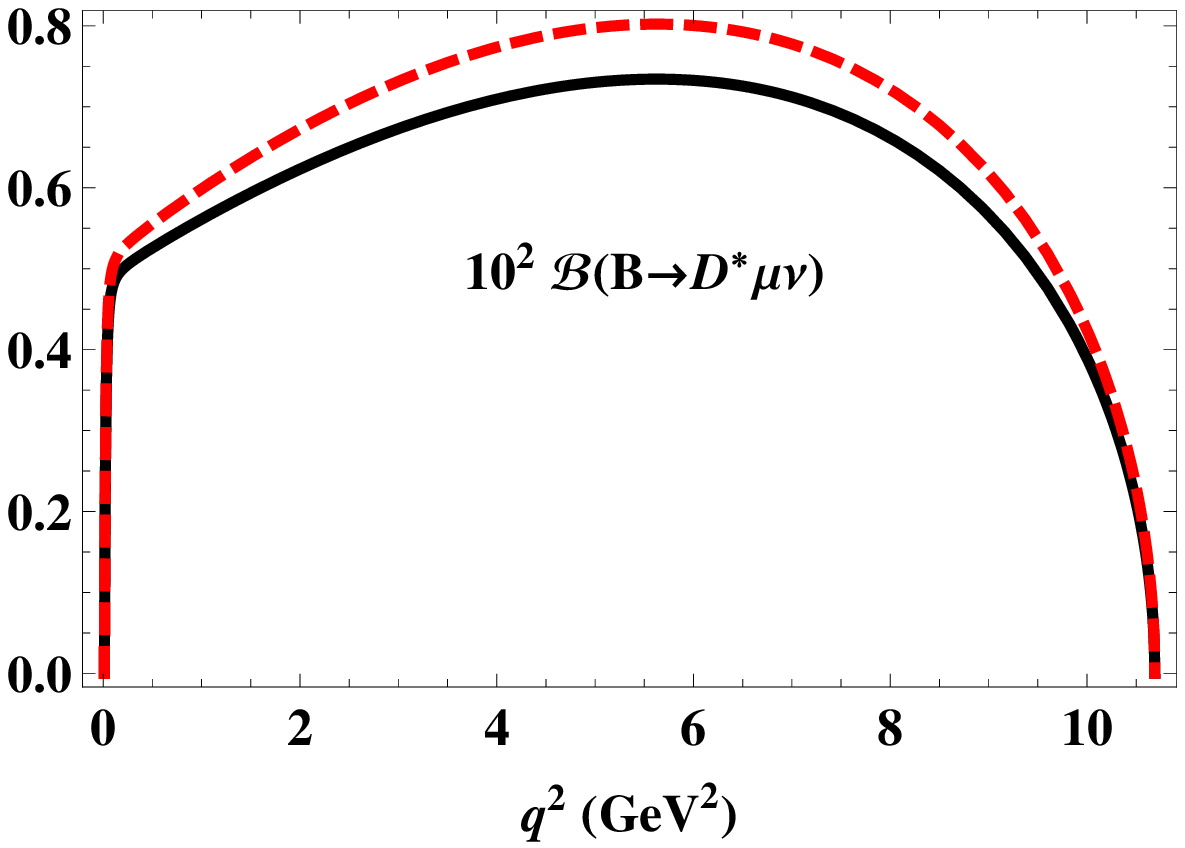}\\
\includegraphics[scale=0.6]{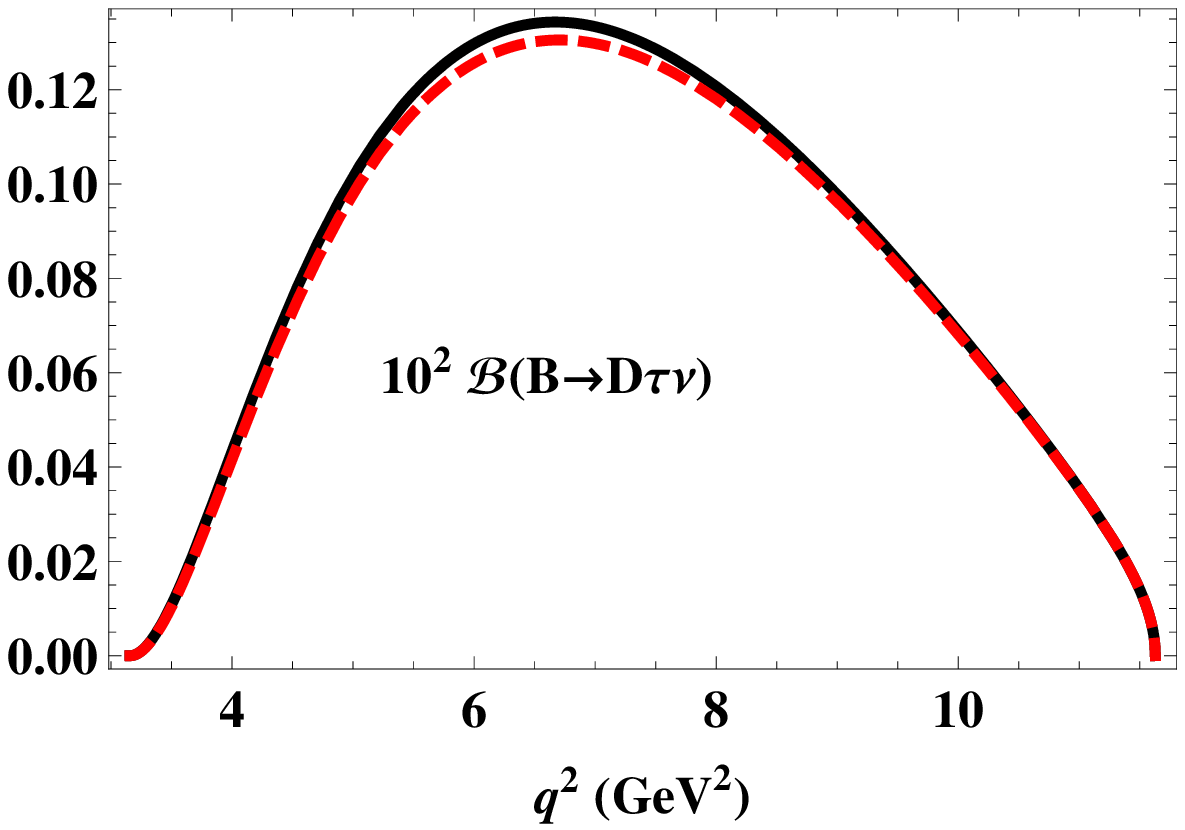}&
\includegraphics[scale=0.6]{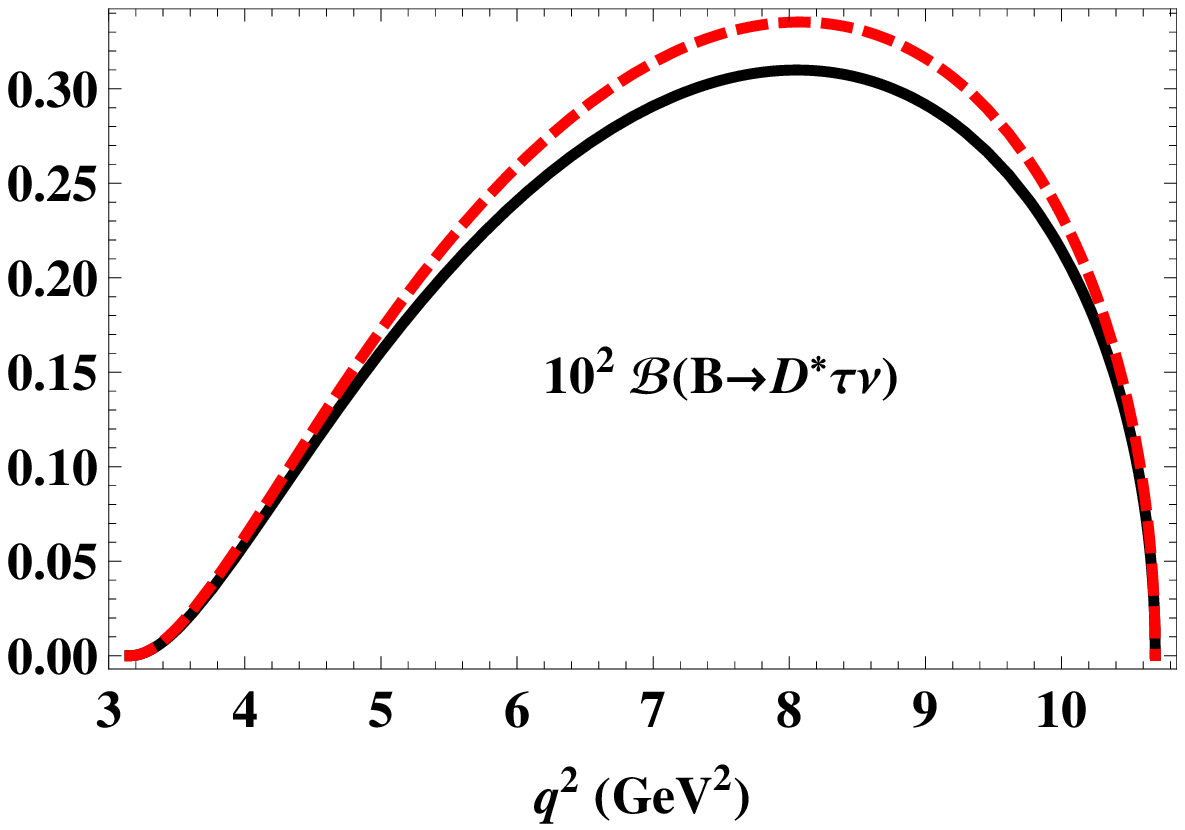}\\
\end{tabular}
\caption{Differential branching fractions of the decays 
$B \to D^{(*)} \ell \nu$. 
The solid lines are the results of exact calculations in our approach,
the dashed lines are the form factors obtained in the heavy quark limit.}
\label{fig:difwidths}
\end{figure}

The values of the lepton and meson masses and their lifetimes are taken from 
Ref.~\cite{Agashe:2014kda}. We also adopt the following values for the 
CKM matrix elements $|V_{ub}|=0.00413$ and $|V_{bc}|=0.0411$.
In Fig.~\ref{fig:difwidths} we represent our results for the differential 
branching fractions of the decays $B \to D^{(*)} \ell \nu$ within the full 
range of the momentum transfer squared. For comparison, we also display
the form factors calculated in heavy quark limit. It is readily seen that
both forms are very close to each other. It confirms that HQET 
works very well in the leading order for $b-c$~transitions. In what follows
we will not display the curves for observables obtained in the HQL.

\begin{figure}[htbp]
\begin{tabular}{lr}
\includegraphics[scale=0.6]{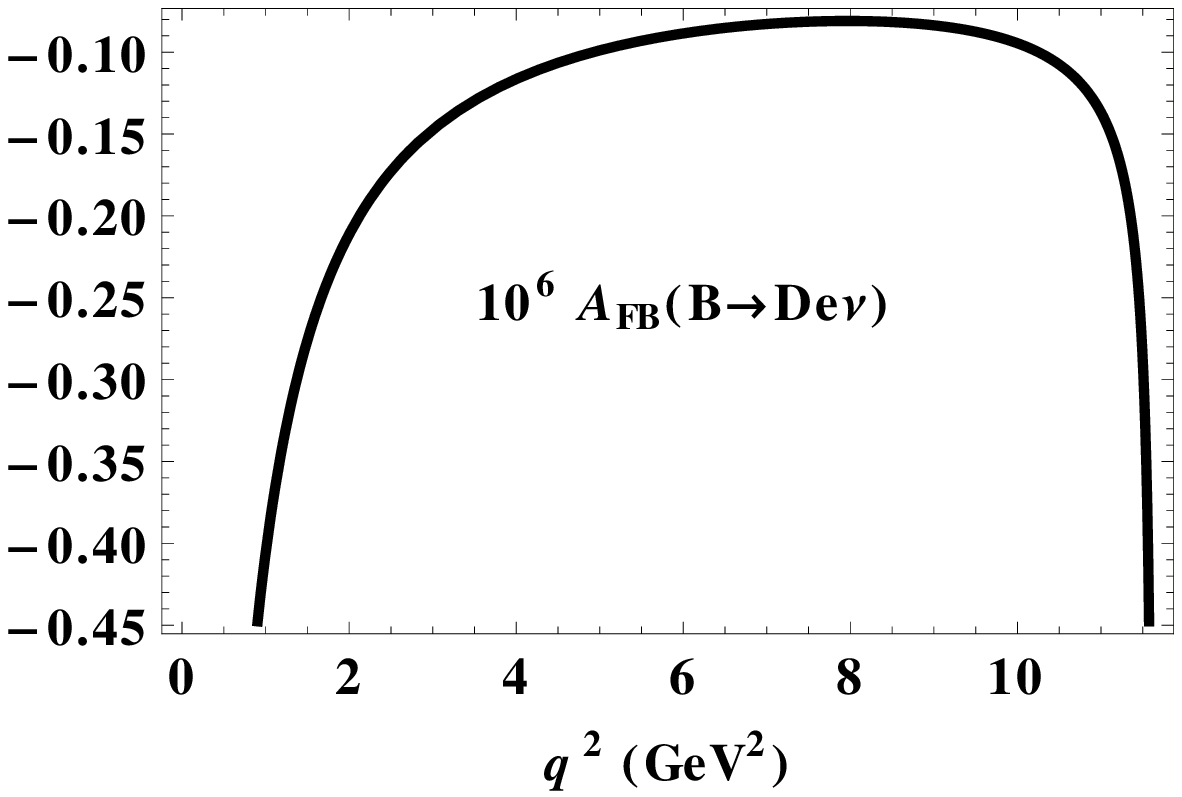} &
\includegraphics[scale=0.6]{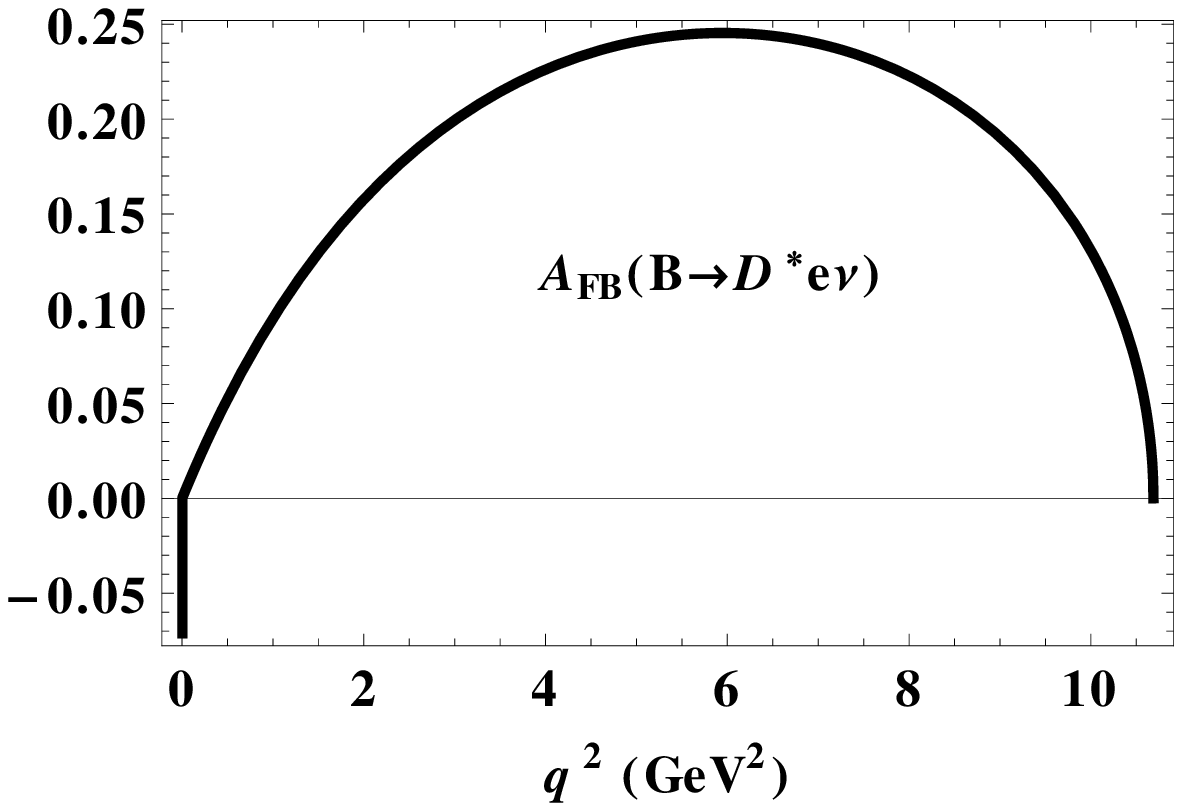}\\
\includegraphics[scale=0.6]{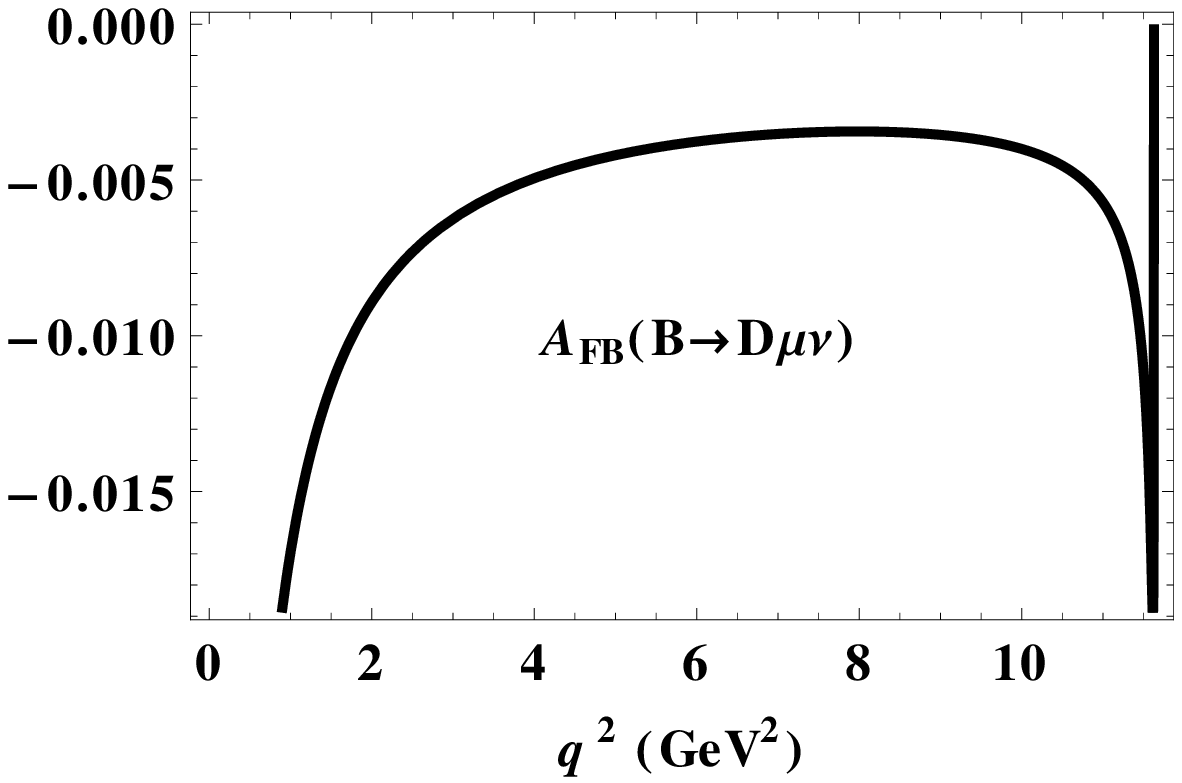}&
\includegraphics[scale=0.6]{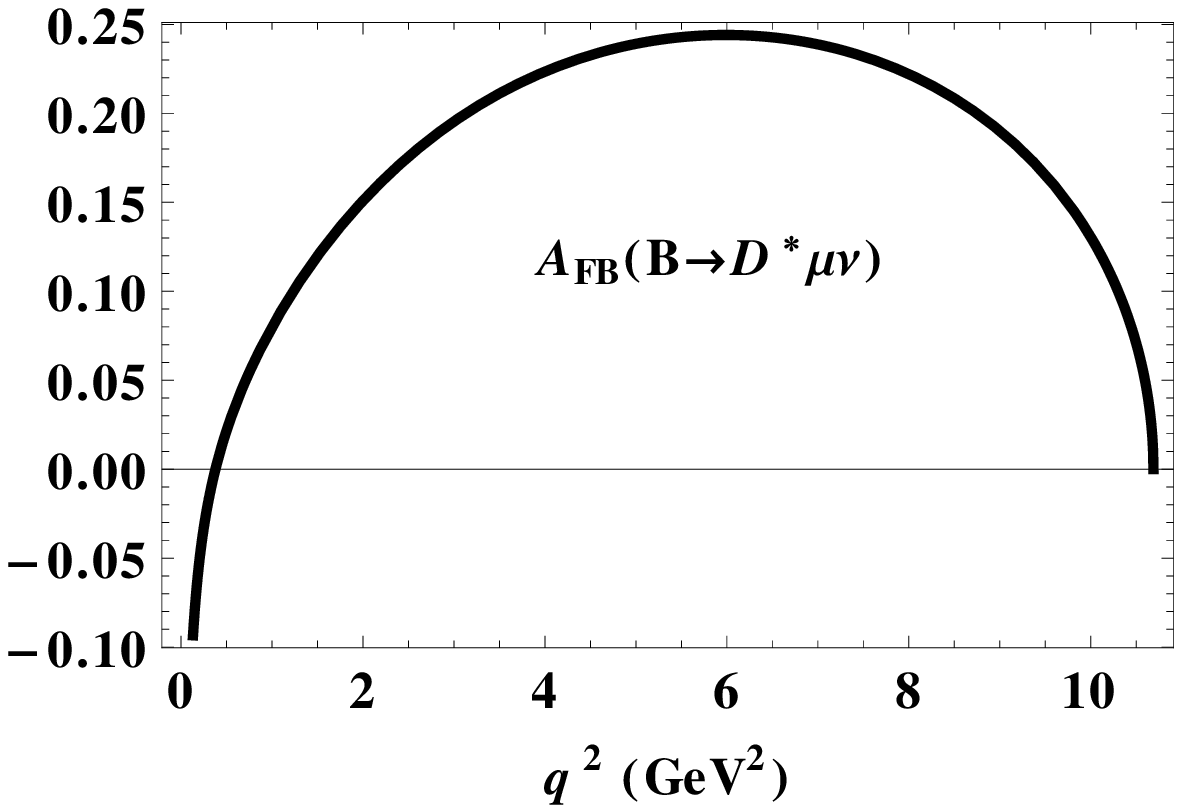}\\
\includegraphics[scale=0.6]{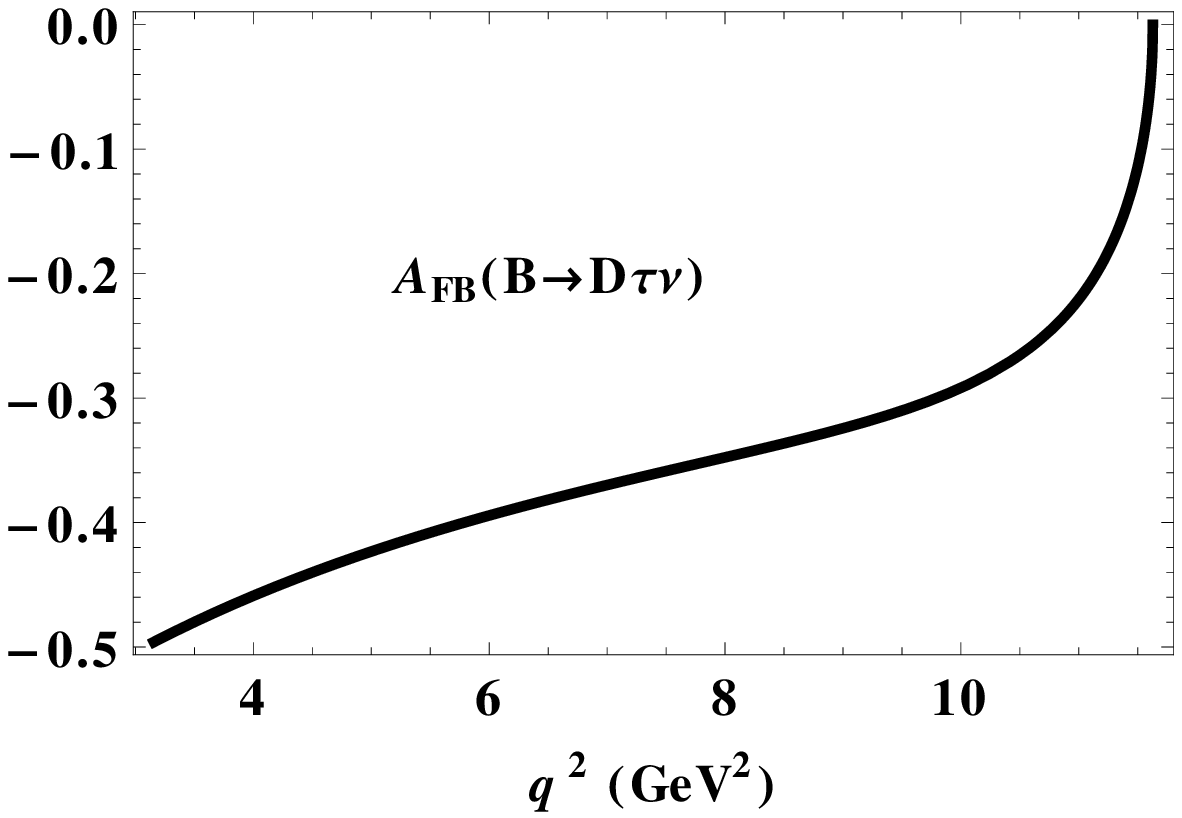}&
\includegraphics[scale=0.6]{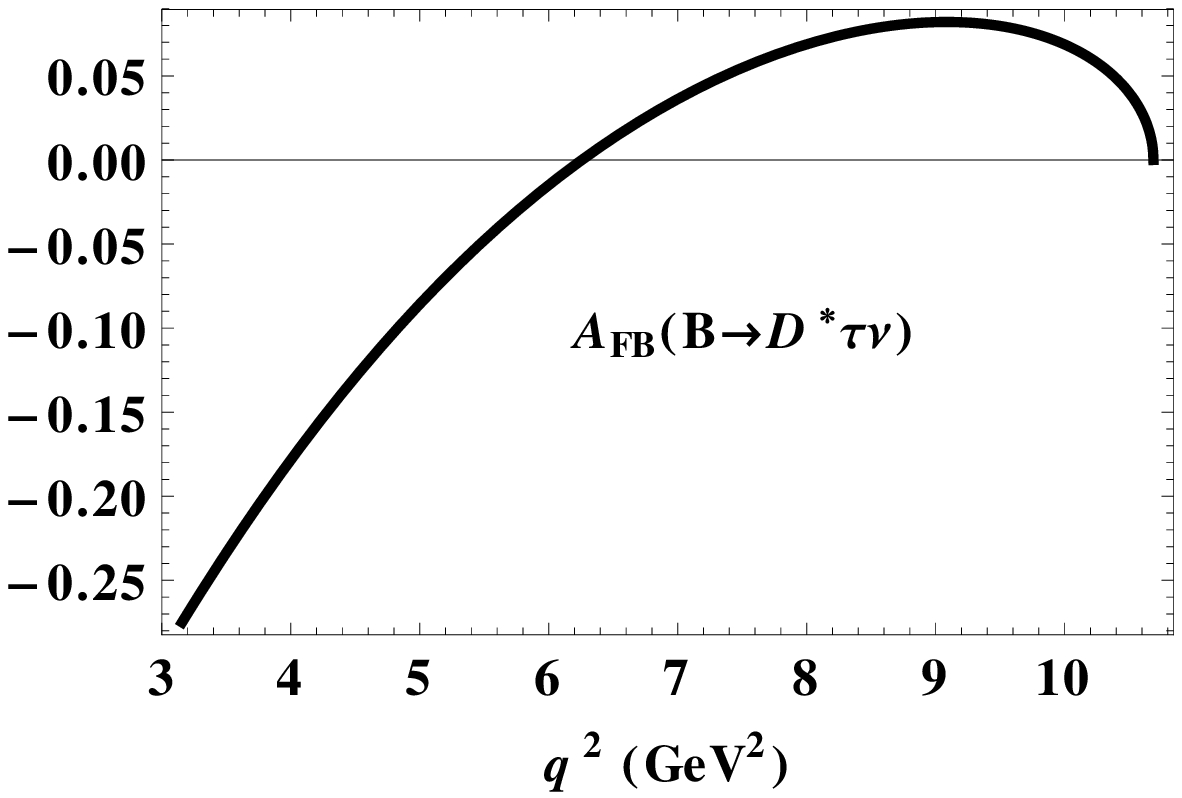}\\
\end{tabular}
\caption{Forward-backward asymmetries of the decays $B \to D^{(*)} \ell \nu$.}
\label{fig:asym}
\end{figure}

In Fig.~\ref{fig:asym} we  represent our results for 
the forward-backward asymmetries of the decays  
$B \to D^{(*)} \ell \nu$ within the full range of the momentum 
transfer squared. The forward-backward asymmetry
for the decay $B \to D\tau^{-}\bar \nu_{\tau}$ is quite large in the lower
half of the $q^{2}$ spectrum which can be understood from the fact that 
$A_{FB}=-3\delta_{\ell}{\cal H}_{SL}$ and that $3\delta(q^{2})$ is large
in the threshold region. It is quite interesting that the forward-backward 
asymmetry for the decay $B \to D^{\ast}\tau^{-}\bar \nu_{\tau}$ goes through
zero at $q^{2}=6.25$ GeV$^{2}$.

The branching fractions of the decays $B \to \ell^-\bar\nu$, 
$B \to D^{(*)} \ell^-\bar\nu$, and
 $B \to \pi\ell^-\bar\nu$, as well as the ratios of branching fractions 
$R(D^{(*)})$ are presented in Tables~\ref{tab:leptBr}, 
\ref{tab:semileptbran} and \ref{tab:ratebran}. The branching fractions 
$\mathcal{B}(B \to \ell^-\bar\nu)$, ($\ell = e,\mu$), satisfy the experimental 
constraints and show good agreement with the CKMfitter results, while 
the branching fraction $\mathcal{B}(B \to\tau^-\bar\nu_\tau)$ is consistent 
with experimental data, giving more constraints on NP effects that may 
contribute to the transitions. The situation is different for the semileptonic 
decays. 
The results for $\mathcal{B}(B \to D^{(*)} \ell^-\bar\nu)$ are slightly larger, 
while the results for $\mathcal{B}(B \to D^{(*)} \tau^-\bar\nu_\tau)$ are 
slightly smaller in comparison with experimental data. As a result, 
the calculated ratios $R(D^{(*)})$ are slightly smaller than the SM 
expectation,
which means they deviate from the experimental values even more. 
This may imply the appearance of NP.
\begin{table}
\begin{tabular}{l c c c l}
\hline\hline
  &  \qquad Unit \qquad & \qquad This work\qquad &\qquad Data\qquad & 
\qquad Ref.\qquad \\
\hline
$\bar{B}^0 \to D^+ \ell^- \bar\nu$ 
 & \qquad  $10^{-2}$\qquad &\qquad $2.74\,(2.65)$ \qquad &
\qquad $2.17\pm0.12$ \qquad  & 
\qquad HFAG~\cite{Asner:2010qj} 
\\ 
 &            &       &\qquad $2.21\pm0.16$   & 
\qquad BABAR~\cite{Aubert:2007qw}
\\[1.2ex]
$\bar{B}^0 \to D^+ {\tau}^- \bar\nu_\tau$ 
 & \qquad $10^{-2}$ \qquad  & \qquad $0.73\,(0.71)$\qquad  & 
\qquad $1.02\pm0.17$ \qquad   & 
\qquad BABAR~\cite{Lees:2012xj}
\\[1.2ex]
$\bar{B}^0 \to D^{\ast\,+} \ell^- \bar\nu$ 
 & \qquad $10^{-2}$ \qquad & \qquad $6.64\,(7.21)$  \qquad & 
\qquad $5.05 \pm 0.12$ \qquad & \qquad HFAG~\cite{Asner:2010qj}
\\ 
 &          &       & \qquad $5.49 \pm 0.30$ \qquad & 
\qquad BABAR~\cite{Aubert:2007qw}
\\[1.2ex]
$\bar{B}^0 \to D^{\ast\,+} \tau^- \bar\nu_\tau$ 
 & \qquad $10^{-2}$ \qquad & \qquad $1.57\,(1.70)$ \qquad & 
\qquad $1.76\pm0.18$ \qquad & \qquad BABAR~\cite{Lees:2012xj}
\\[1.2ex]
\hline
$\bar{B}^0 \to \pi^+ \ell^- \bar\nu$ 
 & \qquad $10^{-4}$ \qquad & \qquad 1.69 \qquad & 
\qquad $1.41 \pm 0.09$ \qquad & 
\qquad BABAR~\cite{delAmoSanchez:2010af}
\\
 &          & & \qquad $1.49 \pm 0.08 $ \qquad & \qquad Belle~\cite{Ha:2010rf}
\\
$\bar{B}^0 \to \pi^+ \tau^- \bar\nu_\tau$ 
 & \qquad $10^{-4} $ \qquad & \qquad 1.01 \qquad & \qquad $\ldots$ \qquad & 
\qquad $\ldots$ \\[1.2ex]
\hline\hline
\end{tabular}
\caption{Semileptonic decay branching fractions of $B$ meson.
The values obtained in the HQL are given in brackets.
The experimental errors are combined in quadrature.}
\label{tab:semileptbran}
\end{table}
\begin{table}[htbp]
\begin{tabular}{cccc}
\hline\hline
\qquad\qquad & \qquad This work \qquad & \qquad SM \qquad  
& \qquad  Data \qquad   
\\
\hline
\qquad $R(D)$ \qquad   & \qquad $ 0.265\,(0.268) $ \qquad & 
\qquad $ 0.297 \pm 0.017 $ \qquad & \qquad $ 0.388 \pm 0.047 $ \qquad
\\[1.2ex]
\qquad $R(D^\ast)$ \qquad &\qquad  $ 0.237\,(0.235) $ \qquad &
\qquad $ 0.252 \pm 0.003 $ \qquad & \qquad $ 0.321 \pm 0.021 $\qquad
 \\[1.2ex]
\hline\hline
\end{tabular}
\caption{Ratios of branching fractions $R(D)$  and $R(D^\ast)$ calculated
in our model (the values obtained in the HQL are given in brackets) 
and compared with the SM expectations and experimental data.
}
\label{tab:ratebran}
\end{table}

\begin{figure}[htbp]
\begin{center}
\vspace*{.25cm}
\epsfig{figure=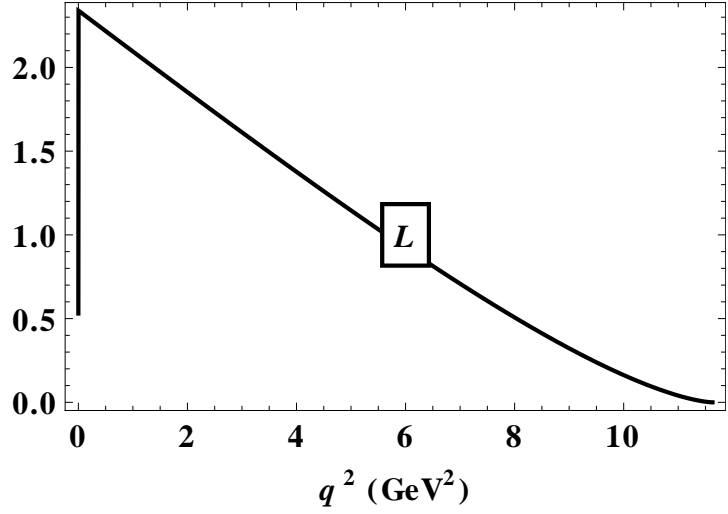,scale=.8}
\caption{ {\boldmath\bf $B\to D$~transition:} 
the  $q^{2}$ dependence of the partial rate
$d\Gamma_{L}/dq^{2}$ 
for the $e^-$ mode (in units of $10^{-15}$~GeV$^{-1}$).
\label{fig:dL.BD}
}
\end{center}
\end{figure}

\begin{figure}[htbp]
\begin{center}
\begin{tabular}{lr}
        \epsfig{figure=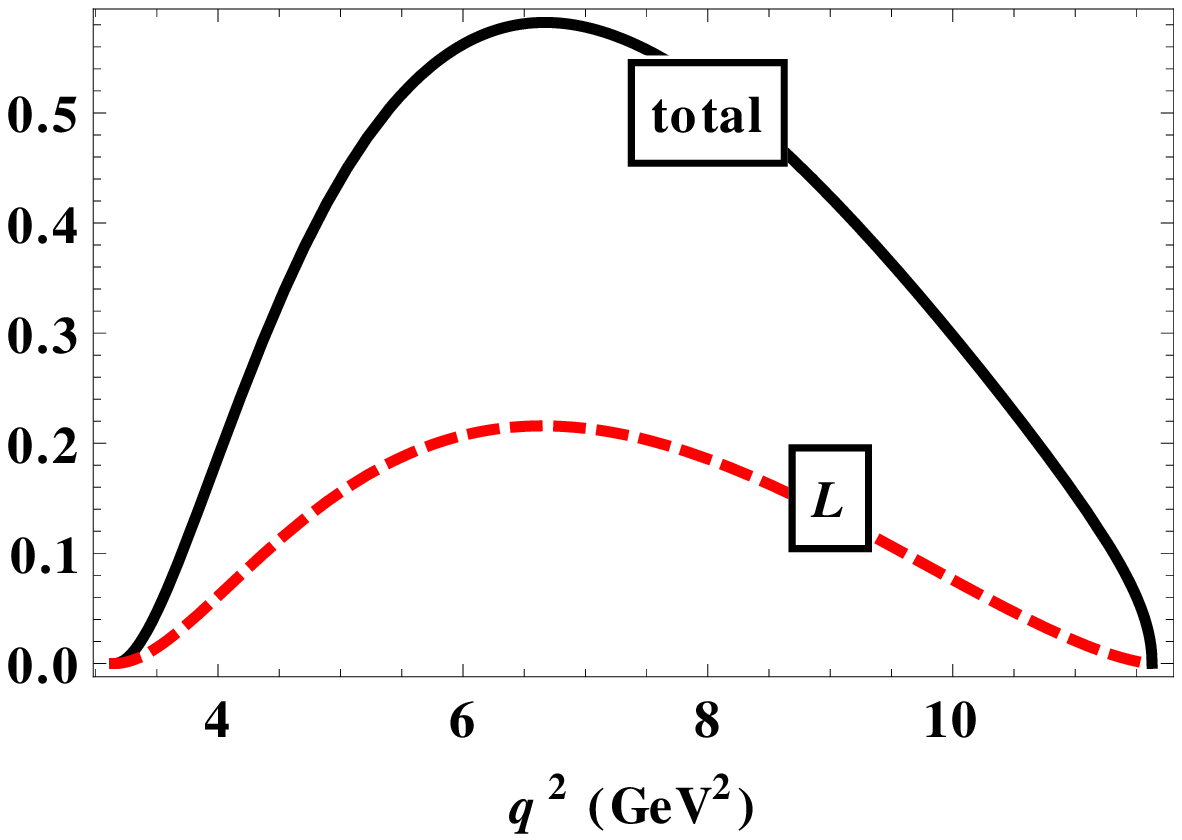,scale=0.55} 
\qquad &  \qquad  
        \epsfig{figure=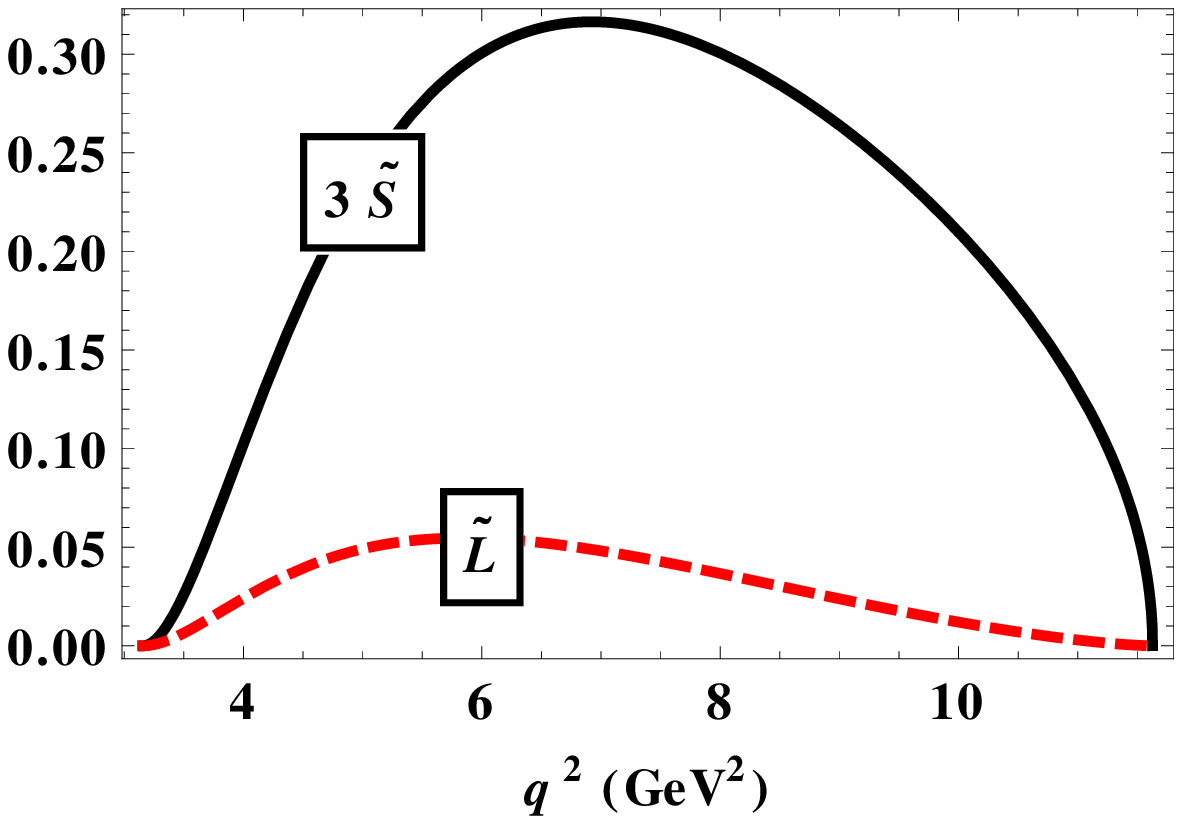,scale=0.55} 
\end{tabular}
\caption{ {\boldmath\bf $B\to D$~transition:} 
the  $q^{2}$ dependence of the partial nonflip rates
$d\Gamma_{L}/dq^{2}$, and the flip rates
$  d\,\widetilde\Gamma_{U,L}/d\,q^{2}$ and 
$3\,d\,\widetilde \Gamma_{S}/d\,q^{2}$ for the $\tau^-$ mode 
(in units of $10^{-15}$~GeV$^{-1}$). Also shown is the total rate 
$d\,\Gamma_{L}/d\,q^{2} 
+  d\,\widetilde\Gamma_{L}/d\,q^{2} + 3\,d\,\widetilde\Gamma_{S}/d\,q^{2}$. 
\label{fig:dLtot.BD}
}
\end{center}
\end{figure}

\begin{figure}[htbp]
\begin{center}
\vspace*{.25cm}
\epsfig{figure=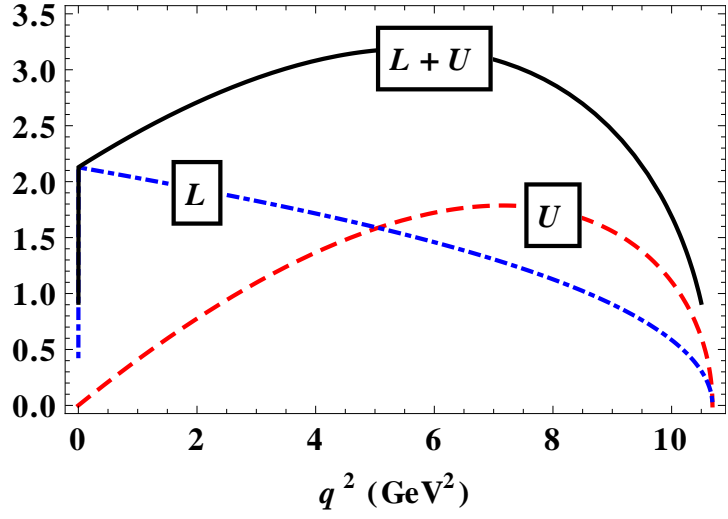,scale=.8}
\caption{ {\boldmath\bf $B\to D^\ast$~transition:} 
the  $q^{2}$ dependence of the partial rates
$d\Gamma_{U}/dq^{2}$ (dashed), $d\Gamma_{L}/dq^{2}$ (dot-dashed) and their 
sum $d\Gamma_{U+L}/dq^{2}$ (solid)
for the $e^-$ mode (in units of $10^{-15}$~GeV$^{-1}$).
\label{fig:dUL.BDv}
}
\end{center}
\end{figure}

\begin{figure}[htbp]
\begin{center}
\begin{tabular}{lr}
        \epsfig{figure=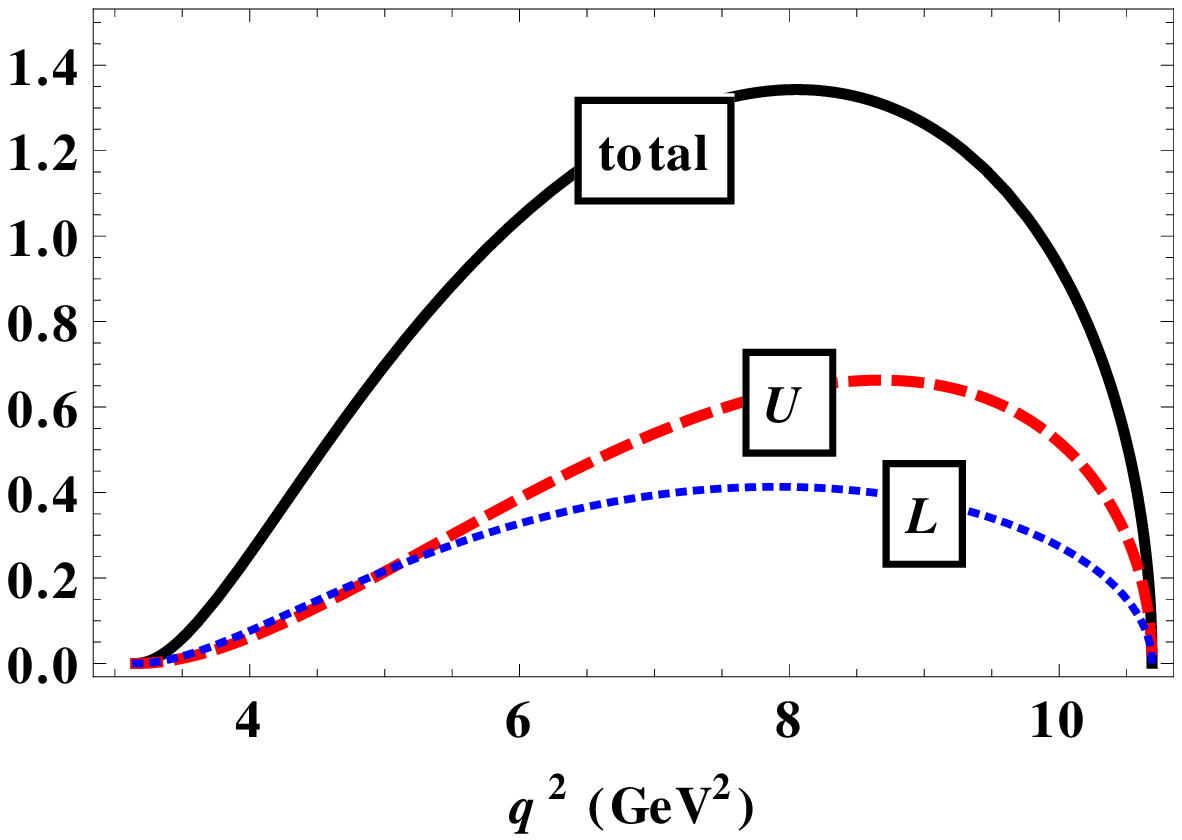,scale=0.5} 
\qquad &  \qquad  
        \epsfig{figure=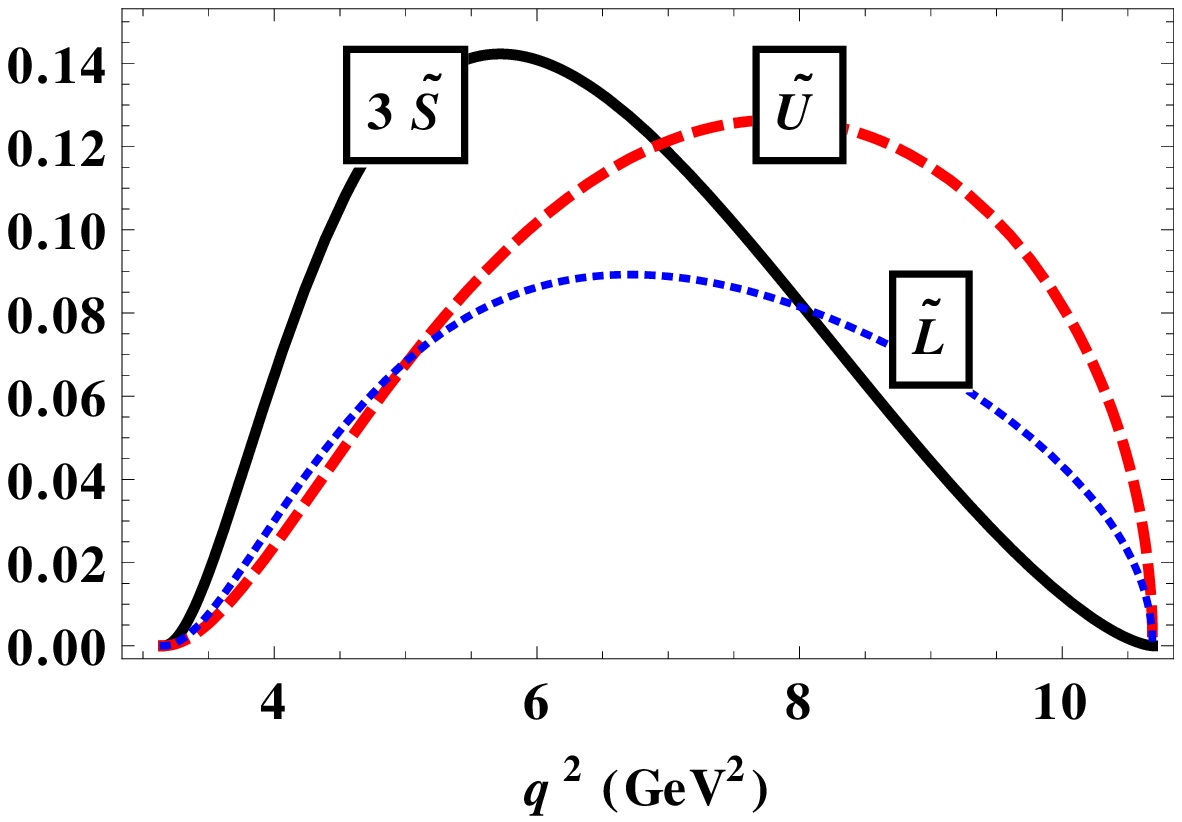,scale=0.5} 
\end{tabular}
\caption{ {\boldmath\bf $B\to D^\ast$~transition:} 
the  $q^{2}$ dependence of the partial nonflip rates
$d\Gamma_{U,L}/dq^{2}$, and the flip rates
$d\,\widetilde\Gamma_{U,L}/d\,q^{2}$ and 
$3\,d\,\widetilde \Gamma_{S}/d\,q^{2}$ for the $\tau^-$ mode 
(in units of $10^{-15}$~GeV$^{-1}$). Also shown is the total rate 
$d\,\Gamma_{U+L}/d\,q^{2}+ d\,\widetilde\Gamma_{U+L}/d\,q^{2}+
3\,d\, \widetilde\Gamma_{S}/d\,q^{2}$. 
\label{fig:dULS.BDv}
}
\end{center}
\end{figure}

Next we define the partial helicity rates by
\be
\frac{d\Gamma_X}{dq^2} =
\frac{G_F^2}{(2\pi)^3}\frac{ |V_{cb}|^2 |{\bf p_2}| q^2 v^2}{12 m_1^2}\,
{\cal H}_X, \qquad
\frac{d\widetilde\Gamma_X}{dq^2}=\delta_\ell\,\frac{d\Gamma_X}{dq^2},
\label{eq:hel_rates}
\en
where $X=U,L,P,\ldots$
In Figs.~\ref{fig:dL.BD} and \ref{fig:dUL.BDv} we display 
the $q^{2}$ dependence of the partial differential rates 
$d\Gamma_{U}/dq^{2}$, $d\Gamma_{L}/dq^{2}$, and 
the total differential rate $d\Gamma_{U+L}/dq^{2}$ for the $e$ mode.
The transverse rate dominates in the low recoil region while the longitudinal 
rate dominates in the large recoil region. The longitudinal and thereby the
total rate show a step-like behavior near the threshold $q^{2}=m^{2}_{e}$.
Figs.~\ref{fig:dLtot.BD} and \ref{fig:dULS.BDv} show the corresponding plots 
for the $\tau$ mode including the partial flip rates  
$d\,\widetilde\Gamma_{U,L}/d\,q^{2}$ and $3\,d\,\widetilde\Gamma_{S}/d\,q^{2}$. 
We also show the total differential rate 
$d\Gamma_{U+L}/dq^{2} + d\,\widetilde\Gamma_{U+L+3S}/d\,q^{2}$.
The helicity flip rates are smaller than the helicity nonflip rates but 
contribute significantly to the total rate.

\begin{figure}[htbp]
\begin{center}
\vspace*{.25cm}
\epsfig{figure=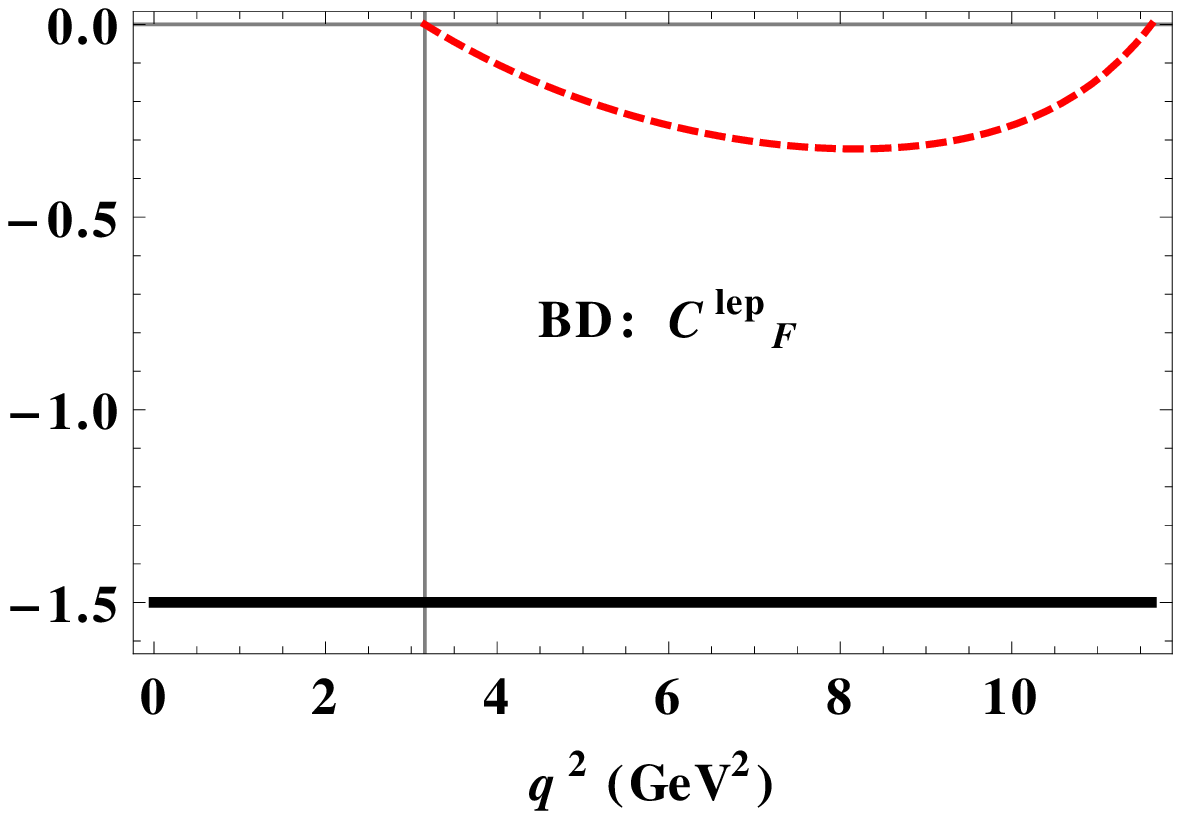,scale=.9}
\caption{ {\boldmath\bf $B\to D$~transition:} 
the  $q^{2}$ dependence of the lepton convexity parameter
$C^\ell_{F}(q^{2})$ for the $e^-$- (solid) and $\tau^-$-mode (dashed).
\label{fig:dCFL.BD}
}
\end{center}
\end{figure}

\begin{figure}[htbp]
\begin{center}
\vspace*{.25cm}
\epsfig{figure=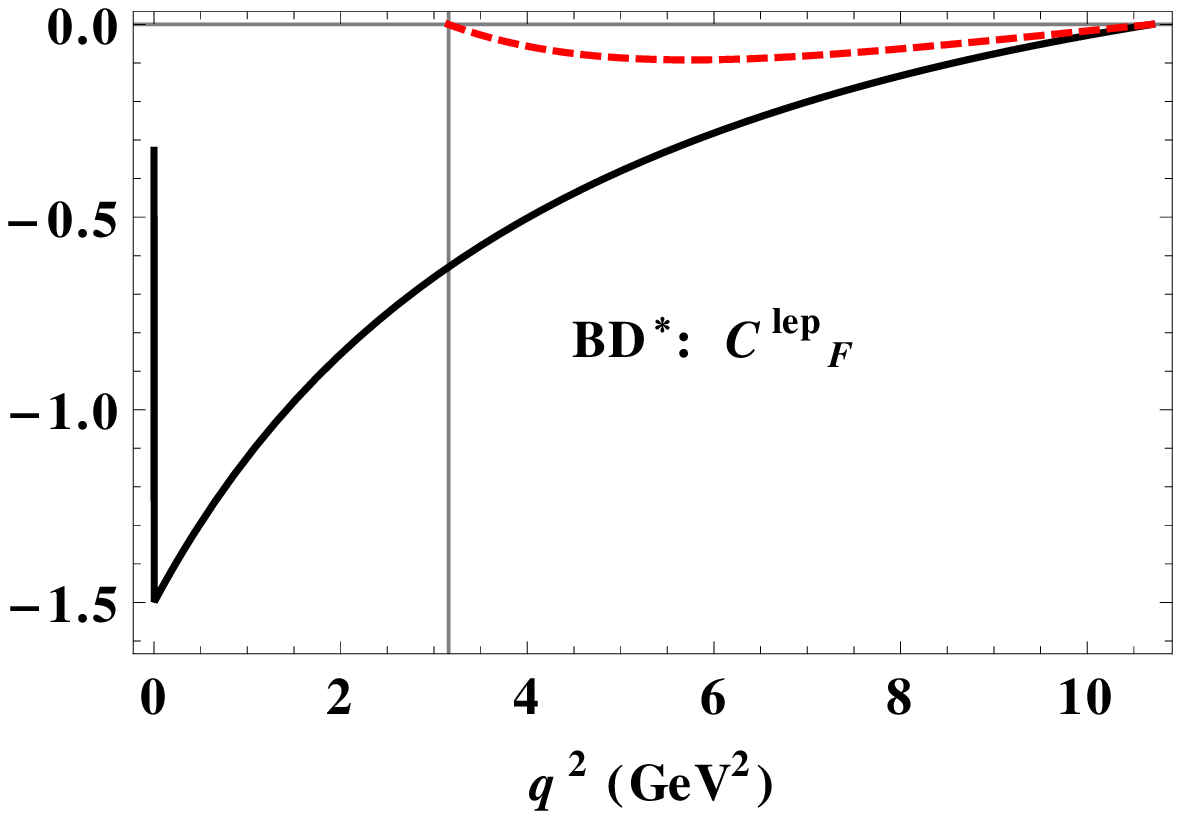,scale=.9}
\caption{ {\boldmath\bf $B\to D^\ast$~transition:} 
the  $q^{2}$ dependence of the lepton convexity parameter
$C^\ell_{F}(q^{2})$ for the $e^-$- (solid) and $\tau^-$-mode (dashed).
\label{fig:dCFL.BDv}
}
\end{center}
\end{figure}

\begin{figure}[htbp]
\begin{center}
\vspace*{.25cm}
\epsfig{figure=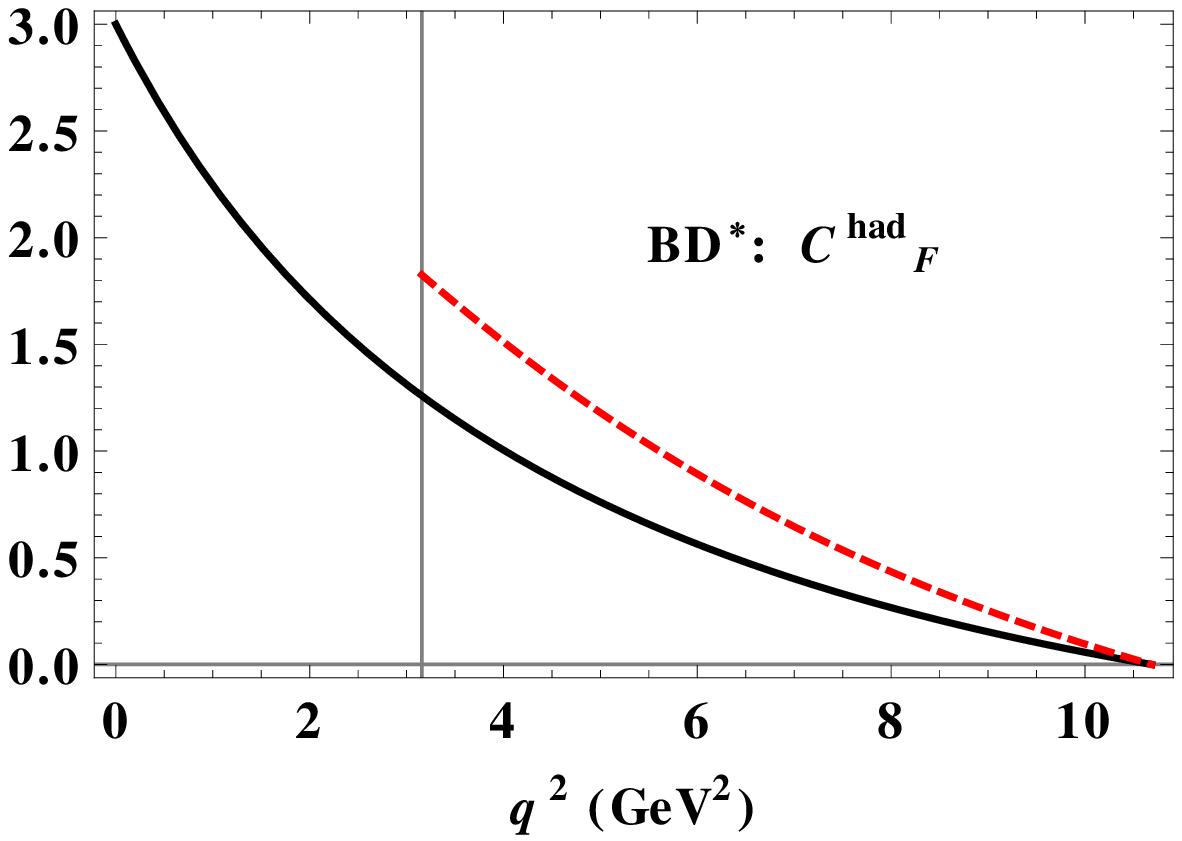,scale=.9}
\caption{ {\boldmath\bf $B\to D^\ast$~transition:} 
the  $q^{2}$ dependence of the hadron convexity parameter
$C^h_{F}(q^{2})$ for the $e^-$- (solid) and $\tau^-$-mode (dashed).
\label{fig:dCFH.BDv}
}
\end{center}
\end{figure}

In Figs.~\ref{fig:dCFL.BD},  \ref{fig:dCFL.BDv}, 
and \ref{fig:dCFH.BDv} we display the $q^{2}$ dependence 
of the convexity parameters $C^{\ell}_{F}$ and $C^{h}_{F}$ for the lepton 
and hadron sides defined in Eqs.~(\ref{eq:convex_lep}) and
(\ref{eq:convex_had}). In the $B \to D$ case the $\cos\theta$ distribution
is described by a downward open parabola which becomes much flatter for
the $\tau$ mode.
We do not plot the hadron-side convexity parameter $C_F^h(q^{2})$ for the 
$B\to D$~transition since it trivially reads $C_F^h=3$ following from the definition~(\ref{eq:convex_had}). For the $B\to D^{\ast}$
transition the lepton-side $\cos\theta$ distribution is again described by
a downward open parabola which becomes almost flat for the $\tau$ mode.
The hadron-side $\cos\theta^{\ast}$ distribution is described by an upward
open parabola which does not become flat at the zero-recoil point.
Lepton mass effects are not very pronounced.

\begin{figure}[htbp]
\begin{center}
\vspace*{.25cm}
\epsfig{figure=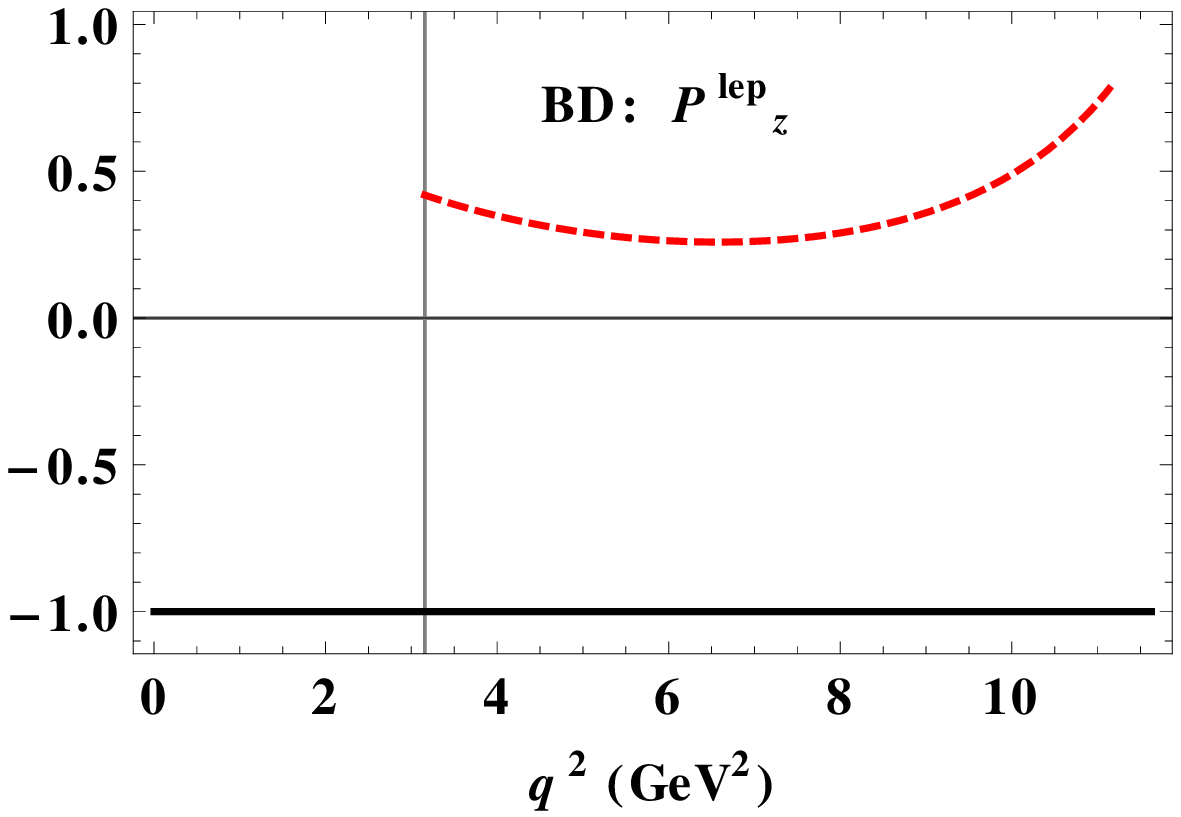,scale=.9}
\caption{{\boldmath\bf $B\to D$~transition:} 
The  $q^{2}$ dependence of the longitudinal polarization component
$P^\ell_z(q^{2})$ for the charged leptons $e^-$- (solid) and $\tau^-$-mode 
(dashed).
\label{fig:dPz_Lept.BD}
}
\end{center}
\end{figure}

\begin{figure}[htbp]
\begin{center}
\vspace*{.25cm}
\epsfig{figure=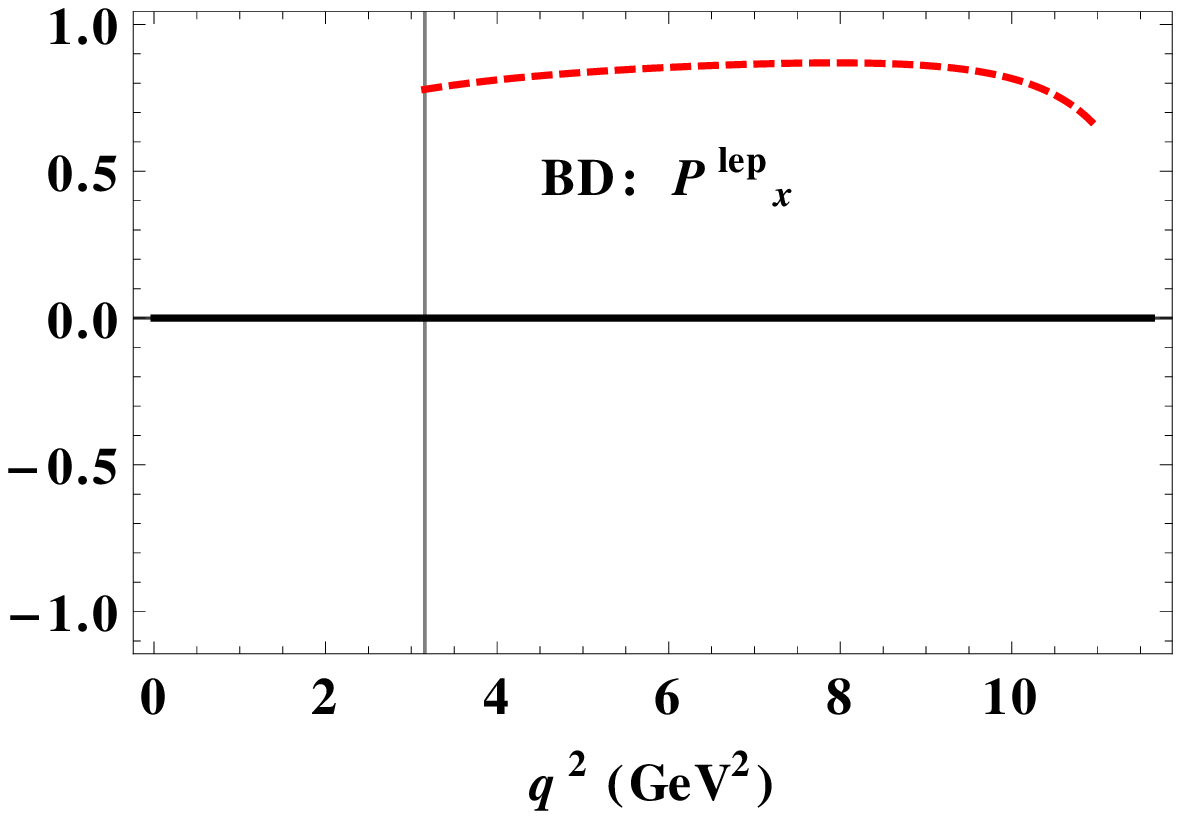,scale=.9}
\caption{{\boldmath\bf $B\to D$~transition:}
the  $q^{2}$ dependence of the transverse polarization component
$P^\ell_x(q^{2})$ for the charged leptons $e^-$- (solid) and $\tau^-$-mode 
(dashed).
\label{fig:dPx_Lept.BD}
}
\end{center}
\end{figure}

\begin{figure}[htbp]
\begin{center}
\vspace*{.25cm}
\epsfig{figure=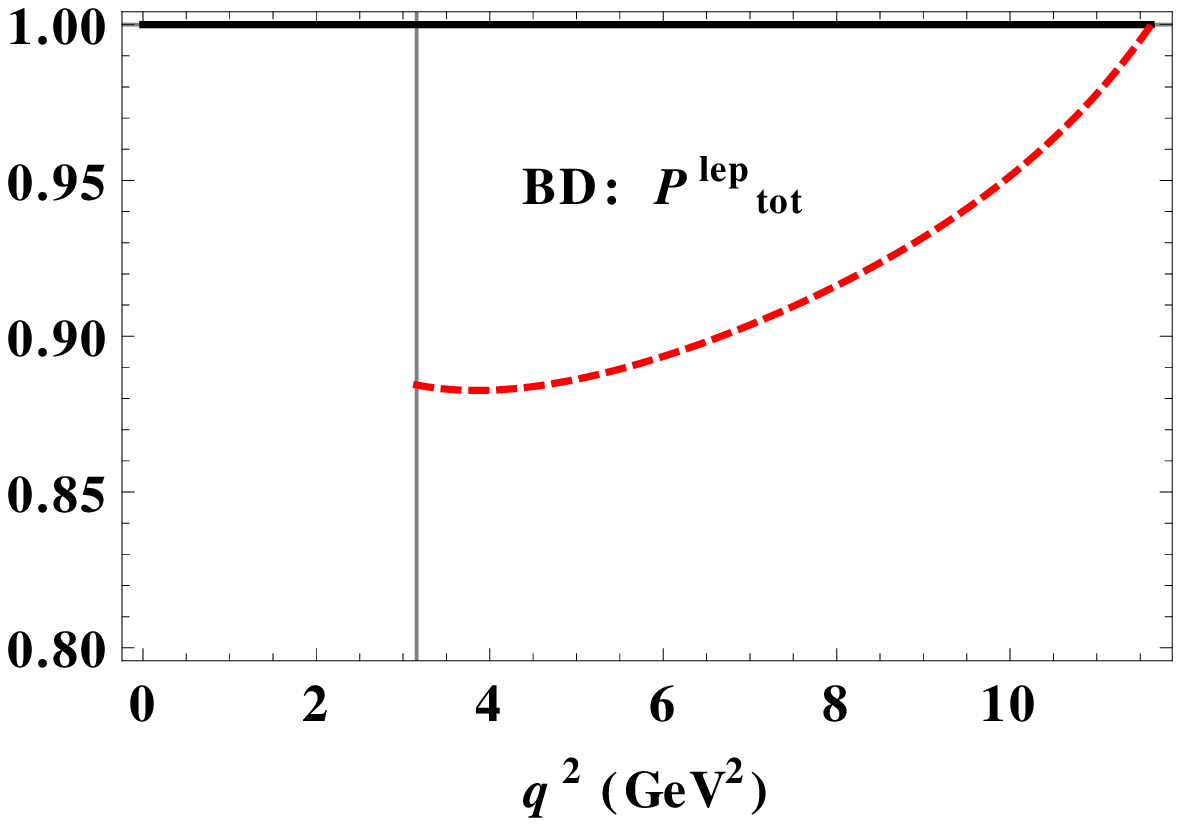,scale=.9}
\caption{{\boldmath\bf $B\to D$~transition:}
the  $q^{2}$ dependence of the total lepton polarization
$|\vec{P}\,^\ell|(q^{2})=\sqrt{(P^\ell_x)^2+(P^\ell_z)^2}$
for the $e^-$- (solid) and $\tau^-$-mode (dashed).
\label{fig:dPLtot.BD}
}
\end{center}
\end{figure}

In Figs.~\ref{fig:dPz_Lept.BD}, \ref{fig:dPx_Lept.BD}, and \ref{fig:dPLtot.BD}
we show plots of the $q^{2}$ dependence of the longitudinal, transverse
and total polarization of the lepton for the 
$B \to D \ell^{-}\bar \nu_{\ell}$ transition. In the case of
the electron the curves reflect the chiral limit of a massless lepton 
in which the lepton is purely left-handed, i.e. one has $P^{\ell}_{z}=-1$,
$P^{\ell}_{x}=0$, and $|\vec P^{\ell}|=1$. For $\ell=\tau$ the transverse
polarization is large and positive and dominates the total polarization.
The transverse polarization of the $\tau$ drops out after
the appropiate azimuthal averaging, as has been done in~\cite{Tanaka:2010se}.
Note that the transverse polarization in the $\tau$ mode results solely
from the scalar-longitudinal interference contribution ${\cal H}_{SL}$.
The longitudinal polarization has switched its sign relative to the
$m_{\ell}=0$ case. 

\begin{figure}[htbp]
\begin{center}
\vspace*{.25cm}
\epsfig{figure=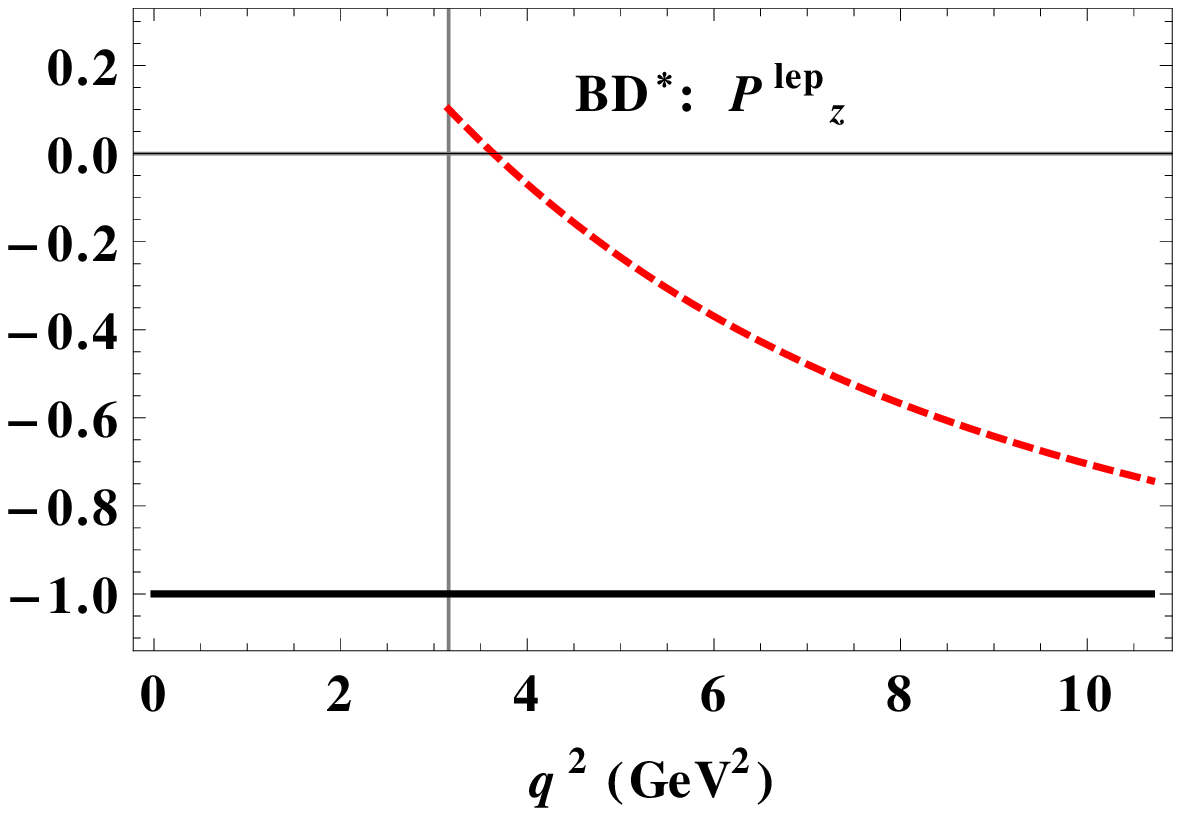,scale=.9}
\caption{{\boldmath\bf $B\to D^\ast$~transition:} 
the  $q^{2}$ dependence of the longitudinal polarization component
$P^\ell_z(q^{2})$ for the charged leptons $e^-$- (solid) and $\tau^-$-mode 
(dashed).
\label{fig:dPz_Lept.BDv}
}
\end{center}
\end{figure}

\begin{figure}[htbp]
\begin{center}
\vspace*{.25cm}
\epsfig{figure=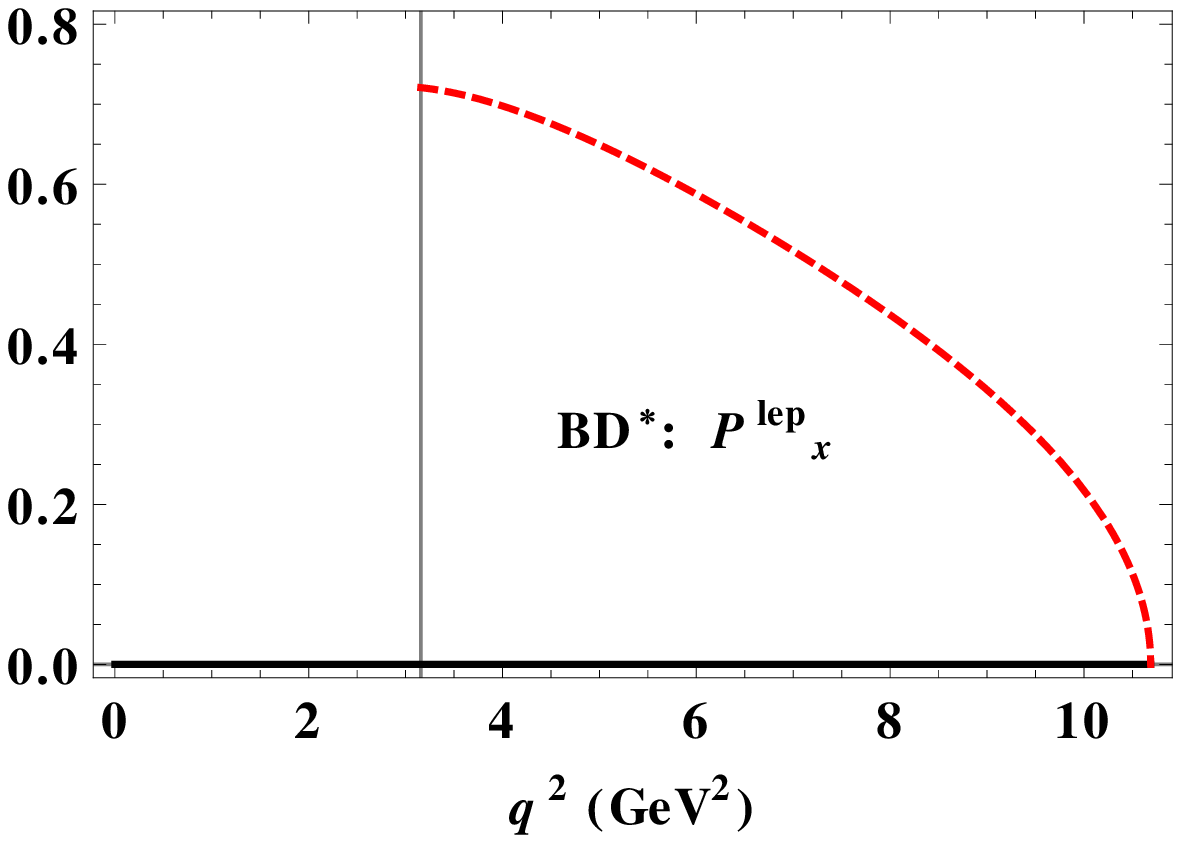,scale=.9}
\caption{{\boldmath\bf $B\to D^\ast$~transition:}
the  $q^{2}$ dependence of the transverse polarization component
$P^\ell_x(q^{2})$ for the charged leptons $e^-$- (solid) and $\tau^-$-mode 
(dashed).
\label{fig:dPx_Lept.BDv}
}
\end{center}
\end{figure}

\begin{figure}[htbp]
\begin{center}
\vspace*{.25cm}
\epsfig{figure=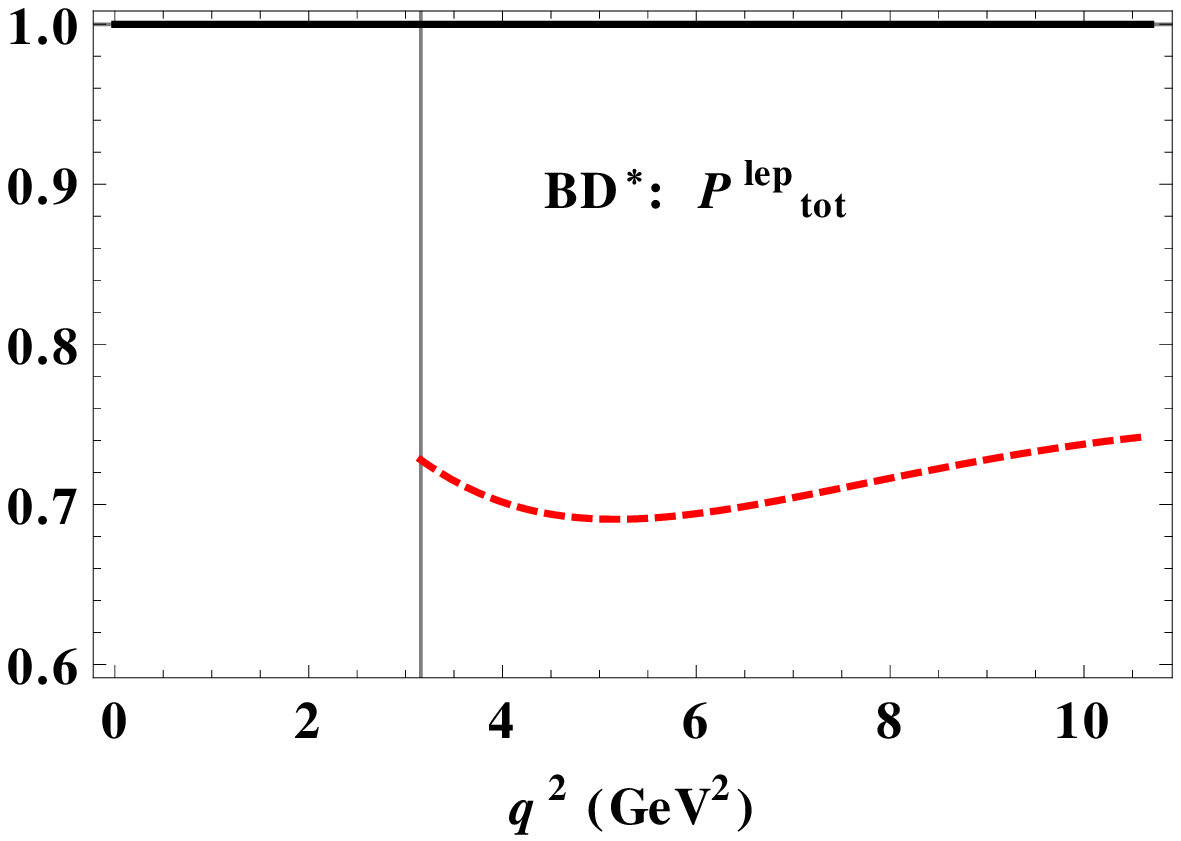,scale=.9}
\caption{ {\boldmath\bf $B\to D^\ast$~transition:} 
the  $q^{2}$ dependence of the total lepton polarization
$|\vec{P}\,^\ell|(q^{2})=\sqrt{(P^\ell_x)^2+(P^\ell_z)^2}$
for the $e^-$- (solid) and $\tau^-$-mode (dashed).
\label{fig:dPLtot.BDv}
}
\end{center}
\end{figure}

The corresponding curves for the 
$B \to D^{\ast}\ell^{-}\bar \nu_{\ell}$ transition are shown in 
Figs.~\ref{fig:dPz_Lept.BDv}, \ref{fig:dPx_Lept.BDv}, 
and \ref{fig:dPLtot.BDv}. The longitudinal and transverse polarization 
components are distinctly different from their $m_{\ell}=0$ values
$P^{\ell}_{z}=-1$ and $P^{\ell}_{x}=0$. The
longitudinal component becomes larger in magnitude when $q^{2}$
increases while the transverse polarization becomes smaller as $q^{2}$
increases. At zero recoil
the transverse polarization of the charged lepton $P^{\tau}_{x}$ tends to 
zero in agreement with the vanishing of ${\cal H}_{P}$ and ${\cal H}_{SL}$ 
at zero recoil. The total polarization of the $\tau$ shown 
in Fig.~\ref{fig:dPLtot.BDv} has an almost flat behavior with 
$|\vec P^{\ell}|\sim 0.7$. The overall picture is that the polarization is 
mostly transverse at threshold and turns to longitudinal as $q^{2}$ reaches 
the zero-recoil point.

\begin{figure}[htbp]
\begin{center}
\vspace*{.25cm}
\epsfig{figure=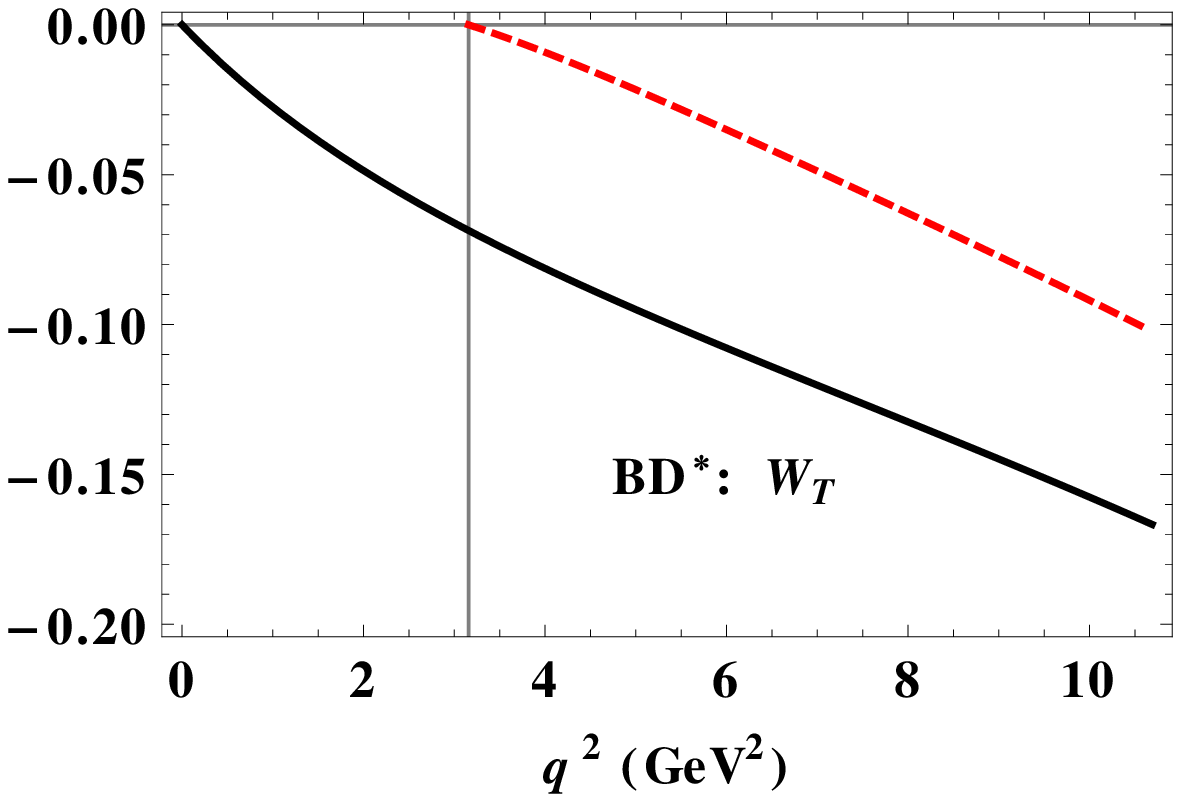,scale=.9}
\caption{ {\boldmath\bf $B\to D^\ast$~transition:} the  $q^{2}$ dependence 
of the trigonometric moment 
$W_T$ defined in Eq.~(\ref{eq:W}) for the 
$e^-$- (solid) and $\tau^-$-mode (dashed).
\label{fig:W1}
}
\end{center}
\end{figure}

\begin{figure}[htbp]
\begin{center}
\vspace*{.25cm}
\epsfig{figure=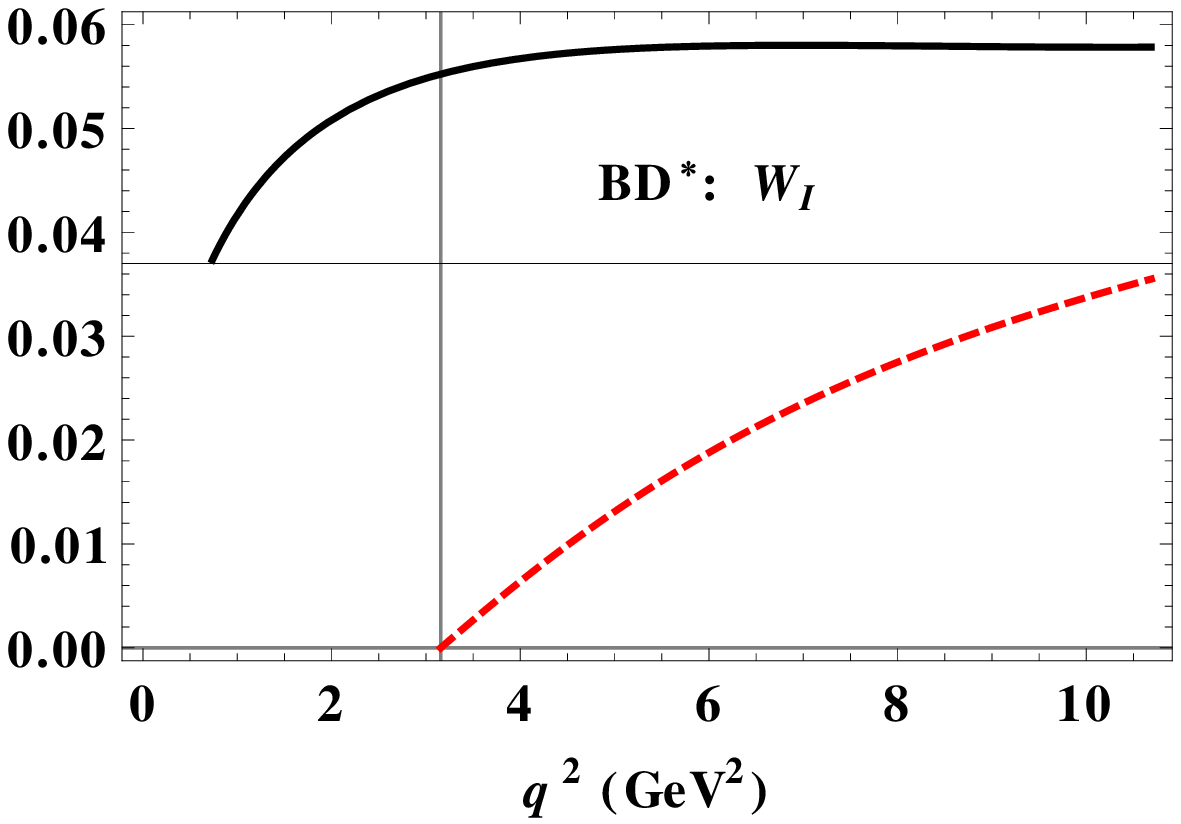,scale=.9}
\caption{ {\boldmath\bf $B\to D^\ast$~transition:} the  $q^{2}$ dependence 
of the trigonometric moment  $W_I$ defined in Eq.~(\ref{eq:W})
for the $e^-$- (solid) and $\tau^-$-mode (dashed).
\label{fig:W2}
}
\end{center}
\end{figure}

\begin{figure}[htbp]
\begin{center}
\vspace*{.25cm}
\epsfig{figure=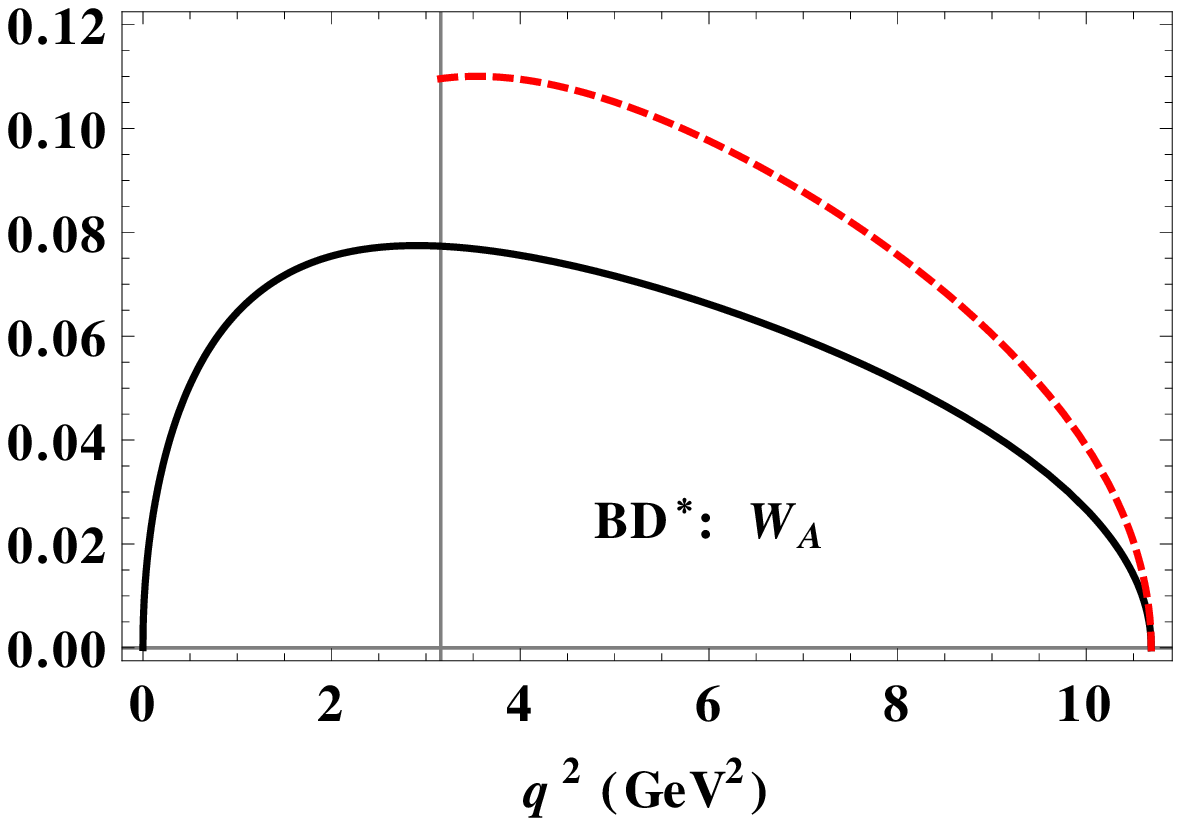,scale=.9}
\caption{ {\boldmath\bf $B\to D^\ast$~transition:} the  $q^{2}$ dependence 
of the trigonometric moment $W_A$ defined in Eq.~(\ref{eq:W}) 
for the $e^-$- (solid) and $\tau^-$-mode (dashed).
\label{fig:W3}
}
\end{center}
\end{figure}

In Figs.~\ref{fig:W1},  \ref{fig:W2},
and \ref{fig:W3} we display 
the  $q^{2}$ dependence  of the three trigonometric moments  $W_i\,(i=T,I,A)$
of the normalized three-fold angular function $\widetilde W(\theta^{\ast},
\theta,\chi)$ defined in Eq.~(\ref{eq:W}). Lepton mass effects can be seen
to be quite large for all three moments.

\begin{figure}[htbp]
\begin{center}
\vspace*{.25cm}
\epsfig{figure=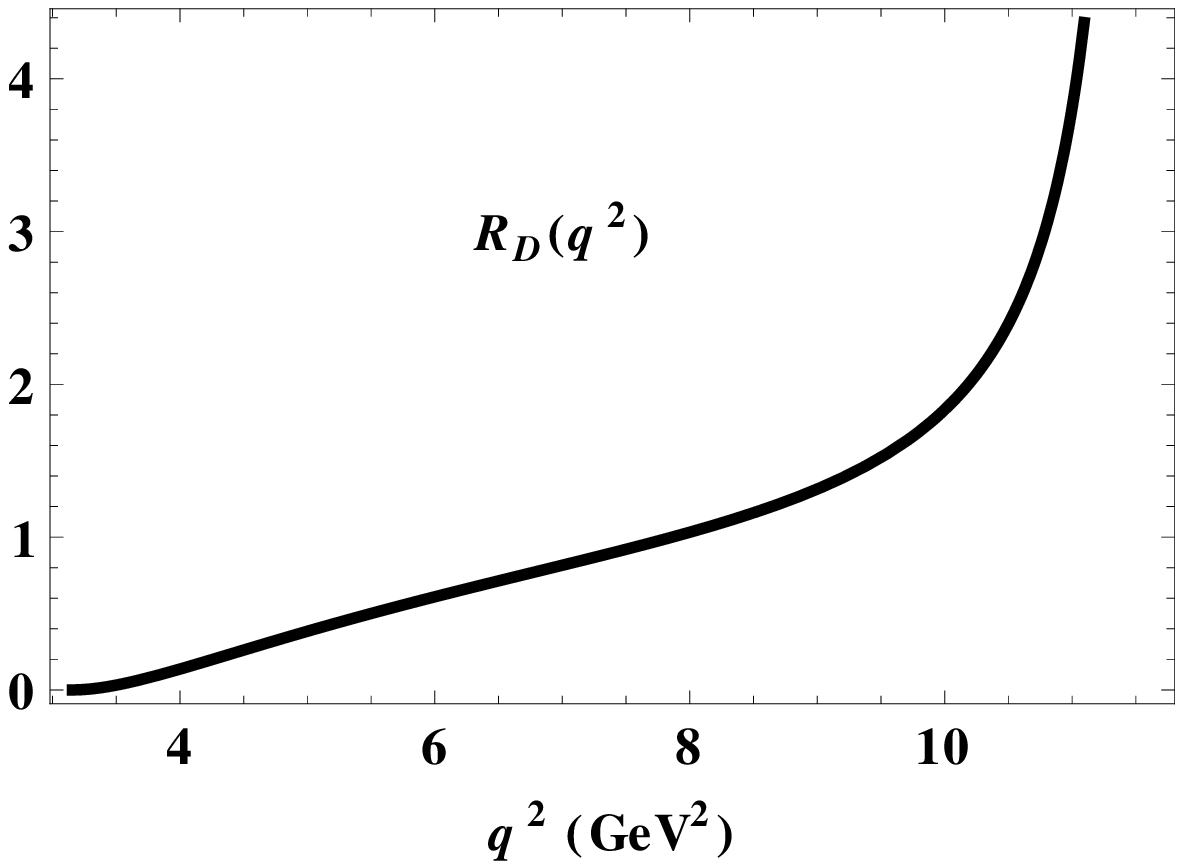,scale=.9}
\caption{The  $q^{2}$ dependence of the ratio $R(D)$.
\label{fig:RD}
}
\end{center}
\end{figure}

\begin{figure}[htbp]
\begin{center}
\vspace*{.25cm}
\epsfig{figure=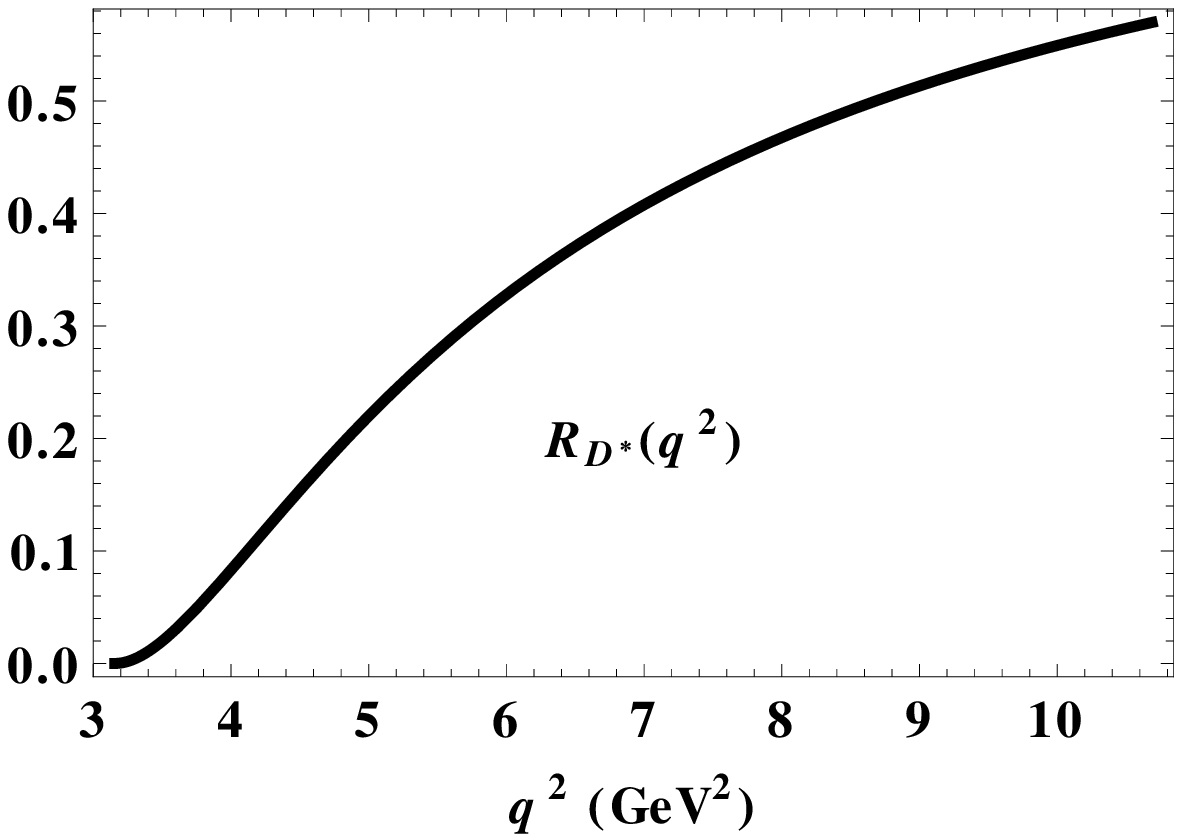,scale=.9}
\caption{The  $q^{2}$ dependence of the ratio $R(D^\ast)$.
\label{fig:RDv}
}
\end{center}
\end{figure}

Finally, in Figs.~\ref{fig:RD} and \ref{fig:RDv} we present the 
$q^{2}$ dependence
of the rate ratios ($\ell=e,\mu$)
\be
R_{D^{(\ast)}}(q^{2})=\frac{d\Gamma(B\to D^{(\ast)}\tau^{-}\bar \nu_{\tau})}
{dq^{2}}
\bigg/\,\frac{d\Gamma(B\to D^{(\ast)}\ell^{-}\bar \nu_{\ell})}{dq^{2}}.
\en
Hopefully there will be enough data in the future to explore the apparent 
flavor violation in the tauonic semileptonic $B \to D^{(\ast)}$ transitions in 
more detail by measuring the rate ratios in different $q^{2}$ bins.

Next we present our model results for the average values of the polarization 
observables:
the forward-backward asymmetry $<A_{FB}>$,  
the convexity parameter $<C_F>$, 
the leptonic  $<P^\ell_{x,z}>$ polarization components, 
and the three trigonometric moments $<W_i\,\,(i=T,I,A)$. Lepton mass effects
can be seen to be quite large for the average values
of the polarization observables. 

\clearpage

\begin{table}[htbp] 
\begin{center}
\vspace*{0.5cm}
\begin{tabular}{c|ccccccc}
\hline
\multicolumn{8}{c}{ $B\to D$ }\\
\hline
 \qquad \qquad 
&\quad $\Gamma_{L}$    \qquad   
&\quad \qquad 
&\quad  \qquad   
&\quad \qquad 
&\quad  \qquad   
&\quad \qquad 
&\quad  \qquad   
\\
\hline
\qquad $e$\qquad  \quad   
&\quad $ 11.9 $  \qquad  
&\quad  \qquad 
&\quad  \qquad  
&\quad  \qquad 
&\quad  \qquad  
&\quad  \qquad 
&\quad  \qquad  
\\
\qquad $\tau$ \qquad \quad   
&\quad $ 1.05 $ \qquad  
&\quad \qquad 
&\quad \qquad  
&\quad  \qquad 
&\quad  \qquad  
&\quad  \qquad 
&\quad  \qquad  
\\
\hline
 \qquad\qquad 
&\quad $\widetilde{\Gamma}_{L}$    \qquad 
&\quad $\widetilde{\Gamma}_{S}$    \qquad 
&\quad $\widetilde{\Gamma}_{SL}$    \qquad  
&\quad  \qquad 
&\quad  \qquad  
&\quad  \qquad 
&\quad  \qquad  
\\
\hline
\qquad $\tau$ \qquad \quad   
&\quad $ 0.25 $ \qquad  
&\quad $ 0.62 $ \qquad 
&\quad $ 0.38 $ \qquad  
&\quad  \qquad 
&\quad  \qquad  
&\quad  \qquad 
&\quad  \qquad  
\\
\hline\hline
\multicolumn{8}{c}{ $B\to D^\ast$ }\\
\hline
 \qquad \qquad 
&\quad $\Gamma_{U}$    \qquad   
&\quad $\Gamma_{L}$    \qquad 
&\quad $\Gamma_{T}$    \qquad   
&\quad $\Gamma_{I}$   \qquad   
&\quad $\Gamma_{P}$   \qquad 
&\quad $\Gamma_{A}$   \qquad 
&  \qquad \qquad 
\\
\hline
\qquad $e$\qquad  \quad   
&\quad $ 13.2 $  \qquad  
&\quad $ 15.6 $  \qquad 
&\quad $ 5.35 $  \qquad  
&\quad $ 8.94 $  \qquad 
&\quad $ -7.42$ \qquad  
&\quad $  -3.01 $  \qquad 
&\quad $ $ \qquad  
\\
\qquad $\tau$ \qquad \quad   
&\quad $ 3.02 $ \qquad  
&\quad $ 2.08 $ \qquad 
&\quad $ 1.32 $ \qquad  
&\quad $ 1.70 $ \qquad 
&\quad $-1.42 $ \qquad  
&\quad $ -0.44 $ \qquad 
&\quad $  $ \qquad  
\\
\hline
 \qquad\qquad 
&\quad $\widetilde{\Gamma}_{U}$    \qquad 
&\quad $\widetilde{\Gamma}_{L}$    \qquad 
&\quad $\widetilde{\Gamma}_{T}$    \qquad  
&\quad $\widetilde{\Gamma}_{I}$   \qquad
&\quad $\widetilde{\Gamma}_{S}$   \qquad   
&\quad $\widetilde{\Gamma}_{SL}$   \qquad
&\quad $\widetilde{\Gamma}_{ST}$   \qquad   
\\
\hline
\qquad $\tau$ \qquad \quad   
&\quad $ 0.64 $ \qquad  
&\quad $ 0.46 $ \qquad 
&\quad $ 0.27 $ \qquad  
&\quad $ 0.37 $ \qquad
&\quad $ 0.20 $ \qquad 
&\quad $ 0.29 $ \qquad
&\quad $ 0.22 $ \qquad 
\\
\hline
\end{tabular}
\caption{$q^{2}$ averages of the rate functions in units of 
$10^{-15}$~GeV. We do not display the helicity flip results for the $e$ mode 
because they are of the order of $10^{-6}-10^{-7}$ in the above units.} 
\label{tab:bilinears-numerics}
\end{center}
\end{table}

\begin{table}[htbp] 
\begin{center}
\vspace*{0.5cm}
\begin{tabular}{c|ccc}
\hline\hline
\multicolumn{4}{c}{ $B\to D$ }\\
\hline
 \qquad \qquad 
&\quad $<A_{FB}^\ell>$    \qquad   
&\quad $<C_{F}^\ell>$\qquad 
&\quad $<C_{F}^h>$ \qquad   
\\
\hline
\qquad $e$\qquad  \quad   
&\quad $ -1.17\,(-1.16)\cdot 10^{-6} $  \qquad  
&\quad $ -1.5\,(-1.5) $ \qquad 
&\quad $  3\,(3) $ \qquad  
\\
\qquad $\tau$ \qquad \quad   
&\quad $ -0.36\,(-0.36) $ \qquad  
&\quad $ -0.26\,(-0.26) $\qquad 
&\quad  $  3\,(3) $ \qquad  
\\
\hline
&\quad  $<P_{z}^\ell>$\qquad 
&\quad  $<P_{x}^\ell>$ \qquad   
&\quad  $<|\vec P^\ell |>$ \qquad  
\\
\hline
\qquad $e$\qquad  \quad   
&\quad $ -1\,(-1) $ \qquad 
&\quad $ 0\,(0) $  \qquad  
&\quad $ 1\,(1) $  \qquad  
\\ 
\qquad $\tau$ \qquad \quad   
&\quad $ 0.33\,(0.33) $  \qquad 
&\quad $ 0.84\,(0.84) $ \qquad  
&\quad $ 0.91\,(0.91) $ \qquad  
\\
\hline
\multicolumn{4}{c}{ $B\to D^\ast$ }\\
\hline
 \qquad \qquad 
&\quad $<A_{FB}^\ell>$    \qquad   
&\quad $<C_{F}^\ell>$\qquad 
&\quad $<C_{F}^h>$ \qquad   
\\
\hline
\qquad $e$\qquad  \quad   
&\quad $ 0.19\,(0.18) $  \qquad  
&\quad $ -0.47\,(-0.44) $ \qquad 
&\quad $ 0.93\,(0.88) $ \qquad  
\\
\qquad $\tau$ \qquad \quad   
&\quad $ 0.027\,(0.021) $ \qquad  
&\quad $ -0.062\,(-0.057) $ \qquad 
&\quad $ 0.58\,(0.52) $  \qquad  
\\
\hline
&\quad  $<P_{z}^\ell>$\qquad 
&\quad  $<P_{x}^\ell>$ \qquad   
&\quad  $<|\vec P^\ell |>$ \qquad   
\\
\hline
\qquad $e$\qquad  \quad   
&\quad $ -1\,(-1) $\qquad 
&\quad $ 0\,(0) $ \qquad  
&\quad $ 1\,(1) $ \qquad  
\\
\qquad $\tau$ \qquad \quad   
&\quad  $ -0.50\,(-0.51) $ \qquad 
&\quad  $ 0.46\,(0.43) $ \qquad  
&\quad  $ 0.71\,(0.71) $ \qquad  
\\
\hline
\qquad \qquad 
&\quad $<W_T>$    \qquad   
&\quad $<W_I>$\qquad 
&\quad $<W_A>$\qquad   
\\
\hline
\qquad $e$\qquad  \quad   
&\quad $ -0.093\,(-0.098) $  \qquad  
&\quad $  0.054\,(0.055) $ \qquad 
&\quad $  0.062\,(0.059) $ \qquad  
\\
\qquad $\tau$ \qquad \quad   
&\quad $ -0.057\,(-0.059) $ \qquad  
&\quad $  0.025\,(0.025) $\qquad 
&\quad $  0.077\,(0.074) $ \qquad  
\\
\hline\hline
\end{tabular}
\caption{$q^{2}$ averages of polarization observables. For comparison with 
results from the HQL, we add in brackets the corresponding HQL values. 
}
\label{tab:obs-numerics}
\end{center}
\end{table}

\clearpage

\section{Summary and conclusions}
\label{sec:summary}
We have provided a detailed analysis of the pure leptonic and semileptonic decays 
$B \to \ell^{-}\bar \nu_{\ell}$ and
$B \to D^{(\ast)} \ell^{-}\bar \nu_{\ell}$ $(\ell=e,\mu,\tau)$ within the
SM in the framework of our covariant quark model with built-in quark 
confinement. We have described in some detail how to compute the  
one-loop quark contributions needed for the calculation of the transition form factors including a discussion of how the confinement of the
constituent quarks is achieved in the covariant quark model.
In the light of the recent experimental indications for a possible breaking of 
lepton universality in the $\tau$ sector we have put
particular emphasis on how to isolate heavy lepton mass effects in the semileptonic 
decays.

We have described how to obtain the full angular decay
distributions for $B \to D\ell^{-}\bar \nu_{\ell}$ and the cascade decay
process $B \to D^{\ast}(\to D\pi) \ell^{-}\bar \nu_{\ell}$ as well as the 
corresponding
angular decay distributions for their charge-conjugate processes.
The coefficients multiplying the angular factors in the angular decay 
distributions have been given in terms of
helicity structure functions for which we have provided simple expressions
for the maximal recoil $q^{2}=0$ and the minimal (zero) recoil 
$q^{2}=(m_{1}-m_{2})^{2}$.
Starting from the angular decay distributions we have defined a multitude of 
polarization observables for which we have provided numerical results on their
$q^{2}$ spectra and their $q^{2}$ averages for zero and nonzero lepton masses.
The polarization observables include the transverse and
longitudinal polarizations of the charged $\tau^{-}$ which considerably deviate
from their simple $m_{\ell}=0$ left-chiral structure.

We are looking forward to a wealth of data on these
decays expected in the near future which will allow one to
deeply probe into their decay structure, in particular for the tauonic mode.
Such an analysis will reveal possible deviations
from the SM predictions not only in the branching fractions of the
processes but also in the 
multitude of polarization observables and their $q^{2}$ spectra.

\begin{acknowledgments}

M.A.I.\ acknowledges the support from Mainz Institute for Theoretical 
Physics (MITP). M.A.I. and J.G.K. thank the Heisenberg-Landau Grant for providing support for their collaboration.  

\end{acknowledgments}

\appendix

\section{Spin kinematics} 

In this Appendix we provide a synopsis of how to obtain the angular
decay distributions for the decays $B \to D^{(\ast)} \ell^{-} \bar\nu_{\ell}$
following the description in~\cite{Gutsche:2015mxa,Berge:2015jra}.
The covariant representation of the angular decay distribution is given
by
\begin{equation}
\label{angdistcov}
W'(\theta^{\ast},\theta,\chi)=h_{\alpha'\beta'}\,P_{1}^{\alpha'\alpha}
P_{1}^{\beta'\beta}\,H_{\alpha\mu}H_{\beta\nu}^{\dagger}\,
P_{0\oplus 1}^{\mu\mu'}P_{0\oplus 1}^{\nu\nu'}\,L_{\mu'\nu'},
\end{equation}
where $h_{\alpha'\beta'}$ is the hadronic decay tensor for the decay
$D^{\ast} \to D \,+\pi$ 
(with $h^{\alpha\beta}=p_{3}^{\alpha}p_{3}^{\beta}/|\mathbf{p_3}|^{2}$), 
$H_{\alpha\mu}H_{\beta\nu}^{\dagger}$ is the
tensor describing the decay $B \to D^{\ast}+W^{-}_{off-shell}$, and
$L_{\mu'\nu'}$ is the lepton tensor describing the decay 
$W^{-}_{\rm off-shell}\to \ell^{-}+\bar \nu_{\ell}$. The tensors are connected
by the appropriate spin 1 and spin $(0\oplus1)$ propagator projectors 
$P_{1}^{\mu\nu}(q)= -g^{\mu\nu}+q^{\mu}q^{\nu}/q^{2}$ and
$P^{\mu\nu}_{0\oplus1}(q)$ which, in the unitary gauge, reads \footnote{
The choice of the unitary gauge is dictated by electroweak gauge invariance.} 
\begin{eqnarray}\label{split}
P_{0\oplus1}^{\nu\beta}(q)&=&-g^{\nu\beta}+\frac{q^{\nu}q^{\beta}}{m_{W}^{2}}
=\bigg(\underbrace{-g^{\nu\beta}+\frac{q^{\nu}q^{\beta}}{q^{2}}}_{\rm spin\,1}
\bigg)-\underbrace{\frac{q^{\nu}q^{\beta}}{q^{2}}
(1-\frac{q^{2}}{m^{2}_{W}})}_{\rm spin\,0}, \\
&=& P_{1}^{\nu\beta}(q) - (1-\frac{q^{2}}{m^{2}_{W}}) P_{0}^{\nu\beta}(q),
\end{eqnarray} 
where  $P_{0}^{\mu\nu}(q)=q^{\mu}q^{\nu}/q^{2}$ is the spin 0 propagator.
The factor $(1-q^{2}/m^{2}_{W})$ multiplying the spin 0 propagator in
Eq.(\ref{split}) is usually set to 1 in low 
energy applications as in the decay 
$B \to D^{\ast} \ell^{-} \bar\nu_{\ell}$. For example, at the highest 
$q^{2}$ value at
$q^{2}=(m_{B}-m_{D^{\ast}})^{2}$ the correction amounts to a mere $0.17\,\%$
and will therefore be dropped in the following.

In order to convert the covariant representation of the angular
decay distribution Eq.~(\ref{angdistcov}) to the helicity representation 
one makes use of the completeness relations 
\bea
P_{0\oplus1}^{\mu\nu}(q)&=&-g^{\mu \nu}+\frac{q^{\nu}q^{\beta}}{m_{W}^{2}} \,
= -\sum_{m,m'=t,\pm,0} 
\varepsilon^\mu(m) \varepsilon^{\dagger\, \nu }(m') g_{m m'}, \\
P_{1}^{\mu\nu}(q)&=&-g^{\mu \nu}+q^{\mu}q^{\nu}/q^{2} \,= \sum_{m,m'=\pm,0} 
\varepsilon^\mu(m) \varepsilon^{\dagger\, \nu }(m'),
\label{eq:completeness}
\ena 
where the tensor $ g_{mm'} = \mbox{diag} (+,-,-,-) $ is the spherical 
representation
of the metric tensor whose components are ordered in the sequence 
$m,m'=t,+,0,-$. 
With the help of the completeness relation (\ref{eq:completeness}) one can
convert the covariant form of the angular decay 
distribution Eq.~(\ref{angdistcov}) into the helicity form 
\begin{equation}
\label{angdisthel}
W'(\theta^{\ast},\theta,\chi)=\sum_{J,J',\lambda_{W},\lambda'_{W},
\lambda_{2},\lambda'_{2}}
(-1)^{J+J'}\delta_{\lambda_{2}\lambda_{W}}
\delta_{\lambda_{2}'\lambda_{W}'} h_{\lambda_{2}\lambda_{2}'}(\theta^{\ast})
H_{\lambda_{2}\lambda_{W}}(J)H_{\lambda_{2}'\lambda_{W}'}^{\ast}(J')
L_{\lambda_{W}\lambda_{W}'}(J,J',\theta,\chi).
\end{equation}
In~(\ref{angdisthel}) we have chosen a representation of the helicity
amplitudes which is particularly well suited for
computer processing. Compared to the helicity amplitudes introduced in the
main text we have used 
$H_{0\,\lambda_{W}=0}\,(J=0)\equiv\,H_{0\,t}$
and 
$H_{0,\pm1\,\,\lambda_{W}=0,\pm}(J=1)
\equiv\,H_{0,\pm1 \,\,0,\pm1}$.

The helicity representation of the hadronic decay tensor 
$h_{\alpha\,\beta}(\theta^{\ast})$
describing the decay $D^{\ast}\to D \pi$ is given by
\be
h_{\lambda_{2}\lambda'_{2}}(\theta^{\ast}) = 
d^{1}_{0\,\lambda_{2}}(\theta^{\ast}) 
d^{1}_{0\,\lambda'_{2}}(\theta^{\ast})  
=\left( \begin{array}{ccc} 
\frac 12\sin^{2}\theta^{\ast} & \frac{1}{2\sqrt{2}}\sin2\theta^{\ast} & 
-\frac 12 \sin^{2}\theta^{\ast} \\
 +\frac{1}{2\sqrt{2}}\sin2\theta^{\ast} &  \cos^{2}\theta^{\ast} &
 -\frac{1}{2\sqrt{2}}\sin2\theta^{\ast}  \\
-\frac 12\sin^{2}\theta^{\ast} &  -\frac{1}{2\sqrt{2}}\sin2\theta^{\ast} 
& \frac 12\sin^{2}\theta^{\ast}
 \\
\end{array} \right).
\en

For the helicity representation of the lepton tensor 
one obtains $(v=1-m^{2}_{\ell}/q^{2})$~\cite{Gutsche:2015mxa} 
\bea 
\label{lt1}
&&(2q^2v)^{-1} L_{\lambda_{W}\lambda'_{W}}(1,1,\theta,\chi) = \nn 
&&\hspace{-2.0cm}\left( \begin{array}{ccc} 
 (1\mp\cos\theta)^2 & \mp \frac{2}{\sqrt{2}} (1\mp\cos\theta) \sin\theta 
e^{i\chi} & \sin^{2}\theta e^{2i\chi} \\
 \mp \frac{2}{\sqrt{2}} (1\mp\cos\theta) \sin\theta 
e^{-i\chi} & 2\sin^2\theta 
& \mp \frac{2}{\sqrt{2}} (1\pm\cos\theta) \sin\theta 
e^{i\chi} \\
 \sin^{2}\theta e^{-2i\chi} & \mp \frac{2}{\sqrt{2}} (1\pm\cos\theta) \sin\theta 
e^{-i\chi} & (1\pm\cos\theta)^2 \\
\end{array} \right)\nonumber\\
&+&\delta_\ell 
\left( \begin{array}{ccc} 
2 \sin^2\theta & 
- \frac{2}{\sqrt{2}} \sin 2\theta e^{i\chi} & -2\sin^{2}\theta e^{2i\chi} \\
 
- \frac{2}{\sqrt{2}} \sin 2\theta e^{-i\chi} & 
4\cos^2\theta & 
\frac{2}{\sqrt{2}} \sin 2\theta e^{i\chi} \\
 -2\sin^{2}\theta e^{-2i\chi} & 
\frac{2}{\sqrt{2}} \sin 2\theta e^{-i\chi} & 
2\sin^2\theta \\
\end{array} \right).
\ena
The upper/lower signs in the nonflip part of~(\ref{lt1}) stand for the 
$(\ell^{-}\bar \nu_{\ell})$ case relevant for the decays 
$\bar B^{0} \to D^{(\ast)\,+}\ell^{-}\bar \nu_{\ell}$ 
and $B^{-} \to D^{(\ast)\,0}\ell^{-}\bar \nu_{\ell}$,
and the $(\ell^{+} \nu_{\ell})$ case relevant for the decays
$B^{+} \to \bar D^{(\ast)\,0}\ell^{+}\nu_{\ell}$ and
$B^{0} \to \bar D^{(\ast)\,-}\ell^{+}\nu_{\ell}$ . 
The spin 0/spin 1 interference contribution is given by
\bea
\label{lt2}
(2q^2v)^{-1} L_{0\,\lambda_{W}}(0,1,\theta,\chi)=
&&(2q^2v)^{-1} L_{\lambda_{W},0}^{\ast}(1,0,\theta,\chi)= \nn
&&\hspace{1.0cm}\delta_{\ell}\,\left( \begin{array}{ccc} 
-\frac{4}{\sqrt{2}}\sin\theta e^{-i\chi} &\quad4\cos\theta&
\quad\frac{4}{\sqrt{2}}\sin\theta e^{i\chi} \\
\end{array}
\right)\,,
\ena
($\lambda_{W}=1,0,-1$) and
\be
\label{lt3}
(2q^2v)^{-1} L_{0\,0}(0,0,\theta,\chi)=\,4\,\delta_{\ell}.
\en 
For the $\cos\theta$ distribution of the decay 
$B \to D^\ast  \ell^- \bar\nu_\ell$ written down in 
Eq.~(\ref{eq:distr2}) one
needs the integrated form of Eq.~(\ref{angdisthel}). One obtains
\bea
\label{distheta}
&&W'(\theta)
=\int d\cos\theta^{\ast}d\chi /2\pi W'(\theta^{\ast},\theta,\chi)= \nn
&&2\,\sum_{J,J',\lambda_{W},
\lambda_{2}}
(-1)^{J+J'}\delta_{\lambda_{2}\lambda_{W}}
H_{\lambda_{2}\lambda_{W}}(J)H_{\lambda_{2}\lambda_{W}}^{\ast}(J')
L_{\lambda_{W}\lambda_{W}}(J,J',\theta),
\ena
where
\be
\label{intlep}
L_{\lambda_{W}\lambda_{W}}(J,J',\theta)=
\int d\chi/2\pi L_{\lambda_{W}\lambda_{W}}(J,J',\theta,\chi).
\en
The integration~(\ref{intlep}) is easily done. The result is
given by Eqs.~(\ref{lt1},\ref{lt2})
where all terms proportional to $e^{\pm i\chi},e^{-\pm 2i\chi}$ have
been dropped. The $\cos\theta$ distribution for 
$B \to D \ell^{-}\bar \nu_{\ell}$ written down in Eq.~(\ref{eq:distr2}) is
obtained from~(\ref{distheta}) by omitting $\delta_{\lambda_{2}\lambda_{W}}$
and dropping the label~$\lambda_{2}$.

 \ed